%% file: MAIN.tex
\documentclass[manuscript,screen]{acmart}

\usepackage[utf8]{inputenc} 
\usepackage{csquotes}
\usepackage{multirow}
\usepackage[most]{tcolorbox}
\usepackage{tabularx}
\usepackage{xspace}
\usepackage[inline]{enumitem}
\usepackage{standalone}
\usepackage{booktabs}

\usepackage{tikz}
\usetikzlibrary{positioning, backgrounds, arrows.meta}
\usepackage{multicol} 
\usepackage{calc} 

\usepackage{booktabs}
\usepackage{tabularx}

\usepackage{listings}
\usepackage{pifont}
\usepackage{cleveref}

\usepackage[table]{xcolor}
\usepackage{comment}

\newcommand{\ai}{GenAI\xspace}
\newcommand{\sdlc}{SDLC\xspace}

\newcommand{\copilot}{\ai{} Copilot\xspace}
\newcommand{\robot}{\ai{} Robot\xspace}
\newcommand{\aiware}{\ai{}ware\xspace}
\newcommand{\teammate}{\ai Teammate\xspace}

\definecolor{myblue}{RGB}{70,100,170}
\definecolor{mygreen}{RGB}{0,150,130}
\definecolor{myorange}{RGB}{223,155,27}

\usepackage{ulem}
\newboolean{showcomments}
\setboolean{showcomments}{true} 

\ifthenelse{\boolean{showcomments}}
{

}

\newcommand{\cmark}{\ding{51}}

\AtBeginDocument{%
  }

\setcopyright{acmlicensed}
\copyrightyear{2025}
\acmYear{2025}
\acmDOI{XXXXXXX.XXXXXXX}
\acmConference[TOSEM-SI]{"Software Engineering 2030"}{NN}

\begin{document}



\title{A Research Roadmap for Augmenting Software Engineering Processes and Software Products with Generative AI}

\author{Domenico Amalfitano}
\email{domenico.amalfitano@unina.it}
\orcid{0000-0002-4761-4443}
\affiliation{%
  \institution{University of Naples "Federico II"}
  \city{Napoli}
  \country{Italy}
}

\author{Andreas Metzger}
\email{andreas.metzger@paluno.uni-due.de}
\orcid{0000-0002-4808-8297}
\authornote{Corresponding author.}
\affiliation{%
  \institution{Ruhr Institute for Software Technology (paluno), University of Duisburg-Essen}
  \city{Essen}
  \country{Germany}
}


\author{Marco Autili}
\email{marco.autili@univaq.it}
\orcid{0000-0001-5951-1567}
\affiliation{%
  \institution{Università degli Studi dell'Aquila}
  \city{L'Aquila}
  \country{Italy}
}

\author{Tommaso Fulcini}
\email{tommaso.fulcini@polito.it}
\orcid{0000-0001-8765-6501}
\affiliation{%
  \institution{Politecnico di Torino}
  \city{Turin}
  \country{Italy}
}

\author{Tobias Hey}
\email{hey@kit.edu}
\orcid{0000-0003-0381-1020}
\affiliation{%
  \institution{Karlsruhe Institute of Technology (KIT)}
  \city{Karlsruhe}
  \country{Germany}
}

\author{Jan Keim}
\email{jan.keim@kit.edu}
\orcid{0000-0002-8899-7081}
\affiliation{%
  \institution{Karlsruhe Institute of Technology (KIT)}
  \city{Karlsruhe}
  \country{Germany}
}

\author{Patrizio Pelliccione}
\email{patrizio.pelliccione@gssi.it}
\orcid{0000-0002-5438-2281}
\affiliation{%
  \institution{Gran Sasso Science Institute (GSSI)}
  \city{L'Aquila}
  \country{Italy}
}

\author{Vincenzo Scotti}
\email{vincenzo.scotti@kit.edu}
\orcid{0000-0002-8765-604X}
\affiliation{%
  \institution{Karlsruhe Institute of Technology (KIT)}
  \city{Karlsruhe}
  \country{Germany}
}


\author{Anne Koziolek}
\email{anne.koziolek@kit.edu}
\orcid{0000-0002-1593-3394}
\affiliation{%
  \institution{Karlsruhe Institute of Technology (KIT)}
  \city{Karlsruhe}
  \country{Germany}
}

\author{Raffaela Mirandola}
\email{raffaela.mirandola@kit.edu}
\orcid{0000-0003-3154-2438}
\affiliation{%
  \institution{Karlsruhe Institute of Technology (KIT)}
  \city{Karlsruhe}
  \country{Germany}
}

\author{Andreas Vogelsang}
\email{andreas.vogelsang@uni-due.de}
\orcid{0000-0003-1041-0815}
\affiliation{%
  \institution{Ruhr Institute for Software Technology (Paluno), University of Duisburg-Essen}
  \city{Essen}
  \country{Germany}
}

\renewcommand{\shortauthors}{D. Amalfitano, A. Metzger, M. Autili, T. Fulcini, T. Hey, J. Keim, P. Pelliccione, V. Scotti, A. Koziolek, et al.}

\begin{abstract}
\textbf{Abstract.} Generative AI (GenAI) is rapidly transforming software engineering (SE) practices, influencing how SE processes are executed, as well as how software systems are developed, operated, and evolved. This paper applies  design science research to build a roadmap for GenAI-augmented SE. The process consists of three cycles that incrementally integrate multiple sources of evidence, including collaborative discussions from the FSE 2025 “Software Engineering 2030” workshop, rapid literature reviews, and external feedback sessions involving peers. McLuhan’s tetrads were used as a conceptual instrument to systematically capture the transforming effects of GenAI on SE processes and software products.
The resulting roadmap identifies four fundamental forms of GenAI augmentation in SE and systematically characterizes their related research challenges and opportunities. These insights are then consolidated into a set of future research directions.
By grounding the roadmap in a rigorous multi-cycle process and cross-validating it among independent author teams and peers, the study provides a transparent and reproducible foundation for analyzing how GenAI affects SE processes, methods and tools, and for framing future research within this rapidly evolving area.
\end{abstract}






\maketitle

\section{Introduction}
\label{sec:intro}

Software engineering (SE) is experiencing a transformation of unprecedented speed and scale, driven by the growing augmentation in software engineering \textit{processes} and software \textit{products} with Generative Artificial Intelligence (\ai{})~\cite{TOSEMPezze2025,PezzeCGPQ24,AhmedACCHHPPX25}.
Such \ai{}-augmentation offers unprecedented opportunities for efficiently and effectively developing and operating novel kinds of software systems and applications, but at the same time challenges established principles, practices, tasks, activities and processes of software engineering~\cite{AhmedACCHHPPX25,Terragni24,10.1016/j.infsof.2025.107751,10.1145/3712005}.
It requires rethinking existing software development lifecycle models (SDLC models), particularly questioning long-held assumptions about the roles of software engineers, the dynamics of hybrid human-AI teams, and the very nature of software artifacts, which in addition to code and data, now include \ai models and prompts.

We systematically identify those challenges and opportunities by starting from a structuring of the main forms \ai{} -augmentation in SE can take.
This structuring consists of two dimensions that answer the following questions.

\textbf{"What is augmented by \ai{}?"}
    This dimension creates a distinction between using \ai{} to augment SE \textit{processes} versus using \ai{} to augment software \textit{products}:
    \begin{itemize}
        \item \ai{}-augmented SE \textit{processes} mean that \ai{} is used to automate software engineering activities and tasks.
        Examples range from requirements analysis, through automatic code completion and code generation, via test case generation and test case prioritization, to optimized CI/CD pipelines and continuous code refactoring and maintenance~\cite{AhmedACCHHPPX25,10.1145/3695988,10.1145/3696630.3730538,10.1145/3695988,10.1145/3696630.3728718,10.1007/s10515-024-00426-z,10.1145/3715003}.
        
        \item \ai{}-augmented software \textit{products} mean that parts of the functionality of a software system or application are not explicitly programmed but are realized with \ai.
        Such \ai{}-augmented software products can generate novel content, power sophisticated conversational interfaces, or create adaptive user experiences~\cite{HassanLRGC00TOL24,DBLP:conf/cain/00010XLX024,11121725,weber2024}.
    \end{itemize}
    
\textbf{"How autonomous is the \ai{} augmentation?"} 
    This dimension creates a distinction of whether \ai{} plays a \textit{passive} or an \textit{active} role:
    \begin{itemize}
        \item In a \textit{passive} role, \ai does not have goals or act on its own. 
        Instead, \ai responds reactively to user prompts or inputs provided via an API. 
        Or, in other words, the human-AI interaction is triggered by the human~\cite{weber2024,HassanLRGC00TOL24}.
        A simple example is an FAQ chat interface included in a website.
        
        \item In an \textit{active} role, \ai{} possesses its own thread of control and makes (semi)autonomous decisions about which actions to perform and when~\cite{Vu25,li2025,LiZH25}.
        This means \ai{} not only processes information or generates content, but can also proactively make decisions and perform activities and tasks.
        Or, in other words, the human-AI interaction is triggered by AI.
        A typical manifestation of such an active role is Agentic AI, which refers to a collection of (semi)autonomous AI agents that jointly aim to achieve higher-level goals~\cite{Wooldridge2009,Sapkota25,Jennings00}.
   \end{itemize}

The intersection of these two dimensions results in four distinct forms of \ai{} augmentation in SE.
On the one hand, these four forms provide us with a clear and concise structure to analyze the state-of-the-art and provide a research roadmap for form-specific research challenges and opportunities.
Hence, these four forms provide more rigor than loose terms such as ``AI for SE'', which may refer to using AI in any form to enhance SE processes, or ``SE for AI'', which may refer to extending established SE principles and techniques to the unique challenges posed by developing AI models or AI-augmented systems and applications.
On the other hand, these four forms help us achieve a more systematic and comprehensive coverage of \ai{} augmentation in SE, thereby also allowing us to identify cross-form research challenges and opportunities.

Overall, we make the following main contributions to SE research:
\begin{itemize}
    \item \textbf{Classification of \ai Augmentation}: We provide a classification of \ai in SE based on the aforementioned two dimensions, resulting in four distinct forms: GenAI Copilot, GenAIware, GenAI Teammate, and GenAI Robot.

    \item \textbf{Analysis of \ai Effects:} We systematically capture the effects of \ai on SE for each of the above forms, determining what \ai enhances, reverses, retrieves from the past, and makes obsolete.

    \item \textbf{Comprehensive Research Roadmap:} We distill those findings into a detailed roadmap identifying form-specific and cross-cutting research challenges and opportunities. 
    Key areas include prompt engineering, accountability, hybrid human-AI team dynamics, and the coordination of process-level versus product-level agents.

    \item \textbf{Predictions for 2030:} We conclude with ten strategic predictions for how SE may look like in the year 2030. 
    This includes the death of manual coding for routine tasks and the shift from the role of SE developer to AI agent orchestrator.
\end{itemize}

We structure the remainder of the article as follows.
In Section~\ref{sec:methodology}, we detail the methodology followed during our review and analysis.
In Section~\ref{sec:foundations}, we elaborate on the four forms of \ai{} augmentation and discuss related research roadmaps.
In Sections~\ref{sec:copilot} through~\ref{sec:robots}, we present the state of the art and a detailed analysis of the impact of \ai{} augmentation for each of the four forms.
In Section~\ref{sec:commonalities}, we distill a roadmap composed of a set of research challenges and opportunities.
In Section~\ref{sec:validity} we discuss validity risks.
In Section~\ref{sec:conclusion}, we conclude the paper.

\section{Methodology}
\label{sec:methodology}
\begin{figure}[htb]
    \centering
    \includegraphics[width=0.9\linewidth]{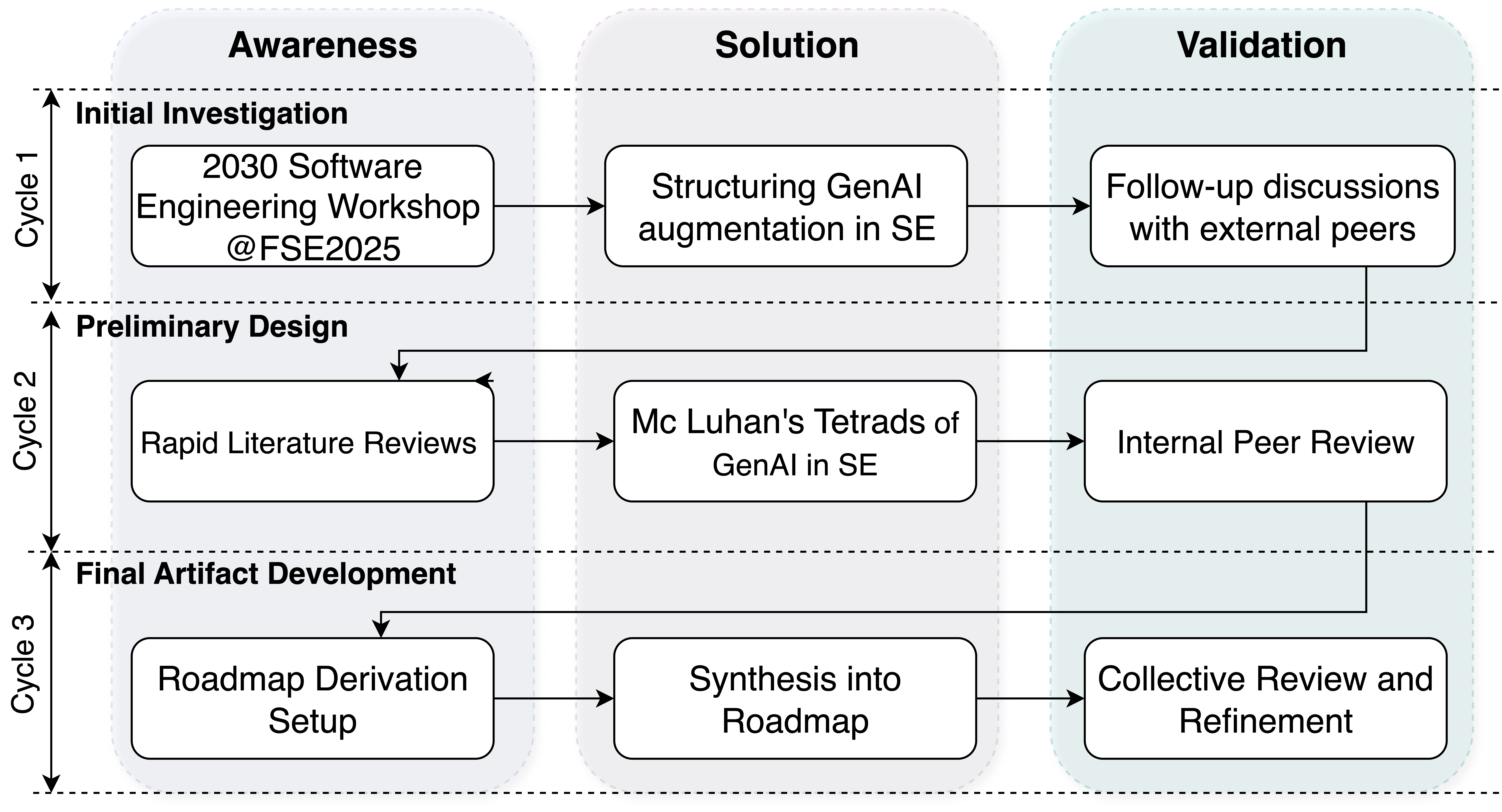}
    \caption{The design science approach for building a roadmap  on \ai{}-augmented processes and products in SE}
     \Description{The design science process structured into three iterative cycles.}
    \label{fig:designscience}
\end{figure}

This section presents the methodology followed to perform this study.
We adopted the Design Science Research (DSR) approach~\cite{wieringa:2009, baskerville:2018, hevner:2021}, as our primary goal was to design a well-founded \textit{artifact} that supports understanding and future development of the GenAI-driven Software Development Life Cycle (SDLC). 
In our case, the artifact is the \textit{Roadmap for the SDLC in the GenAI era}, which was progressively derived through a sequence of McLuhan Tetrads capturing challenges, limitations, and opportunities observed in current software engineering practices.
Figure~\ref{fig:designscience} visualizes the overall DSR process, which was structured into three cycles.
Each cycle is characterized by distinct tasks and produces specific outputs that progressively shape the final roadmap.

\subsection{Cycle 1 -- Initial Investigation}
\label{sec:cycle1}
The first cycle was intended to create initial awareness and structure the problem space.

\subsubsection{Awareness - 2030 Software Engineering Workshop @FSE2025}
\label{sec:workshop}
The inspiration and basis for this special issue article was the ``2030 Software Engineering'' workshop\footnote{\url{https://conf.researchr.org/home/2030-se-2025}},
co-located with the ACM SIGSOFT FSE 2025 conference (June 26--27, Trondheim, Norway), in which most of the authors of this article participated.
The workshop adopted the format of ``liberating structures''\footnote{\url{https://www.liberatingstructures.com/}},
leading to two full days of intensive discussions on topics including AI for software engineering,
software engineering for AI, sustainable software engineering and quantum software engineering.

\subsubsection{Solution - Structuring \ai augmentation in SE}
\label{sec:str_and_limit}
To facilitate a systematic discussion of the impact of \ai{} augmentation, an initial structure was defined for the different forms of \ai{} augmentation in SE.
The structure was informed by discussions and insights from the FSE 2025 workshop and complemented by earlier efforts to classify \ai{} augmentation in SE, notably the taxonomy by CMU's Software Engineering Institute (SEI) in 2023~\cite{ozkaya_2023}.

This initial effort resulted in a structure for \ai{} augmentation in SE along the two dimensions into four distinct forms as introduced in Section~\ref{sec:intro}.
These four forms helped us analyze related roadmaps and understand how our article may refine these.
Additionally, these four forms provided the structure for Cycle 2, where rapid literature reviews were conducted to collect evidence and characterize each of these four forms in detail.

\subsubsection{Validation - Follow-up Discussions with External Peers}
\label{sec:external_discussion}
The resulting structure was validated through discussions with seven fellow SE professors from the Ruhr Institute for Software Technology (paluno\footnote{\url{https://paluno.uni-due.de/en/}}) during a two-day research retreat.

\subsection{Cycle 2 -- Preliminary Design}
\label{sec:cycle2}
Cycle 2 focused on the collection of evidence through rapid literature reviews and the determination of the impact of \ai augmentation using McLuhan tetrads.

\subsubsection{Awareness - Rapid Literature Reviews}
\label{sec:rlr}

When the research focus is well defined and unambiguous, as in our case, rapid literature reviews (RLRs) constitute pragmatic evidence synthesis methods designed to deliver timely and resource-efficient insights compared to full-fledged Systematic Literature Reviews (SLRs) or Systematic Mapping Studies (SMSs)~\cite{cartaxo2018}. 
Unlike SLRs and SMSs, RLRs typically relax some methodological requirements, such as triangulation among multiple researchers and systematic validation of study selection and data extraction steps. 
Instead, these activities are often conducted by one or at most two researchers, which reduces the possibility of cross-checking but allows the review to be completed within a short time frame (usually one or two months). 
Despite this lower degree of formality, prior work in software engineering has shown that RLRs still provide valuable evidence to support decision-making and efficiently characterize emerging research domains~\cite{cartaxo2018, Amalfitano_rapidreview}. 
We adopted this approach because, after the FSE workshop held at the end of June (see Section~\ref{sec:workshop}), only a few months remained to obtain an initial evidence-based overview of the four fundamental forms of \ai{} augmentation in SE. 
Consequently, small teams of one or two authors conducted focused RLRs, each team addressing one fundamental form, and completed the activity within two months. 
This pragmatic choice allowed us to align the literature analysis with the design science process, ensuring that each fundamental form was grounded on an initial evidence base despite the short time frame.
To provide additional transparency and replicability of the RLR process, we provide open access to the data of our literature reviews~\cite{Zenodo}. 

\paragraph{Research Protocol.}
All rapid literature reviews followed a shared baseline protocol, which was then specialized by each group to fit the specific characteristics of the form under investigation. 
The protocol consisted of the following steps: 
\begin{enumerate}
    \item \textbf{Selection of bibliographic databases:} Relevant digital libraries were selected as sources for each form. In a pragmatic manner, and depending on the coverage of the topic, in addition to standard databases such as Scopus, Web of Science, IEEE Xplore, and the ACM Digital Library, we also used Google Scholar to capture recently published papers not yet indexed elsewhere. 
    \item \textbf{Search string definition.} Search strings were constructed from form-specific constructs and iteratively refined by the responsible authors. 
    \item \textbf{Inclusion and exclusion criteria (IC/EC):} Explicit criteria were defined to determine whether publications should be included or excluded, ensuring that only relevant contributions were retained. The following core inclusion and exclusion criteria (applied consistently across all reviews) were defined, while some groups complemented them with additional form-specific criteria:
    \begin{itemize}[leftmargin=1cm]
        \item \textbf{IC:} (i) studies employ techniques of generative \ai{} (e.g., Large Language Models such as GPT, Generative Adversarial Networks, or Variational Autoencoders); (ii) studies explicitly address the SDLC and/or software development and operations processes. 
        \item \textbf{EC:} (i) studies not written in English; (ii) duplicated studies; (iii) studies authored by the same group without introducing new contributions. 
    \end{itemize}
    \item \textbf{Search execution:} The search strings were executed in the selected databases and the retrieved records aggregated. 
    \item \textbf{Screening and selection:} Depending on the group, records were screened either in two phases (titles/abstracts followed by full-text) or directly at full-text level, by applying the IC/EC to identify the final set of primary studies. 
    \item \textbf{Analysis and synthesis:} The selected publications were analyzed, and their relevant content synthesized into the entries of the tetrads, i.e., the constituents of each quadrant for the corresponding fundamental form. 
\end{enumerate}

\subsubsection{Solution - McLuhan's Tetrads for \ai augmentation in SE}
\label{sec:tetrad_intro}
Marshall McLuhan's Tetrad of media effects\footnote{\url{https://en.wikipedia.org/wiki/Tetrad_of_media_effects}} provides a framework for analyzing and visualizing the effects of a technology by categorizing them into four interrelated dimensions as explained below.
We chose McLuhan's Tetrads for our analysis, as they were successfully used in previous studies.
In addition to being used during the FSE 2025 workshop (see Section~\ref{sec:workshop}) and its predecessor at FSE 2024~\cite{SE2020V1} to examine the impact of \ai{} on SE, they were also used to examine the effects of LLMs on SE research~\cite{TrinkenreichEtAl2025}.

For each form of \ai, a preliminary McLuhan's Tetrad was created based on the analysis of the selected publications and insights from the FSE 2025 workshop (see Section~\ref{sec:workshop}), thereby synthesizing the findings.
Each tetrad was independently developed by a team of authors, with no author contributing to more than one tetrad, to avoid bias and facilitate diversity of perspectives.

The four interrelated dimensions of a McLuhan's Tetrad are depicted in Figure~\ref{fig:tetrad}.
Together, these dimensions offer a holistic perspective on the consequences of technology adoption.
\begin{enumerate}
    \item "What does the technology \textbf{enhance} or intensify?"
    This aims to understand how the technology augments or amplifies certain capabilities.
    For example, the advent of the printing press intensified the dissemination of information, revolutionizing communication and education.
    \item "What does the technology \textbf{reverse} or flip into when pushed to its extreme?"
    This examines how certain technology, when pushed to its extreme, undergoes a reversal or transformation.
    As an example, the ubiquity of smartphones, offering constant connectivity, has the potential to reverse into isolation and disconnection.
    \item "What does the technology \textbf{retrieve} or recover from the past?"
    This aims to understand how far technology makes it possible to retrieve or recover concepts and ideas from the past.
    As an example, the advent of the Internet retrieved the decentralized and participatory nature of oral communication in the digital realm.
    \item "What does the technology make \textbf{obsolete} or displace?"
    This explores the obsolescence or displacement caused by the technology in question.
    As an example, the rise of television displaced radio as the primary source of news and entertainment.
    \end{enumerate}

\input{figures/mcluhans_tetrad}

\subsubsection{Validation - Internal Peer Review}
\label{sec:peer_disc}
The preliminary results, including the structuring into the four forms, the findings of the rapid literature review, and the derived tetrads, were subjected to internal peer reviews between the authors.
Each tetrad was validated by the authors who had not participated in the derivation of the respective tetrad, ensuring a collective assessment and reducing the risk of confirmation bias.
Feedback was collected to refine the structure, clarify ambiguities and improve the overall consistency of intermediate results.

\subsection{Cycle 3 -- Final Artifact Development}
\label{sec:cycle3}
Cycle 3 focused on the development of the final artifact, that is, the roadmap for the augmentation of \ai{} in SE, and its collective validation.

\subsubsection{Awareness - Roadmap Derivation Setup}
In this phase, two authors, who were not involved in the construction of individual tetrads, analyzed the consolidated tetrads from Cycle 2 with the aim of synthesizing an overall research roadmap on \ai augmentation in SE.
Their external position with respect to the earlier cycles was intended to ensure a fresh and unbiased view on the consolidated tetrads.

\subsubsection{Solution - Synthesis into Roadmap}
To synthesize the roadmap, the two authors systematically analyzed all validated tetrads, identifying transversal patterns, recurring themes, and complementary insights. 
The results of this synthesis were distilled into a comprehensive roadmap that covers the form-specific and cross-form research challenges and opportunities.

\subsubsection{Validation - Collective Review and Refinement}
The roadmap draft was subjected to a validation process involving all remaining authors. 
Feedback was provided asynchronously through comments and revisions and was further discussed in three one-hour meetings. 
This iterative process allowed for clarifications, refinements, and resolution of disagreements, ultimately leading to the final version of the roadmap.

\section{Preliminaries}
\label{sec:foundations}
Resulting as an outcome of Cycle 1 of our methodology (see Section~\ref{sec:cycle1}), we provide the foundations for a systematic discussion of the impact of \ai{} augmentation in SE by elaborating on the four different forms of \ai augmentation, and discussing related research roadmaps.

\subsection{Structuring of \ai{} in SE}
\label{sec:taxonomy}

As introduced in Section~\ref{sec:intro}, we define the following two dimensions: 
\begin{itemize}
    \item \textbf{"What is augmented by \ai{}?"}, distinguishing between \ai{}-augmented SE \textit{processes} (meaning that \ai{} is used to automate software engineering activities and tasks) and \ai{}-augmented software \textit{products} (meaning that parts of the functionality of a software system or application are not explicitly programmed, but are implemented with \ai).
    \item \textbf{"How autonomous is the \ai{} augmentation?"}, distinguishing between \textit{passive} roles (meaning that the human-AI interaction is triggered by the human~\cite{weber2024,HassanLRGC00TOL24} and that the behavior of \ai{} is determined by the methods invoked upon the \ai{} model) and \textit{active} roles (meaning that the human-AI interaction is triggered by \ai{} and that \ai{} has its own thread of control and makes (semi)autonomous decisions about which actions to perform and when; e.g., Agentic AI).
\end{itemize}

The intersection of these two dimensions forms a 2x2 matrix, as shown in Table~\ref{tab:forms}, which defines four distinct forms of \ai{} augmentation in SE.

\begin{table}[htb]
    \centering
        \caption{Structuring of \ai{} in SE: Four distinct forms of \ai{} augmentation}
    \label{tab:forms}
    {\renewcommand{\arraystretch}{1.3}
    \begin{tabular}{cc|cc}
        & & \multicolumn{2}{c}{\textit{"What is augmented by \ai{}?"}}\\
        & & {Process} & {Product} \\
        \hline
       \textit{"How autonomous is} & {Passive Role}  & \textbf{\copilot} & \textbf{\aiware} \\
        \textit{the  \ai{} augmentation?"} & {Active Role}  & \textbf{\teammate} & \textbf{\robot{}} \\
        \hline


    \end{tabular}
    \vspace{.5em}

}\end{table}

To be able to refer to these forms succinctly in this article, we use the following labels throughout the paper:

 \textbf{\copilot :} \ai{} is used as a tool -- either standalone or integrated into an IDE or CI/CD pipeline -- to automate various SE tasks.
    Here, one typically can find solutions from the field of Automated Software Engineering.
     For the label \copilot , we took inspiration from Github Copilot~\cite{ErhaborUNA25} and many other AI/\copilot{}s available.
    Examples of tasks supported by \copilot{} include requirement elicitation~\cite{korn2025}, code generation~\cite{jiang2024}, testing~\cite{WangHCLWW24}, and program repair~\cite{ZhangFMSC24}.

    \textbf{\aiware:} \ai{} is used to realize software functionality that otherwise would be impossible to realize or require a significant effort to realize~\cite{weber2024}.
    This means that \ai{} is not used to generate actual code, but that the \ai{} model is invoked from the code to perform specific computations.
    For the label \aiware , we took inspiration from FMware, which refers to ``the type of software that uses foundation models (FMs), such as Large Language Models (LLMs), as one of its building blocks''~\cite{HassanLRGC00TOL24}.
    In this sense, the label ``\aiware{}'' generalizes from FMware to encompass a wider range of \ai models.
    An example of \aiware is a software system that leverages large language models (LLMs) to summarize instructional video transcriptions, thus offering support to teachers and students~\cite{ChomatekSP25}.
    Here, the LLM prompt is far more than a simple text input -- it is a meticulously engineered part of the codebase.
    Another example is augmenting an online store with a chatbot to guide customers in finding products that fit their needs.

    \textbf{\teammate:} \ai{} acts as an agent that proactively participates in the software development process in one or more roles.
    \teammate{}s are (semi)autonomous, goal-driven agents that collaborate with human software engineers in real-world workflows.
    For the label \teammate , we took inspiration from a recent post from the World Economic Forum\footnote{\url{https://www.weforum.org/stories/2025/01/why-you-should-think-of-ai-as-a-teammate-not-a-tool-when-building-a-better-future/}}.
    Examples are autonomous coding agents, which actively initiate, review and evolve code on a scale as part of open source ecosystems~\cite{li2025}.

    \textbf{\robot{}:} \ai{} acts as a (semi)autonomous, goal-driven agent to deliver parts of the functionality of the software system.
    For the label \robot , we took inspiration from the general definition of a robot as a ``machine [...] capable of carrying out a complex series of actions automatically''\footnote{\url{https://en.wikipedia.org/wiki/Robot}}.
    As a simple example, a \robot{} can browse the Web and make online purchases on behalf of a user: compare prices, select items, and complete checkouts~\cite{Gabriel25}.
    As a more complex example, a process-aware information system that facilitates the on-boarding of new suppliers as part of a procurement process may be implemented via \robot{}s~\cite{Fettke2025XABPs}.
    In such a system, there may be a \textit{Buyer} \robot{} and different \textit{Supplier} \robot{}s.
    The Buyer \robot{} identifies the potential Supplier \robot{}s and issues a request for quotes, which is then answered by the Supplier \robot{}s.
    Based on these answers, the Buyer \robot{} can then select a suitable supplier.



Note that while we used the 2x2 matrix to define a set of four distinct forms, in a concrete software systems more than one form may be present simultaneously. 
For instance, a highly advanced \copilot{} could also exhibit some proactive behavior of a \teammate{}.

\subsection{Related Research Roadmaps}
\label{sec:related}
Here we discuss how our article relates to existing efforts concerning research roadmaps on \ai in SE.
On the one hand, we discuss recent roadmap papers (journal articles and conference papers published since 2024).
On the other hand, we assess the scope of research workshops that took place since 2024 at major SE conferences, as they often feature a set of very timely forward-looking papers based on the participants' contributions and onsite discussions.

Concerning related roadmap articles, the first TOSEM special issue on ``2030 Roadmap for Software Engineering''  particularly stands.
This special issue presents the results of intensive 2-day discussions at the 2030 Software Engineering Workshop, co-located with FSE 2024~\cite{SE2020V1}.
The presented roadmap was intended to serve as a living body to be refined based on follow-up workshops and updated during a series of forthcoming TOSEM special issues.
Our article presents such an update for one of the roadmap's seven themes: "artificial intelligence for software engineering"~\cite{AhmedACCHHPPX25}.
This theme was covered in the 2025 TOSEM special issue by nine contributed articles.
Six of those discuss the impact of AI on the SE process (thereby providing initial directions on \copilot{}s and \teammate{}s).
The remaining three articles deal with the integration of AI into systems and applications (thereby providing initial directions on \aiware{} and \robot{}s).
In addition to the aforementioned special issue articles, Table~\ref{tab:se-roadmaps} lists relevant roadmap articles on \ai{} in SE over the past two years and identifies which of the four forms of \ai augmentation they addressed.
The list of roadmap articles indicates a very strong focus on \copilot{}s, or in other words automated SE, while only few address the other forms or even multiple forms at the same time.
Some of the recent roadmap articles touch on the aspect of autonomous \ai, but mainly focus on \teammate{}s.

\begin{table}
\centering
\caption{Coverage of \ai{} augmentation in SE roadmap papers: ++ = main focus of article, + = secondary focus}
\label{tab:se-roadmaps}
   \rowcolors{2}{white}{gray!20} 
\small
\begin{tabularx}{\textwidth}{c l X c c c c}
\toprule
\textbf{Year} & \textbf{Journal} & \textbf{Title} & \textbf{Copilot} & \textbf{AIware} & \textbf{Teammate} & \textbf{Robot} \\
\midrule
\textbf{2025} 
 &  Autom. Softw. Eng.  & Future of software development with generative AI~\cite{10.1007/s10515-024-00426-z} & ++ & & & \\
 & Applied Sc. & AI-driven innovations in software engineering: a review of current practices and future directions~\cite{Alenezi2025} & ++ & & & \\
& Inf. Softw. Technol. &  Copiloting the Future: How Generative AI Transforms Software Engineering~\cite{10.1016/j.infsof.2025.107751} & ++ & & + & \\
 & SP\&E & Generative Artificial Intelligence for Software Engineering---A Research Agenda~\cite{Duc2025} & ++ & & & \\
 & SSRN & Generative AI for Software Architecture, Applications, Trends, Challenges, and Future Directions~\cite{Esposito2025} & ++ & & + & \\
& TOSEM-SI & 2030 Roadmap for Software Engineering~\cite{SE2020V1} & ++ & ++ & + & + \\
& arXiv & Challenges and Paths Towards AI for Software Engineering~\cite{Gu2025} & ++ & & + & \\
 & arXiv & From LLMs to LLM-based Agents for Software Engineering: A Survey of Current, Challenges and Future~\cite{Jin2025} & ++ & & ++ &\\
 \midrule
\textbf{2024} & TAAS & Generative AI for Self-Adaptive Systems: State of the Art and Research Roadmap~\cite{LiZLWJT24} & & ++ & & \\
\bottomrule
\end{tabularx}
\end{table}

Concerning related roadmap workshops, Table~\ref{tab:se-workshops} lists the relevant ones\footnote{Note that for AgenticSE@ASE'25 and Ex-ASE@ASE'25 there was no program available at time of writing.} on \ai{} in SE of the past two years held at the three top-tier (A*) SE conferences and identifies which of the four forms of \ai augmentation they addressed.
The list of workshops allows us to make the following high-level observations.
First, there is a clear increase between 2024 and 2025 in the number of workshops dedicated to the topic of \ai, indicating the increasing relevance of \ai in SE.
Second, similar to the roadmap articles, most workshops focus on \copilot{}s.
Third, some of the recent workshops have started to cover the aspect of autonomous \ai, but mainly focus on \teammate{}s.

\begin{table}
\centering
\small
\caption{Coverage of \ai{} augmentation in SE workshops: ++ = main focus of workshop, + = secondary focus}
\label{tab:se-workshops}
   \rowcolors{2}{gray!20}{white} 

\begin{tabularx}{\textwidth}{llXcccc}
\toprule
\rowcolor{white}
\textbf{Conference} & \textbf{Workshop} & & \textbf{Copilot} & \textbf{AIware} & \textbf{Teammate} & \textbf{Robot} \\
\midrule

{\textbf{ICSE'25}} & AIOps & AI for Cloud Service & & ++ & & \\
& APR & Automated Program Repair & & ++ & & \\
& BotSE & Bots in SE & ++ & + & & \\
& DeepTest & Deep Learning $<>$ Testing & & ++ & ++ & \\
& LLM4Code & Large Language Models for CODE & ++ & & & \\
& NLBSE & Natural Language Based SE & ++ & & & \\
& NSE & Neuro-Symbolic SE & + & & & \\
& RAIE & Responsible AI Engineering & ++ & & & \\
& RAISE & Requirements Eng. for AI-Powered SW & ++ & & & \\
& IWSIB & Software-Intensive Business & & + & & \\
[2pt]\hline

{\textbf{ICSE'24}} & APR & Automated Program Repair & + & & & \\
& DeepTest & Deep Learning $<>$ Testing & & + & & \\
& FTW & Flaky Tests Workshop & & + & & \\
& Intense & Interpretability, Robustness, and Benchmarking in Neural SE & ++ & & & \\
& IWSIB & Software-Intensive Business & & + & & \\
& LLM4Code & Large Language Models for Code & ++ & & & \\
& NLBSE & Natural Language Based SE & ++ & & & \\
& RAIE & Responsible AI Engineering & + & + & & \\
& SATrends & New Trends in Software Architecture & + & & & \\
\midrule[1.2pt]

{\textbf{FSE'25}} & 2030 SE & 2030 Software Engineering & ++ & ++ & ++ & + \\
& AI IDE & Artificial Intelligence for Integrated Development Environments & ++ & & & \\
& AI-SDLC & Envisioning the AI-Augmented Software Development Life Cycle & ++ & & ++ & \\
& BI4LLMC & Benchmark Infrastructure for LLMs for Code & ++ & & & \\
& Human AISE & Human-Centered AI for SE & ++ & ++ & & \\
& LLinMAP & Large Language Model-Oriented Empirical Research & ++ & + & & \\
& LLMApp & LLM App Store Analysis & & + & + & \\
& ResponsibleSE & Engineering Responsible SE & + & & & \\
[2pt]\hline

\textbf{FSE'24} & 2030 SE & 2030 Software Engineering & ++ & ++ & + & + \\
\midrule[1.2pt]

& AISM & AI for Software Modernization & ++ & & & \\
\midrule

{\textbf{ASE'24}} & Intelligent SE & Intelligent Software Engineering & ++ & & & \\
& MAS-GAIN & Multi-Agent Systems using Generative Artificial Intelligence for ASE & & & ++ & \\
& ASYDE & Automated and verifiable Software syStem Development & & ++ & & \\

\bottomrule
\end{tabularx}

\end{table}

As a major  difference from the above roadmap articles and forward-looking workshops, our roadmap covers all four forms, thus providing both individual and cross-form research challenges and opportunities. 

\section{Impact of \copilot{}s in SE}
\label{sec:copilot}

This section presents the outcomes of Cycle 1 conducted on the \copilot{} form. 
As introduced in Section~\ref{sec:taxonomy}, a \copilot{} is a passive element (such as a component, service, or tool) that supports human software engineers in performing development tasks.
We describe how this review was designed and executed using a refinement of the general protocol defined in Section~\ref{sec:cycle2} and discuss the McLuhan tetrad derived from the collected evidence. 

\subsection{Rapid Literature Review}
\label{sec:review1}

To provide an overview of the implications and impacts of \copilot{}s on SE as identified in the literature, we performed a rapid literature review as explained in Section~\ref{sec:cycle2}.

\subsubsection{Review Approach}
\label{sec:review_approach1}
\begin{figure}
    \centering
    \includegraphics[width=1\linewidth, trim=0cm 1cm 0cm 1.2cm, clip=true]{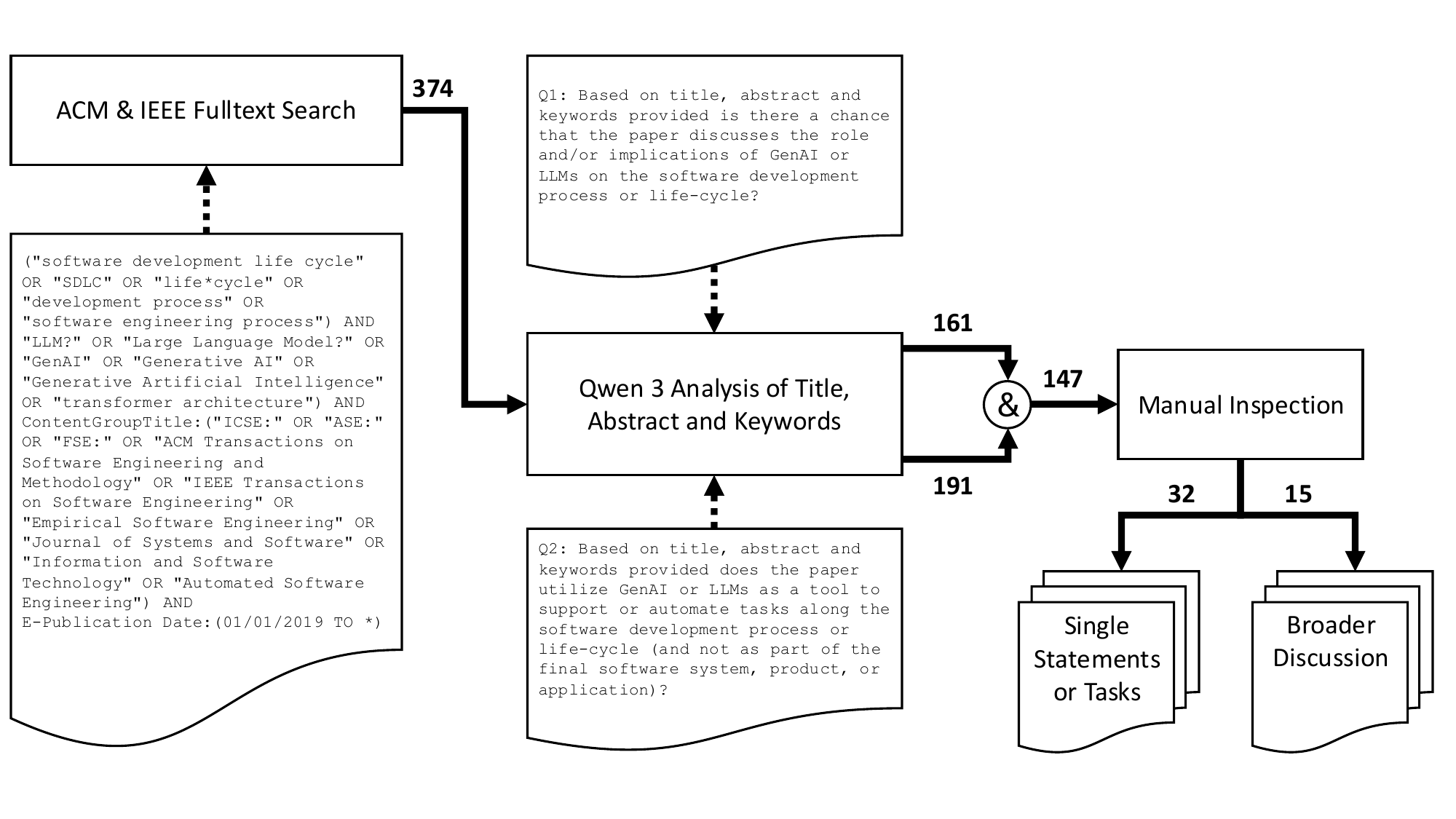}
    \caption{Identifying relevant publications concerning \copilot{}s 
    }
    \label{fig:copilot-review}
    \Description{Image shows the rapid literature review process to determine relevant publications.}
\end{figure}

The procedure we followed to perform the rapid literature review for \copilot{}s is depicted in \autoref{fig:copilot-review}.

As this category of \ai has attracted the most research focus in software engineering to date, we limited our search to top-tier software engineering conferences and journals (ICSE, ASE, FSE, TOSEM, TSE, EMSE, JSS, IST, ASEJ).
We restricted the search to the years 2019--present, as the first Transformer-based models (BERT and GPT) were released in 2018, requiring time for adoption by the SE community.
To ensure comprehensive coverage, we opted for full-text searches using ACM and IEEE as search engines, as we initially found that many papers only discuss the impact on SE in the body of the paper and not in the title, abstract, or keywords.
We narrowed the scope to papers explicitly mentioning the \sdlc, development process, or software engineering process in conjunction with LLMs, \ai, or transformer architecture.
These criteria are reflected in the search string\footnote{The exact search strings can be found in \Cref{sec:copilot-searchstrings}} presented in \autoref{fig:copilot-review}.

To further streamline the analysis process, we leveraged Qwen3\footnote{\url{https://ollama.com/library/qwen3:8b}} to analyze the titles, abstracts, and keywords of the retrieved publications.
We asked Qwen3 to answer two questions about the publication that reflect the scope of our search:
(Q1) Whether the paper explicitly discusses the role and/or implications of \ai on the \sdlc;
(Q2) Whether the paper utilizes \ai or LLMs as a tool to support or automate tasks along the \sdlc.
Q1 focuses on the publication's scope, while Q2 excludes \aiware or \robot publications, as these are addressed in the chapters below.
For each question, Qwen3 provided a yes/no answer along with its reasoning\footnote{The exact prompts can be found in \Cref{sec:copilot-promts}}.
The publications positively evaluated by Qwen3 for both questions were then manually inspected by two authors (split equally).



\subsubsection{Review Results}
\label{sec:review_results1}
The initial search in ACM and IEEE resulted in 374 publications.
Qwen3 classified 161 of these as discussing the role or implications of \ai on the \sdlc and 191 as utilizing \ai as a tool to support or automate tasks along the \sdlc.
Of these, 147 met both criteria and were manually inspected.
During the manual inspection process, the authors identified that most publications only mention \sdlc as motivation or in the related work, but do not describe the implications of \ai on it (100 articles).
Furthermore, many publications made only single statements about the impact or role of \ai in \sdlc, or single tasks of the \sdlc that are improved (32 articles).
Only a few publications provided a broader discussion of the implications and role of \copilot{}s on the \sdlc (15 papers, see Table \ref{tab:copilot-sota}).

The latter two categories were used to inform our McLuhan's tetrad (see Section~\ref{sec:roadmap1}), either by deriving insights from the single statements or analyzing the full discussion in the 15 other papers.
Furthermore, we included our proposal for the FSE 2025 workshop on \enquote{Automated Knowledge Management in the Software Life Cycle}~\cite{keim_automated_2025}, as it was one of the starting points for this contribution and fits into the category \copilot.

\begin{table}
    \centering
      \caption{Relevant publications concerning \copilot{}s}
   \label{tab:copilot-sota}
   \rowcolors{2}{gray!20}{white} 
    \footnotesize
    \begin{tabularx}{\textwidth}{Xlc}

    \toprule
    \rowcolor{white}
    Paper & Authors & Year\\
    \midrule
    Future of software development with generative AI~\cite{10.1007/s10515-024-00426-z} & Sauvola et al. & 2024\\
A disruptive research playbook for studying disruptive innovations~\cite{10.1145/3678172} & Storey et al. & 2024\\
Large language models for software engineering: a systematic literature review~\cite{10.1145/3695988} & Hou et al. & 2024\\
Software testing with large language models: Survey, landscape, and vision~\cite{WangHCLWW24} & Wang et al. & 2024\\
Formal requirements engineering and large language models: A two-way roadmap~\cite{10.1016/j.infsof.2025.107697} & Ferrari and Spoletini & 2025\\
Copiloting the future: How generative AI transforms software engineering~\cite{10.1016/j.infsof.2025.107751} & Banh et al. &2025\\
Challenges and opportunities for generative AI in software engineering: a managerial view~\cite{10.1145/3696630.3728718} & Rico and Öberg &2025\\
What do professional software developers need to know to succeed in an age of Artificial Intelligence?~\cite{10.1145/3696630.3727251} & Kam et al.&2025\\
AI in the software development lifecycle: Insights and open research questions~\cite{10.1145/3696630.3730538} & Guimaraes and Nascimento &2025\\
The future of AI-driven software engineering~\cite{10.1145/3715003} & Terragni et al. &2025\\
From today's code to tomorrow's symphony: The AI transformation of developer's routine by 2030~\cite{10.1145/3709353} & Qiu et al. &2025\\
From triumph to uncertainty: The journey of software engineering in the AI era~\cite{10.1145/3709360} & Mastropaolo et al. &2025\\
Automatic programming: Large language models and beyond~\cite{10.1145/3708519} & Lyu et al. &2025\\
The current challenges of software engineering in the era of large language models~\cite{10.1145/3712005} & Gao et al.&2025\\
Artificial intelligence for software engineering: The journey so far and the road ahead~\cite{AhmedACCHHPPX25} & Ahmed et al.&2025\\
Towards Automated Knowledge Management in the Software Life Cycle~\cite{keim_automated_2025} & Keim et al.&2025\\
    \bottomrule
    \end{tabularx}

\end{table}


\input{figures/mcluhans_tetrad_copilot}

\subsection{McLuhan's Tetrad}
\label{sec:roadmap1}
Informed by our rapid literature review, we interpret each quadrant of the tetrad in Figure~\ref{fig:tetrad-copilots} and map the relevant findings of previous work into these four dimensions.

\subsubsection{Enhances}
\label{sec:enhances1}
\begin{itemize}

    \item \textit{Requirements}: Ferrari and Spoletini~\cite{10.1016/j.infsof.2025.107697} describe how the \textit{Requirements} phase can be enhanced by LLMs through improved extraction, analysis, verification, and validation of requirements. Their two-way roadmap demonstrates the mutual influence between formal requirements engineering and advances in \ai. 
    Additionally, Rico and Öberg~\cite{10.1145/3696630.3728718} highlight enhancements in the generation and refinement of requirements.
    
    \item \textit{Design \& Documentation}: \copilot{}s have been reported to improve activities in \textit{Design \& Documentation} by enabling the automated creation of context-sensitive and stakeholder-specific artifacts, thereby reducing manual workload and ensuring that design intent is preserved throughout project iterations (Rico and Öberg~\cite{10.1145/3696630.3728718}, Terragni et al.~\cite{10.1145/3715003}). 
    Also, Storey et al. ~\cite{10.1145/3678172} mention in their tetrad for code generation, that documentation is enhanced by the introduction of \copilot{}s.
    Furthermore, Banh et al.~\cite{10.1016/j.infsof.2025.107751} describe how \copilot{}s support conceptualization and provide contextual knowledge, such as relevant documents.

    \item \textit{Implementation}: \copilot{}s also enhance activities in the \textit{Implementation} phase. For instance, coding itself is enhanced~\cite{10.1007/s10515-024-00426-z,10.1145/3695988,10.1016/j.infsof.2025.107751} through the abilities of the \copilot{}s to translate natural language descriptions into functional code, enabling developers to work more efficiently and reducing the cognitive gap between design intent and implementation.  
    For improving code comprehension~\cite{10.1007/s10515-024-00426-z,10.1016/j.infsof.2025.107751,10.1145/3708519}, these tools can provide explanations of unfamiliar code.
    Additionally, in the refinement and optimization of code~\cite{10.1145/3696630.3728718,10.1016/j.infsof.2025.107751}, the literature identifies potential enhancements, such as \copilot{}s suggesting performance improvements, cleaner syntax, and better algorithmic solutions. 
    Code analysis~\cite{10.1145/3696630.3728718,10.1145/3708519} is elevated with AI-driven insights that catch errors, suggest alternatives, and flag security vulnerabilities as the code is written (Real-time).

    \item \textit{Testing \& Quality Assurance}: \copilot{}s also play a transformative role in \textit{Software Testing \& Quality Assurance} by improving the efficiency, coverage, and intelligence of testing and debugging processes~\cite{10.1007/s10515-024-00426-z,10.1145/3696630.3730538,10.1145/3696630.3728718,10.1145/3695988}.
    Test design and generation~\cite{10.1145/3696630.3728718,WangHCLWW24} is enhanced by the ability to derive test cases from more diverse sources, such as documentation or even natural language requirements.
    \copilot{}s also improve bug detection and fixing~\cite{10.1007/s10515-024-00426-z,10.1016/j.infsof.2025.107751,10.1145/3708519}, by analyzing code context to locate faults and propose fixes, reducing the time developers spend on manual debugging.
    Vulnerability Detection~\cite{10.1145/3695988} is enhanced by automatically identifying potential security risks based on learned patterns and best practices, even in complex codebases and potentially real-time.
    \copilot{}s support or even enable automated debugging~\cite{10.1145/3696630.3728718,WangHCLWW24} by explaining runtime errors, tracing root causes, and suggesting corrective actions in real-time. 
    When bugs or vulnerabilities are found, \copilot{}s can automatically generate fix suggestions or patches, enabling automated program repair~\cite{WangHCLWW24}. 
    In general, the introduction of \copilot{}s can enhance code quality~\cite{10.1145/3678172,10.1145/3696630.3727251,10.1145/3696630.3730538,10.1145/3715003,AhmedACCHHPPX25}, especially by automating code reviews~\cite{10.1145/3695988,10.1145/3708519}, flagging potential errors, style inconsistencies, and design-level issues, allowing human reviewers to focus on broader architectural or strategic concerns. 

    \item \textit{Project \& Process Management}: \copilot{}s are mentioned to enhance \textit{Project \& Process Management} by improving productivity~\cite{10.1145/3696630.3727251,10.1145/3678172,10.1145/3695988}, development time~\cite{10.1016/j.infsof.2025.107751}, effort estimation, completeness and consistency checks, and knowledge management. 
    For Effort Estimation~\cite{10.1145/3696630.3728718}, they can analyze diverse historical data sources to provide more accurate time and resource predictions, helping managers plan and allocate effectively.  
    By identifying gaps, redundancies, or conflicting logic across artifacts, \copilot{}s may be able to ensure completeness \& consistency throughout the \sdlc~\cite{keim_automated_2025}. 
    Finally, Keim et al.~\cite{keim_automated_2025} envision \copilot{}s to enhance knowledge management throughout the \sdlc as they are able to extract, summarize, and deliver contextually relevant information from different artifacts and sources, such as code, documentation, and requirements, thus supporting more effective onboarding, collaboration, and knowledge sharing.

    \item \textit{Human \& Team Factors}: Another area influenced by the emergence of \copilot{}s are \textit{Human \& Team Factors}. 
    They may support the development, collaboration, and well-being of individuals within software engineering teams. 
    For onboarding and training~\cite{10.1145/3696630.3728718}, copilots can reduce ramp-up time by answering technical questions, explaining project contexts, and guiding new developers through unfamiliar codebases.
    In the area of Skills Learning~\cite{10.1145/3709353}, \ai{} tools provide on-demand, personalized learning by offering code explanations, best practice recommendations, and interactive coding suggestions that help developers improve their competencies as they work. 
    Team Coordination~\cite{10.1145/3709353} may benefit from automated task summaries, meeting notes, and synchronization of project status, making collaboration smoother and more transparent.
    Communication-wise~\cite{10.1145/3696630.3727251}, \copilot{}s can facilitate clearer and more efficient interactions between team members and/or stakeholders by, for example, translating technical concepts into simpler language, drafting documentation, or generating summaries for cross-functional stakeholders. 
    Finally, they may also positively impact mental health~\cite{10.1145/3709353} by reducing repetitive tasks or providing more structured and manageable workflows.
\end{itemize}

Across the reviewed literature, enhancements represent the most dominant impact of \copilot{}s.
Hou et al.~\cite{10.1145/3695988} provide quantitative evidence, with most LLM applications enhancing mid-phase SDLC tasks: coding (57\%), maintenance (23\%), and quality assurance (15\%).
Wang et al.\cite{WangHCLWW24} echo this in the domain of software testing, with LLMs improving test generation, bug detection, program repair, and debugging, leading to faster testing cycles and increased coverage.

In summary, \copilot{}s are reported and/or envisioned to enhance the \sdlc by improving efficiency, quality, and collaboration, by enabling automation of tasks such as requirements engineering, coding, testing, documentation, and project management, while also supporting human factors like learning, communication, and well-being.

\subsubsection{Reverses}
\label{sec:reverses1}

\begin{itemize}
        \item \textit{Trustworthiness \& Reliability}~\cite{WangHCLWW24,10.1145/3695988, 10.1145/3696630.3728718,10.1145/3696630.3730538,10.1145/3708519,10.1145/3709360,keim_automated_2025}: Though intended to streamline tasks, a major concern arises from \enquote{hallucinations} or plausible-but-incorrect code suggestions as well as misaligned designs. This lowers trustworthiness in generated artifacts. Moreover, inexperienced developers may accept erroneous suggestions without critical review, introducing errors and subtle bugs that compromise reliability.
        \item \textit{Explainability}~\cite{10.1145/3696630.3730538,keim_automated_2025}: Traditionally, explainability in software development flows from code to developer, meaning that because humans write the code, they inherently understand its logic, structure, and intent. With \copilot{}s, this paradigm is reversed, and the developer must then interpret and validate the \copilot{}'s output. This shift challenges explainability, as the rationale behind a copilot's suggestion is not always transparent or grounded in an easily traceable design decision.
        \item \textit{Fairness}~\cite{AhmedACCHHPPX25}: \copilot{}s may generate code or logic that reflects patterns from biased training data, unintentionally encoding unfair assumptions or discriminatory behavior. Instead of proactively designing for fairness, developers may find themselves reacting to unintentional bias introduced by the copilot, scrutinizing and retroactively correcting outputs whose fairness implications are opaque.
        \item \textit{Compliance}~\cite{10.1145/3709360}: Instead of designing for compliance, developers may unknowingly introduce non-compliant constructs created by the AI. As a result, compliance becomes reactive, and developers must verify and audit AI-generated code post hoc, increasing the burden of compliance validation and introducing new risks if oversight is insufficient.
        \item \textit{Sustainability}~\cite{10.1145/3709360}: Traditionally, sustainability in software development (e.g., energy-efficient code, minimizing computational overhead, designing maintainable systems) is an intentional design goal. With GenAI-powered \copilot{}s, this mindset can be inadvertently reversed: copilots often prioritize functionality and speed of development over resource efficiency or long-term maintainability. They may generate verbose, redundant, or suboptimal code that increases energy consumption or contributes to technical debt, especially if developers accept suggestions without critical evaluation.
        \item \textit{Accountability}~\cite{10.1145/3709360}: Developers and teams are usually directly responsible for the code they write and the decisions they make. With GenAI copilots, this clarity is blurred, effectively reversing accountability dynamics. As copilots generate code and design suggestions, it becomes more difficult to trace the origin of specific decisions or defects, especially when developers treat AI outputs as trustworthy by default. This creates ambiguity around who is responsible when something goes wrong: the developer, the organization, or the AI provider.
        \item \textit{Individual Understanding Capabilities}~\cite{10.1145/3678172,10.1016/j.infsof.2025.107751}: \copilot{}s may reverse the individual understanding capabilities of developers, by encouraging overreliance or abstraction away from low-level detail.
        \item \textit{Strict boundaries between \sdlc stages}~\cite{10.1007/s10515-024-00426-z}: Strict boundaries between \sdlc stages begin to dissolve, creating blurred transitions where coding, testing, and releasing collapse into a continuous loop.
        While this can enhance agility, it may also reverse long-established process controls, thereby increasing risks to traceability, defect localization, and process accountability.
        \item \textit{Clear code ownership}~\cite{10.1145/3696630.3728718,10.1145/3708519,10.1145/3709360}: Copilots produce code without clear attribution or traceable authorship, raising questions of accountability and provenance. Developers risk becoming curators rather than creators, which can weaken their sense of ownership and design rationale. Psychological ownership may likewise decrease -- an effect observed for AI-generated text~\cite{draxler2024ai} -- with yet unknown consequences for developer productivity and long-term maintenance.
        \item \textit{Security}~\cite{10.1145/3695988,10.1145/3708519}: With GenAI, security often becomes an afterthought, as developers may unknowingly incorporate insecure code patterns suggested by the AI, which is trained on a mixture of secure and insecure code.
\end{itemize}

Taken together, these reversed effects highlight the importance of critical oversight and contextual awareness in the deployment of AI within software engineering, to prevent inadvertently undermining productivity rather than enhancing it.

\subsubsection{Retrieves}
\label{sec:retrieves}
\begin{itemize}
\item \textit{Formal Requirements Specification}~\cite{10.1016/j.infsof.2025.107697}: Ferrari and Spoletini argue that \copilot{}s can retrieve the value of formal requirements specifications by lowering the barrier to entry. 
By automating or assisting in transferring classical natural language requirements to formal requirements, even developers or stakeholders without deep expertise in formal methods can retrieve those more rigorous and precise software specifications.

\item \textit{Formal Verification}~\cite{10.1016/j.infsof.2025.107697}: Similarly, \copilot{}s can act as a bridge between informal descriptions and formal verification logic. 
By generating or refining logical assertions, pre- and postconditions, and invariants, they help retrieve the rigor and guarantees offered by formal verification practices, which were once restricted to high-assurance systems and required highly specialized experts.

\item \textit{Rapid Prototyping}~\cite{10.1145/3696630.3728718,10.1145/3715003}: Through their coding capabilities \copilot{}s enable fast, exploratory implementations and, thus, can retrieve rapid prototyping where it was, beforehand, not feasible to perform.

\item \textit{Natural Language Explanation of Code}~\cite{10.1145/3678172}: Storey et al.  note that the widespread use of LLMs retrieves natural language explanations within code, a practice that echoes literate programming traditions. \copilot{}s may generate intuitive summaries and explanations, making codebases more accessible and human-centric once again.
\end{itemize}
Together, these capabilities illustrate that \copilot{}s do not simply retrieve practices, but rather help re-integrate enduring principles of software engineering that have long remained valuable yet were often too difficult or time-consuming to apply widely.

\subsubsection{Obsolesces}
\label{sec:obsolesces1}
\begin{itemize}
    \item \textit{Manual debugging and bug reproduction}~\cite{WangHCLWW24,10.1145/3709353}: Qiu et al.~\cite{10.1145/3709353} and Wang et al.~\cite{WangHCLWW24} highlight that test generation, bug reproduction, and oracle design are increasingly supported by \copilot{}s, replacing manual processes.
    \item \textit{Manual elicitation and formalization of requirements}~\cite{10.1016/j.infsof.2025.107697,10.1145/3715003}: Ferrari and Spoletini~\cite{10.1016/j.infsof.2025.107697} predict that the use of \copilot{}s in requirements engineering and formalization may reduce the need for manual elicitation work, an idea echoed by Terragni et al.~\cite{10.1145/3715003}. 
    \item \textit{Need for Search of Code Snippets}~\cite{10.1145/3678172}: Storey et al. assert that the need for searching code snippets is reduced or disappears by \copilot{}s suggesting or even generating whole parts of the software.
    \item \textit{Need for manually created documentation}~\cite{10.1145/3678172}: They also discuss that the need for manually created documentation is reduced if \copilot{}s are capable enough of generating informative and precise documentation.
    \item \textit{Expert-only use of formal methods}~\cite{10.1016/j.infsof.2025.107697}: Ferrari and Spoletini predict that the exclusive use of formal methods by experts may diminish, as \copilot{}s can assist or even automate the extraction or generation of formal specifications from different sources, such as natural language requirements and code.
    \end{itemize}

As a result of this analysis, we can state that each quadrant of the tetrad reflects deep shifts in software engineering due to the emergence of \copilot{}s.
Enhancements dominate current research, ranging from automation of coding and testing to productivity and team dynamics.
At the same time, developers and organizations must actively manage reversals, including diminishing understanding, blurred responsibility, and ethical concerns.
\copilot{}s retrieve formalism and creativity once undervalued in industry, while also rendering some traditional workflows and manual effort increasingly obsolete.
This complex balance invites continuous investigation as the role of \ai{} in the \sdlc matures.


\section{Impact of \teammate{}s in SE}
\label{sec:teammate}
This section focuses on the \teammate{} form of \ai{} augmentation and examines how (semi-)autonomous and goal-driven agents collaborate with human software engineers within development processes. 
As described in Section~\ref{sec:taxonomy}, \teammate{}s differ from \copilot{}s discussed in Section~\ref{sec:copilot}, as they are not merely invoked by humans but proactively participate in software engineering activities, sharing goals, responsibilities, and contextual awareness. 
This section presents the results of Cycle~2 of our methodology (see Section~\ref{sec:cycle2}) and discusses the McLuhan tetrad derived from the collected evidence.

\subsection{Rapid Literature Review}
\label{sec:review2}
\subsubsection{Review Approach}
\label{sec:review_approach2}

\begin{figure}[htb]
    \centering
    \includegraphics[width=1\linewidth]{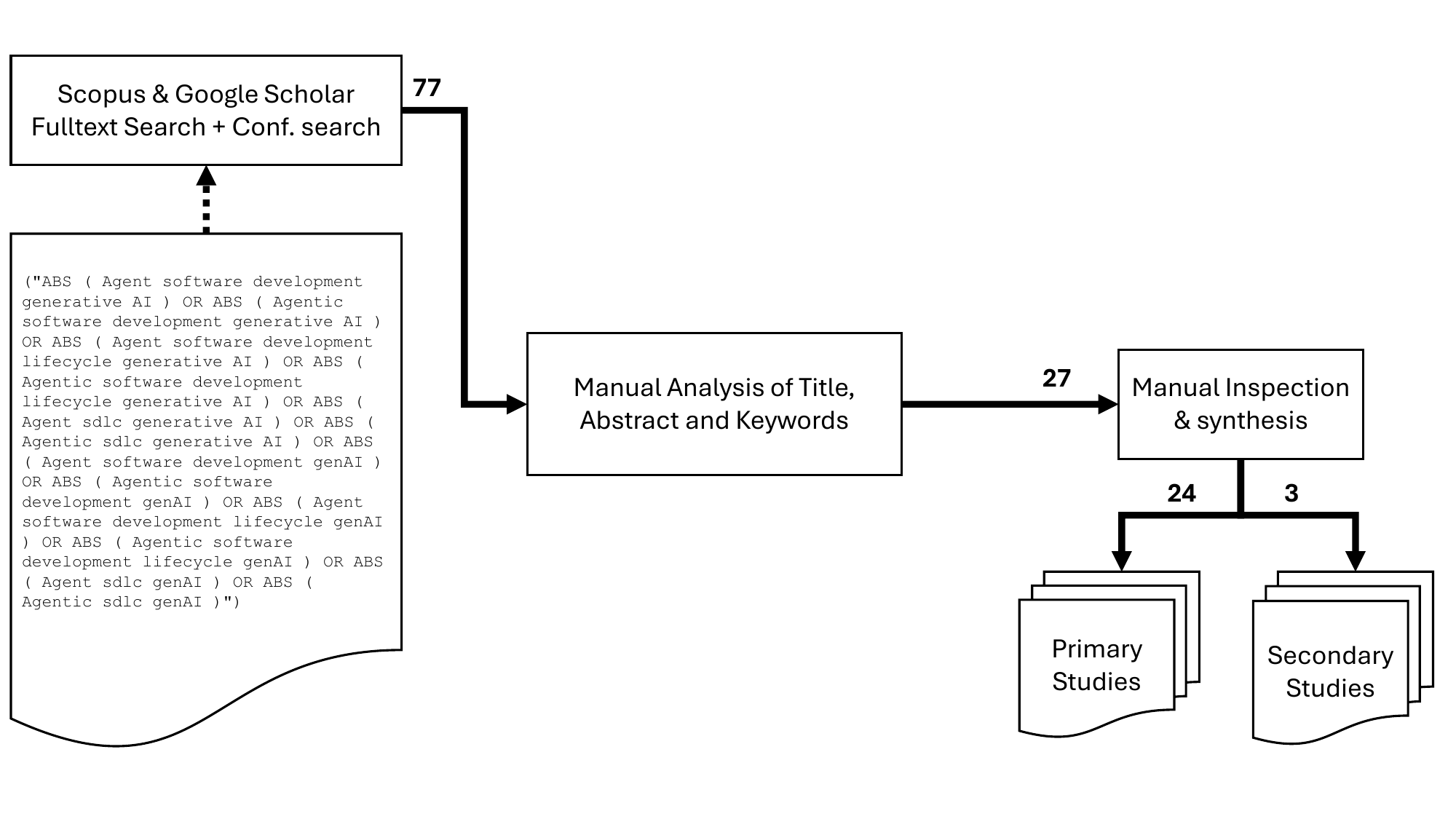}
    \caption{Identifying relevant publications concerning \teammate{}}
    \Description{Image shows the rapid literature review process to determine relevant publications.}
    \label{fig:teaMMATES-review}
\end{figure}

To include papers in the analysis of the perspective of \teammate{}s, a mixed search approach -- as a refinement of the one introduced in Section~\ref{sec:cycle2} -- was conducted by searching among the papers accepted for the top-tier (A*) SE conferences (ICSE, FSE, and ASE), jointly with an effort-bound search on the academic database aggregator Google Scholar and Scopus. This refinement method was chosen to reduce the noise in the search by directly addressing the most important conference about the topic analyzed. The use of academic aggregators was nevertheless employed to complement focused research with extensive research on all possible venues.
Unlike the other RLRs, no AI support was utilized in the analysis of the papers because the total number of primary sources from the initial phase (77) was not sufficiently high to justify the use of automated assistance.

The process for conducting the rapid literature review is depicted in \autoref{fig:teaMMATES-review}.
The search string is reported in Appendix~\ref{sec:teammate-searchstrings} and illustrated in \autoref{fig:teaMMATES-review}. The effort bound was set to 200 results; however, in both cases, this limit was never reached. The search was performed using the popular tool \textit{Publish or Perish}\footnote{\url{https://harzing.com/resources/publish-or-perish}} to facilitate a parameterized search on Google Scholar and Scopus. To include only the relevant sources in the non-peer-reviewed databases, the number of citations ranked the papers, and only those publicly accessible, written in English, with at least one citation were considered for inclusion in the initial pool of sources.
Two inclusion criteria were set to ensure the inclusion of papers only relevant to the topic of Generative AI used as an Agent included in the SDLC: (i) the approach presented in the paper had to discuss multiple agents being able to act autonomously or on demand with additional capabilities beyond purely conversational ones (chatbots); (ii) the paper needed to explicitly discuss implication of the approach on the software development lifecycle.

\begin{table}[!htbp]
    \centering
    \caption{Relevant publications concerning \teammate{}s}
    \label{tab:Teammate-sota}
    \rowcolors{2}{gray!20}{white} 
    \footnotesize
    \begin{tabularx}{\textwidth}{Xlc}
    \toprule
    \rowcolor{white}
    Paper & Authors & Year\\
    \midrule
    ChatDev: Communicative Agents for Software Development~\cite{qian2023chatdev} & Qian et al. & 2023\\
    Toward AI-facilitated Learning Cycle in Integration Course through Pair Programming with AI Agents~\cite{10663037} & Wei et al. & 2024\\
    SOEN-101: Code Generation by Emulating Software Process Models Using Large Language Model Agents~\cite{lin2024soen} & Lin et al. & 2024\\
    From LLMs to LLM-based Agents for Software Engineering: A Survey of Current, Challenges and Future~\cite{Jin2025} & Jin et al. & 2024\\
    Experimenting with Multi-Agent Software Development: Towards a Unified Platform~\cite{sami2024experimenting} & Sami et al. & 2024\\
    DevCoach: Supporting Students in Learning the Software Development Life Cycle at Scale with Generative Agents~\cite{10.1145/3657604.3664663} & Wang et al. & 2024\\
    Autonomous Agents in Software Development: A Vision Paper~\cite{rasheed2023autonomousagentssoftwaredevelopment}& Rasheed et al.& 2024\\
    CodePori: Large-Scale System for Autonomous Software Development Using Multi-Agent Technology~\cite{codepori} & Rasheed et al. & 2024\\

    Self-Evolving Multi-Agent Collaboration Networks for Software Development~\cite{hu2024self}& Hu et al. & 2024\\
    An Autonomous Multi-Agent LLM Framework for Agile Software Development~\cite{manish2024autonomous}& Sanwal et al. & 2024\\
    COMMIT0: Library Generation from Scratch~\cite{zhao2024commit0} & Zhao et al. & 2024\\
    An AI-native application assemble platform for easy-integrating of AIGC based services~\cite{10.1145/3718751.3718843} & Jin et al. & 2024\\
    RepairAgent: An Autonomous, LLM-Based Agent for Program Repair~\cite{bouzenia2024repairagent} & Bouzenia et al.& 2024\\
    SALLMA: A Software Architecture for LLM-Based Multi-Agent Systems~\cite{DBLP:conf/satrends/BecattiniVV25} & Becattini et al & 2025\\
    ProphetAgent: Automatically Synthesizing GUI Tests from Test Cases in Natural Language for Mobile Apps~\cite{prophetAgent} & Kong et al. & 2025\\
    MUARF: Leveraging Multi-Agent Workflows for Automated Code Refactoring~\cite{11024270} & Xu et al. & 2025\\
    Knowledge-Based Multi-Agen~\cite{10.1145/3696630.3728493} & Zhang et al. & 2025\\
    IDE Native, Foundation Model Based Agents for Software Refactoring~\cite{11052708} & Bellur et al. & 2025\\
    Facilitating Trustworthy Human-Agent Collaboration in LLM-based Multi-Agent System oriented Software Engineering~\cite{10.1145/3696630.3728717} & Ronanki & 2025\\
    Evaluating Agent-based Program Repair at Google~\cite{rondon2025evaluating} & Rondon et al.& 2025\\
    Engineering LLM Powered Multi-Agent Framework for Autonomous CloudOps~\cite{11030040}& Parthasarathy et al. & 2025\\
    Developing Multi-Agent LLM Applications through Continuous Human-LLM Co-Programming~\cite{11030030} & Song et al. & 2025\\
    AGILECODER: Dynamic Collaborative Agents for Software Development based on Agile Methodology~\cite{11052788} & Nguyen et al. & 2025\\
    AI-Driven Automation in Agile Development: Multi-Agent LLMs for Software Engineering~\cite{khan2025ai} & Khan et al. & 2025\\
    Multi-Agent Collaboration in AI: Enhancing Software Development with Autonomous LLMs~\cite{multi-agentcollab} & Wasif et al. & 2025\\
    Multi-Agent Collaboration via Cross-Team Orchestration~\cite{du2025multi}& Du et al. & 2025\\
    \bottomrule
    \end{tabularx}
\end{table}

\subsubsection{Review Results}
\label{sec:review_results2}

A total of 77 were identified via direct search to be considered for inclusion in the analysis; among these, 27 passed the screening phase  by applying duplicate removal, inclusion, and exclusion criteria (details listed in Table~\ref{tab:Teammate-sota}). Among these, three secondary studies were found, and they were used to complement the information synthesized by directly analyzing primary sources.

\subsection{McLuhan's Tetrad}
\label{sec:roadmap2}

To summarize the state of the art and the existing challenges in the domain of \teammate{}s on the \sdlc, the corresponding McLuhan's tetrad shown in \autoref{fig:tetrad-teammates} was built to depict the latest advances in the field and identify future directions.

\input{figures/mcluhans_tetrad_teammate}

\subsubsection{Enhances}
\label{sec:enhances2}

\begin{itemize}
    \item \textit{Software Automation}: Most research effort in Agent-based GenAI aimed a fully automating the software development process, relying on the agents to create software artifacts and descriptive documentation~\cite{sami2024experimenting}.
    Some human-in-the-loop proposals emerged, aiming to put human developers, or more generally software engineers, in control of the whole process, characterizing the person with the ability to act as a validator of the generated software~\cite{10.1145/3696630.3728717}. Other architectural proposals place the LLM agent(s) in the role of the verifier, double checking the solutions proposed by the model~\cite{rondon2025evaluating, 10.1145/3718751.3718843}.

    \item \textit{Speed}: These agentic methodologies, relying on full or semi-automation, facilitate acceleration of the \sdlc, improving developer efficiency~\cite{prophetAgent}.

    \item \textit{Fast Prototyping}: The agentic approach can achieve rapid prototyping, allowing the developer to evaluate their artifact in the early stages of the \sdlc{}~\cite{11030030}.

    \item \textit{Time To Market}: LLM generation capabilities, parallelized in an agentic approach, allow developers to quickly produce high-quality code and even entire programs, significantly expediting the \sdlc{}, reducing the development effort, and accelerating project timelines~\cite{sami2024experimenting}.

    \item \textit{Debugging Skills}: Improvements can be made in debugging competencies, which can gain from the expertise of specialized agents in the role of pair programmers, investigating the root causes of errors by elucidating error messages, thereby increasing the effectiveness of debugging~\cite{bouzenia2024repairagent, 10.1145/3657604.3664663}.


\end{itemize}
\subsubsection{Reverses}
\label{sec:reverses2}


\begin{itemize}
    \item \textit{Git Versioning}: The commonly used Git Version control in a fully automated agentic approach may lack any remaining meaning~\cite{11030040}. Its original purpose of tracking versions and facilitating parallel programming human intent in assembling software may be reversed in an environment where every modification is automatically done by an agent, especially because most architectures do not parallelize code generation (mocking human teams' structure), but rather streamline the development pipeline with specialized agents~\cite{sami2024experimenting}.

    \item \textit{Environmental Sustainability}: Heavy reliance on agentic architecture involves different LLM instances working jointly towards a common solution. Among the costs of this autonomy, an important communication overhead has to be taken into account, directly associated with potentially high energy costs, leading to environmental and sustainability concerns~\cite{qian2023chatdev}.

    \item \textit{Accountability \& Code Ownership}: Accountability challenges are partially addressed by frameworks that incorporate human oversight, holding individuals responsible for the software developed under their guidance.
    Specifically, responsibility assignment guidelines mandate that when an LLM-based multi-agent (LMA) system is assigned `Responsible' status for a task, there must be at least one human actor assigned the `Accountable' responsibility, thereby ensuring human control and validation over critical software development activities.
    However, in proposals for entirely autonomous systems, accountability continues to be an unresolved matter~\cite{10.1145/3696630.3728717, 10.1145/3696630.3728493}.

    \item \textit{Code Comprehension}: This landscape also includes situations in which individuals progressively diminish their proficiency in essential programming skills and code comprehension. Extensive codebases that were once developed through profound human expertise may now be constructed merely by depending on requirements definition: this could result in an increasing reliance on the tool, thereby losing control over the knowledge that underpins it. Furthermore, automated unsupervised solutions may prioritize functionality over readability, ultimately impairing the overall capacity for comprehension~\cite{10.1145/3696630.3728493}.

    \item \textit{Ethics and Autonomy}: A core ethical mandate is respecting Human Autonomy and ensuring Human Agency and Oversight in AI systems. This is particularly challenging because agentic systems are designed to autonomously plan and execute actions to achieve goals, invoking tools and making decisions without human initiation for every step, mimicking a human developer's workflow~\cite{bouzenia2024repairagent}. This level of autonomy necessitates structured governance to maintain control, leading to frameworks like the RACI matrix, which explicitly mandates that when an LLM-based multi-agent (LMA) system is assigned the `Responsible' status for a task, at least one human actor must be assigned `Accountable' responsibility~\cite{10.1145/3696630.3728717}. This mechanism ensures that human control is maintained over critical activities, aligning with trustworthy AI guidelines.

    Additionally, multi-agent systems, particularly those trained on large-scale code repositories, risk reinforcing existing biases in software development practices, which necessitate developing bias-mitigation techniques to ensure responsible AI deployment. Multi-agent systems may also pursue conflicting goals that should be taken into account when planning the overall system behavior. Ultimately, achieving trustworthy human-agent collaboration requires explicitly defining roles, ensuring human validation, and addressing the inherent lack of transparency in autonomous decision-making processes~\cite{khan2025ai}.

\end{itemize}

\subsubsection{Retrieves}
\label{sec:retrieves2}


\begin{itemize}

\item \textit{Software Design}: Architectural definition becomes crucial both concerning the definition of the agents' architecture to be used, and the software architecture to be implemented~\cite{sami2024experimenting, 10.1145/3696630.3728493}.
RAG mechanisms are also employed to retrieve knowledge from existing software designs (both from literature and from existing industrial documents), thereby empowering the agents' knowledge related to architectural patterns and design trade-offs, which influences their decision-making ability~\cite{DBLP:conf/satrends/BecattiniVV25}.

\item \textit{Requirement Engineering Elicitation and Modeling}: Multi-agent systems utilize roles such as the Requirement Engineer, Analyst, and Product Manager to replicate human development activities. These specialized agents actively apply knowledge by interpreting natural language user input and translating it into structured artifacts, such as precise user stories containing clear acceptance criteria~\cite{lin2024soen, DBLP:conf/satrends/BecattiniVV25}.
To make the most of the agents' skills, an effective requirements elicitation 

\item \textit{Linguistic Skills, Clarity, and Transparent Communication}: Other non-technical skills, such as linguistic ability and unambiguity, enable a clear and transparent specification of the software to be implemented. Clarity is the key aspect in all the input artifacts provided to any agentic approach to obtain a working implementation consistent with the end-user needs~\cite{qian2023chatdev}.
To enforce clarity and precision across the whole process, agents leverage specialized Prompt Engineering techniques and employ structured mechanisms like communicative dehallucination, which systematically requires the assistant agent to proactively seek more detailed suggestions from the instructor agent before providing a formal solution~\cite{sami2024experimenting}.



Although the Waterfall model is often applied in systems requiring a linear path, the agentic emulation demonstrates how these established software process models can enhance the quality and stability of LLM-generated code by organizing collaborative activities~\cite{lin2024soen}.

\item \textit{Code Quality and Technical Debt}: Given the probabilistic nature of the implementation provided by the agents, assessing software quality becomes essential to minimize code debt~\cite{qian2023chatdev}. Metric-driven assessments, refactoring practices, and code review should be used to maintain an acceptable code quality.
Furthermore, fine-grained analysis of how development activities impact code quality reveals that a dedicated phase like code review is crucial for reducing the density of code smells and improving code reliability~\cite{lin2024soen}.

\item \textit{Static Analysis}: Static quality analysis and functional validation become essential steps in creating robust code. Automatic or manual validation of AI-generated code becomes essential to avoid pitfalls~\cite{lin2024soen, rondon2025evaluating}.
Specialized agents can leverage external tools to detect and categorize code smells, providing objective feedback that guides code refinement. By integrating these specialized quality checks and iterative refinement loops, the agentic approach systematically combats the creation of low-quality, high-technical-debt code, bringing existing software quality knowledge directly into the autonomous execution cycle~\cite{sami2024experimenting}.

\item \textit{Human-Computer Interactions Principles}: As the automated generation of software through agentic AI has demonstrated the ability to create functional solutions with rather naive or no graphical interfaces, a critical challenge that must be tackled by leveraging the existing body of knowledge is the effective design of interactions between humans and technology, drawing upon principles of Human-Computer Interaction~\cite{11052708}.

\item \textit{Dependency Management}: Although some agentic approaches can autonomously handle broader project management tasks related to dependencies, such as migrating a dependency management system, updating dependency files, resolving compatibility issues, and generating necessary lock files~\cite{manish2024autonomous}, in a completely automated setting, the challenge of managing dependencies and ensuring that libraries are up to date is a significant hurdle that requires appropriate expertise to navigate~\cite{zhao2024commit0}.

\end{itemize}

\subsubsection{Obsolesces}
\label{sec:obsolesces2}

\begin{itemize}
    
    \item \textit{Manual Coding}: The obsolescence of manual coding is cemented by the agentic AI's ability to handle iterative development, debugging, and quality assurance autonomously~\cite{multi-agentcollab}. Agents autonomously generate executable code segments and iteratively refine them through collaborative exchanges and internal feedback loops. Since agentic AI effectively automates complex and time-consuming technical tasks, the role of human software engineers shifts from manual coders to that of oversight and strategic decision makers, allowing them to focus on more complex and innovative problem solving activities~\cite{sami2024experimenting, codepori}. 

    \item \textit{Project Tracking Software Tools}: Autonomous agentic environments aim to seamlessly link all phases of development, incorporating a continuous feedback loop where discussions among stakeholders are automatically transformed into actionable items, such as prioritizing features in the project's backlog, generating user stories, and suggesting test cases, all without manual intervention. This integrated and automated approach suggests a paradigm shift in which task and project management artifacts are generated and maintained fluidly within the AI development ecosystem itself, reducing the dependence on separate dedicated project tracking software~\cite{rasheed2023autonomousagentssoftwaredevelopment, sami2024experimenting}.
     
    \item \textit{Traditional IDE editors}: Although some research emphasizes that IDEs remain the "ideal place" for agents to reside because they provide critical safeguards and static analysis APIs that can be utilized as tools by the agent, the agent's ability to automate core code generation, editing, and quality assurance processes inherently reduces the constant necessity for manual file opening and editing within the visual editor interface, ultimately yielding to a chat-oriented development environment~\cite{11052708}.
    
    \item \textit{Conventional Quality Assurance Testing}: Agentic AI significantly obsolesces traditional and conventional Quality Assurance (QA) testing by integrating automated testing and verification directly into autonomous software development ~\cite{sami2024experimenting}. Multi-agent systems, leveraging specialized agents like Developers and Testers, streamline the QA process by autonomously generating and executing comprehensive test suites, thereby reducing the need for continuous manual testing efforts. This autonomous generation and execution of diverse tests, from unit and end-to-end tests to GUI and robustness checks, integrates QA as a continuous, intrinsic element of the multi-agent workflow, rather than a separate, manual phase.

    Human involvement gains a pivotal role in clearly expressing desired goals and tasks for the \teammate{}, currently in the form of prompts. 
    This prompt-based programming obsolesces traditional Integrated Development Environments and their built-in features, such as conventional debugging and testing strategies. 
    Rethinking Quality Assurance and Program Correctness strategy becomes crucial to deliver code without the burden of technical debt~\cite{sami2024experimenting}.

    Agentic Gen-AI research is clearly centered around the automatic software generation, directly from specifications. In this scenario, the creation of ``off-the-shelf'' solutions can largely lose its appeal. This can happen since adapting existing software in a specific process is more complex compared to automatically generating software tailored to users' and companies' needs. The same phenomenon can cause the obsolescence of project tracking software tools, which will likely turn into requirements definition management software to gather user needs and organize requirements to be dispatched to the agentic architecture for software evolution~\cite{10.1145/3657604.3664663}.

\end{itemize}

\section{Impact of \aiware in SE}
\label{sec:aiware}
This section reports the results of Cycle~2 of our methodology (see Section~\ref{sec:cycle2}) for the \aiware{} form of \ai{} augmentation. 
The corresponding literature review investigated how \aiware{} manifests in software systems whose functionality is partially or fully realized through \ai{} components, embedding intelligence directly into the tools and environments that support software engineering activities.

\subsection{Rapid Literature Review}
\label{sec:review3}

\subsubsection{Review Approach}
\label{sec:review_approach3}
\begin{figure}[htb]
    \centering
    \includegraphics[width=1\linewidth]{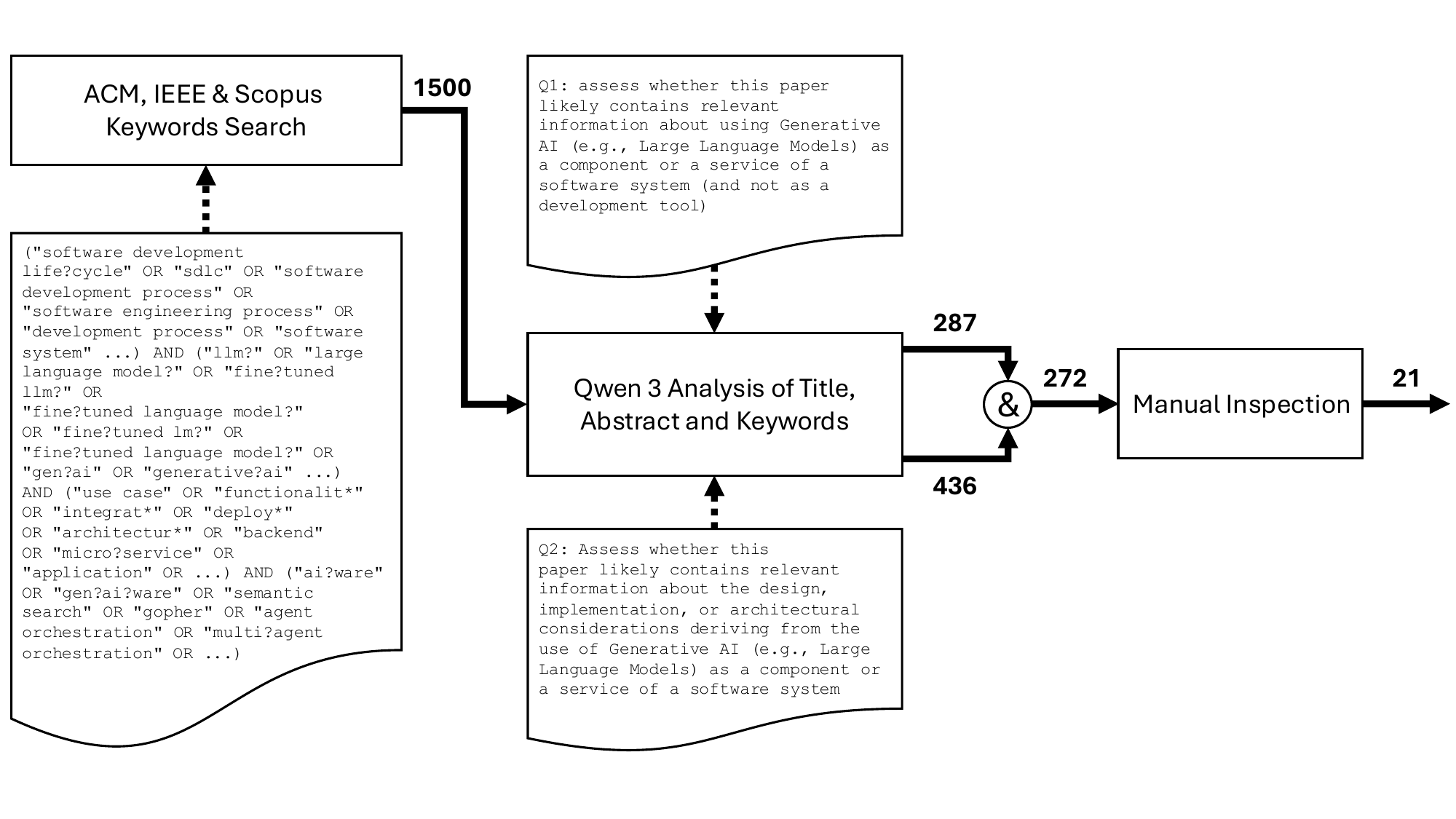}
    \caption{Identifying relevant publications concerning \aiware{}}
    \Description{Image shows the rapid literature review process to determine relevant publications.}
    \label{fig:genaiware-review}
\end{figure}

For the literature review, we followed the process depicted in \Cref{fig:genaiware-review}.
 We report further details on the search strings and the prompts to filter the results for interested readers in \Cref{sec:genaiware-searchstrings,sec:genaiware-prompts}.
 
Due to the topic of \aiware has only recently gained traction, we slightly adapted the process from Section~\ref{sec:rlr}.
We started the rapid literature review searching results on full-text and metadata from ACM and IEEE, and on metadata only on Scopus.
We focused on papers published since January 2019, following the release of the main Transformer-based language models, namely GPT and BERT.
We did not limit ourselves to the main SE venues, because \aiware is often only reflected in specific events dedicated to this topic, such as the ACM International Conference on AI-powered Software (AIware).
Yet, we only considered the top 500 results from each source for a total of maximum 1500 papers.

To find \aiware{}-related work, we adapted the search terms to reflect, besides \sdlc and \ai
\begin{enumerate*}[label=(\roman*)]
  \item software components and processes (e.g., ``deployment'', ``architecture''), and
  \item \aiware{}-specific keywords (e.g., ``prompt engineering'', ``ai-ware'')
\end{enumerate*}.

We automatically filtered the initial results using the Qwen3 LLM to help reduce the burden of manually assessing which of the papers were  relevant to the topic.
Given the title, abstract and keywords of each retrieved paper, we prompted the LLM with two questions, in as many separate contexts, covering the two aspects of \ai{} characterizing \aiware{}.
Specifically, we asked the LLM these two questions:
\begin{enumerate*}[label=(Q\arabic*)]
  \item assess whether the paper likely contains relevant information about using Generative AI (e.g., Large Language Models) as a passive element (e.g., as a component or a service of a software system), and
  \item assess whether this paper likely contains relevant information about the design, implementation, or architectural considerations deriving from the use of Generative AI (e.g., Large Language Models) as a component or a service of a software system.
\end{enumerate*}
Q1 was aimed at highlighting papers where \ai{} is augmenting the developed software system and Q2 was aimed at highlighting papers where \ai{} was integrated as a component or a service of the software system.
We retrieved the papers that received a positive answer to both questions, for a total of 272 papers.

We manually processed these LLM-selected papers by taking care of removing any remaining papers not relevant to the scope of the review.
At this step, we noticed many false positives coming from papers concerning the integration of \copilot{} into software development tools.
Additionally, we discarded all papers presenting or discussing application-specific solutions, which -- despite the fact that they represent a valuable reference for the development of \aiware{} -- are outside the scope of this review.
After this last manual filtering step, we were left with 21 papers.

\subsubsection{Review Results}
\label{sec:review_results3}

\begin{table}[htbp]
    \centering
      \caption{Relevant publications concerning \aiware{}}
   \label{tab:genaiware-sota}
   \rowcolors{2}{white}{gray!20} 
    \footnotesize
    \begin{tabularx}{\textwidth}{Xlccccccc}
    \toprule
    \multirow{2}{*}{Paper} & \multirow{2}{*}{Authors} & \multirow{2}{*}{Year} & \multicolumn{6}{c}{Covered topic(s)} \\ \cmidrule(l){4-9}
    & & & \rotatebox{90}{Requirements} & \rotatebox{90}{Architecture} & \rotatebox{90}{Integration} & \rotatebox{90}{Safety} & \rotatebox{90}{Data} & \rotatebox{90}{Prompt} \\
    \midrule
    Seven Failure Points When Engineering a Retrieval Augmented Generation System~\cite{DBLP:conf/cain/BarnettKTB024} & Barnett et al. & 2024 &  &  &  &  & \cmark &  \\
    CoPrompt: Supporting Prompt Sharing and Referring in Collaborative Natural Language Programming~\cite{DBLP:conf/chi/FengYY0ZL24} & Feng & 2024 &  &  &  &  &  & \cmark \\
    Secure Software Architecture for Enterprise Generative Artificial Intelligence~\cite{DBLP:conf/csit/Joshi24} & Joshi & 2024 &  & \cmark &  & \cmark &  &  \\
    Securing Applications of Large Language Models: A Shift-Left Approach~\cite{DBLP:conf/eit/LanKPPP24} & Lan et al. & 2024 &  &  &  & \cmark &  &  \\
    A Taxonomy of Foundation Model based Systems through the Lens of Software Architecture~\cite{DBLP:conf/cain/00010XLX024} & Lu et al. & 2024 &  & \cmark &  &  &  &  \\
    Towards Responsible Generative AI: A Reference Architecture for Designing Foundation Model Based Agents~\cite{DBLP:conf/icsa/LuZXXHW24} & Lu et al. & 2024 &  & \cmark &  &  & \cmark & \cmark \\
    Feature Model-based Integration of Machine Learning in Software Product Lines~\cite{DBLP:conf/icsa/KholkarTPR24} & Kholkar et al. & 2024 &  &  & \cmark &  &  &  \\
    Requirements Elicitation for Machine Learning Applications: A Research Preview~\cite{10771003} & Elvira at al. & 2024 & \cmark &  &  &  &  &  \\
    Tutorial on Landing Generative {AI} in Industrial Social and E-commerce Recsys~\cite{DBLP:conf/www/XuZRZYYXW25} & Xu et al. & 2025 &  &  & \cmark &  &  &  \\
    Iterative Proof-Driven Development LLM Prompt~\cite{DBLP:conf/www/Bakharia25} & Bakharia & 2025 &  &  &  &  &  & \cmark \\
    Prompts Are Programs Too! Understanding How Developers Build Software Containing Prompts~\cite{DBLP:journals/pacmse/LiangLRM25} & Liang et al & 2025 &  &  &  &  &  & \cmark \\
    Supporting Students in Prototyping AI-backed Software with Hosted Prompt Template APIs~\cite{DBLP:conf/iticse/Aveni0FH25} & Aveni et al. & 2025 &  &  &  &  &  & \cmark \\
    SALLMA: A Software Architecture for LLM-Based Multi-Agent Systems~\cite{DBLP:conf/satrends/BecattiniVV25} & Becattini et al. & 2025 &  & \cmark &  &  &  &  \\
    Beyond the Comfort Zone: Emerging Solutions to Overcome Challenges in Integrating LLMs into Software Products~\cite{11121725} & Nahar et al. & 2025 &  &  & \cmark &  &  &  \\
    Verification and Validation of LLM-RAG for Industrial Automation~\cite{11127266} & Min et al. & 2025 &  &  &  & \cmark &  &  \\
    An Architecture and Protocol for Decentralized Retrieval Augmented Generation~\cite{DBLP:conf/icsa/HeckingSF25} & Hecking et al. & 2025 &  & \cmark &  &  & \cmark &  \\
    A Functional Software Reference Architecture for LLM-Integrated Systems~\cite{DBLP:conf/icsa/BucaioniWHL025} & Bucaioni et al. & 2025 &  & \cmark & \cmark &  &  &  \\
    Are LLMs Correctly Integrated into Software Systems?~\cite{DBLP:conf/icse/Shao0S00025} & Shao et al. & 2025 &  &  & \cmark &  &  &  \\
    Towards Retrieval-Augmented Large Language Models: Data Management and System Design~\cite{11113067} & Fan et al. & 2025 &  & \cmark &  &  & \cmark &  \\
    Development of AI Agent Based on Large Language Model Platforms~\cite{11082094} & Chen et al. & 2025 &  &  & \cmark &  &  &  \\
    DAWN: Designing Distributed Agents in a Worldwide Network~\cite{DBLP:journals/access/AminiranjbarTWPV25} & Aminiranjbar et al. & 2025 &  & \cmark &  &  &  &  \\
    \bottomrule
    \end{tabularx}
\end{table}

We report the main information about the papers retrieved from the rapid literature review in \Cref{tab:genaiware-sota}.
The most covered topics were architectures for and integration of \ai{} in a software system.
This was followed by topics related to prompt engineering and information retrieval.
Finally, the few remaining papers discuss requirements, security and safety of \ai{}-augmented software.

From the architecture perspective, several patterns have been presented to exploit existing models depending on the target task, often with an agent-oriented perspective~\cite{DBLP:conf/icsa/LuZXXHW24}.
These patterns for architectures, mainly thought for LLM, prescribe how to manage memory (short- and log-term), schedule model queries and execution, possibly with accessory plugins (like guardrails).
Frameworks like DAWN~\cite{DBLP:journals/access/AminiranjbarTWPV25} or SALLMA~\cite{DBLP:conf/satrends/BecattiniVV25} focus on distribution and multi-agent use cases.
Finally, security is becoming a concern already at architectural-level when the software is backed by \ai{}, especially when moving to enterprise-level solution~\cite{DBLP:conf/csit/Joshi24}.

The challenges concerning the integration are the lack of detailed interface specifications and the variability of the model's output~\cite{DBLP:conf/icse/Shao0S00025}, which often cause defects affecting functionality, efficiency and security~\cite{DBLP:conf/icse/Shao0S00025}.
As a result, traditional software engineering practices are exposed to new failure modes and the subsequent need for testing, hence requiring manual effort, which in turn leads to inherent subjectivity in evaluating LLM outputs~\cite{11121725}.
Emerging solutions are pushing for clean component interfaces, as well as input preprocessing and output postprocessing~\cite{DBLP:conf/icse/Shao0S00025}.
On top of these issues, there are those related to the integration of the Machine Learning processes into the pipeline, which require accounting for data dependencies management and Machine Learning model lifecycle~\cite{DBLP:conf/icsa/KholkarTPR24}.

As mentioned before, memory plays a special role~\cite{DBLP:conf/icsa/LuZXXHW24}.
Long-term memory plays a central role in dealing with challenges given by hallucinations, outdated knowledge, and domain-specific gaps, thus the introduction of Retrieval Augmented Generation (RAG)-powered applications for information retrieval~\cite{DBLP:conf/cain/BarnettKTB024}.
These applications enable better reliability and trustworthiness of LLM-powered software. Building these applications implies the choice of the appropriate data store (e.g., vector databases for semantic search, that is, replacing traditional lexical search) and the correct preprocessing of the stored documents~\cite{DBLP:conf/cain/BarnettKTB024}.
User data can be stored and retrieved as well, introducing the need for privacy-related requirements.

Finally, we are witnessing the development of prompt programming, as prompts for LLMs are becoming  a meticulously engineered part of the code-base, functioning as programs themselves~\cite{DBLP:journals/pacmse/LiangLRM25}.
What makes prompting challenging is their ever-evolving nature as they strongly depend on the LLM they are given to, which may change, and on the current knowledge about that LLM's capabilities, which is again changing through time.
Only recently, approaches for rigorous prompt engineering~\cite{DBLP:conf/www/Bakharia25} and prompt sharing~\cite{DBLP:conf/chi/FengYY0ZL24,DBLP:conf/iticse/Aveni0FH25} are emerging, leaving otherwise the task subject to a naive trial and error approach~\cite{DBLP:journals/pacmse/LiangLRM25}.

Additionally, emerging topics concerning include requirements engineering, safety and security.
Requirements are affected by the stochastic and probabilistic nature of Machine Learning models in general, leading to the need to define a ``window of acceptable behavior'' or ``soft-goals'' rather than traditional hard constraints~\cite{10771003}.
These soft constraints are also applied in terms of the safety of the generated content, as challenges about the generation of toxic, untruthful or unfaithful~\cite{DBLP:journals/csur/JiLFYSXIBMF23} content do not allow for 100\% solutions~\cite{DBLP:conf/csit/Joshi24,DBLP:conf/eit/LanKPPP24}.
Nevertheless, some use cases still require hard constraints which result in tailored Verification and Validation frameworks~\cite{11127266}.

\subsection{McLuhan's Tetrad}
\label{sec:roadmap3}

\input{figures/mcluhans_tetrad_genaiware}

Our rapid literature review informed each quadrant of the tetrad as shown in Figure~\ref{fig:tetrad-genaiware}.

\subsubsection{Enhances}
\label{sec:enhances3}
\begin{itemize}
    \item \textit{Prompt Programming}: The introduction of \ai{} has altered the programming paradigm, extending it to include dynamic prompts and data, along with code.  
    Prompts allow for the programming of the behavior of LLMs, resulting in a form of natural language coding or prompt programming~\cite{DBLP:journals/pacmse/LiangLRM25}.
    As a result, prompt engineering developed into an actual research area, providing structured approaches to create and (re-)use prompts and strategies to manage the context~\cite{DBLP:conf/chi/FengYY0ZL24,DBLP:conf/www/Bakharia25,11082094}.
    At the same time, data are necessary to prompt and evaluate these generative models.
    RAG and few-shot learning rely on input data to construct the prompt and manage the context as well.
    \item \textit{Data Management}: Data management, especially information retrieval, has undergone a strong makeover~\cite{DBLP:conf/icsa/HeckingSF25,11113067}. 
    Semantic search approaches are now integrated into search engine architectures, pushing for the enhancement of vector databases to support the new indexing approach.
    LLMs have become the new interface to unstructured and semi-structured data stores and, at the same time, LLMs result in more reliable and trustworthy having access to updated and trusted data sources~\cite{11127266,DBLP:conf/csit/Joshi24}.
    \item \textit{User Experience}: In addition to search engines, the integration of AI assistants in existing applications enhances the user experience, providing an interactive interface that can search the application and respond in a human-understandable language~\cite{DBLP:conf/icsa/BucaioniWHL025,11121725}.
    This experience is further improved by improved customization.
    In fact, user-specific or domain-specific data can be plugged into the generative model without needing any training or fine-tuning.
    \item \textit{Unified Software Design}: On the design and architecture side, we are finally observing steps toward unified interfaces for these models~\cite{DBLP:conf/icsa/BucaioniWHL025,DBLP:journals/access/AminiranjbarTWPV25}.
    This unification has multiple positive effects.
    It helps the modularization of software architectures using these models, it enhances the automation of deployment and testing and it helps the interoperability with and among agentic approaches~\cite{DBLP:conf/satrends/BecattiniVV25,DBLP:journals/access/AminiranjbarTWPV25}.
    \item \textit{Responsible and Secure AI}: The integration of AI modules in software systems and applications is raising awareness towards the risks connected to its use.
    As a result, we are observing interests in non-functional aspects like responsible AI and security~\cite{DBLP:conf/icsa/LuZXXHW24,11121725}.
    Security is not only limited to data privacy, but also to the safety of user inputs and generated content, leading to the development of dedicated testing approaches and guardrails~\cite{11121725}.
\end{itemize}

\subsubsection{Reverses}
\label{sec:reverses3}

\begin{itemize}
    \item \textit{Reliability}: \ai{} models are de facto probabilistic tools; this nature clashes with the control, predictability, and replicability of traditional software.
    For example, LLMs are stochastic~\cite{DBLP:conf/cain/BarnettKTB024,10771003,11127266}, which introduce new challenges for validating \ai{}-augmented software~\cite{DBLP:conf/cain/BarnettKTB024,DBLP:conf/icse/Shao0S00025}.
    At the same time, the issue of hallucinations may compromise trust in \aiware{}~\cite{DBLP:conf/cain/BarnettKTB024,DBLP:conf/eit/LanKPPP24,11127266}.
    Moreover, prompts, being fragile, model-dependent, and context-dependent, often push the development toward rapid, unsystematic, trial-and-error practices rather than structured processes~\cite{DBLP:journals/pacmse/LiangLRM25}. 
    \item \textit{Standardization}: The existence of multiple, independent, and different interfaces to \ai{} models challenges modularity, creating difficulties in system composition and management due to insufficient specifications and varied contextual requirements~\cite{DBLP:conf/icse/Shao0S00025,DBLP:conf/icsa/BucaioniWHL025,DBLP:conf/csit/Joshi24}. Several solutions, such as the MCP (Model Context Protocol) and A2A (Agent-to-Agent Protocol), have been proposed; however, the absence of a (yet) universally adopted framework enforcing consistent interface definitions and interoperability across platforms and vendors has resulted in fragmented ecosystems.
    \item \textit{Transparency and Trustworthiness}: Quality attributes and the development process are also affected.
    The opaque nature of \ai{} models obscures transparency and lack of control, complicating fault localization and scattering responsibility across multiple components~\cite{DBLP:conf/icsa/LuZXXHW24,DBLP:journals/pacmse/LiangLRM25,DBLP:conf/cain/00010XLX024,DBLP:conf/icsa/HeckingSF25,DBLP:conf/icsa/KholkarTPR24}.
    Furthermore, biases arising from training data manifest in outputs that can be discriminatory or unfair, making ethics an ongoing challenge and a fixed requirement~\cite{DBLP:conf/eit/LanKPPP24,DBLP:conf/icsa/LuZXXHW24,DBLP:conf/cain/00010XLX024}. 
    \item \textit{Efficiency and Sustainability}: From an operational standpoint, efficiency and sustainability are reversed in exchange for escalating costs and more difficult maintenance.
    Relying on external APIs may introduce latency and high costs, while local deployment of large \ai models may require significant computational and human resources~\cite{DBLP:conf/cain/BarnettKTB024,DBLP:journals/pacmse/LiangLRM25,11121725}.
    As these models evolve, their behavior drifts over time, demanding continuous calibration, monitoring, and adaptation.
    This dynamism disrupts established workflows and erodes long-term stability, reversing the traditional pursuit of cost-effectiveness and reliability in software systems~\cite{DBLP:conf/cain/BarnettKTB024,DBLP:journals/pacmse/LiangLRM25,DBLP:conf/icsa/LuZXXHW24,DBLP:conf/cain/00010XLX024}.
\end{itemize}

\subsubsection{Retrieves}
\label{sec:retrieves3}
\begin{itemize}
    \item \textit{Manual evaluation}: Development once again becomes grounded in human-intensive, iterative practices, with prompt engineering emerging as a central trial-and-error activity where prompts are often crafted and revised to enforce the desired agent behavior~\cite{DBLP:conf/icsa/LuZXXHW24,DBLP:conf/cain/00010XLX024,DBLP:journals/pacmse/LiangLRM25}.
    Much of the manual effort that was thought to be lost resurfaces in new forms, particularly in data annotation and curation, which become essential for building domain-specific, high-quality datasets~\cite{DBLP:journals/pacmse/LiangLRM25,11121725}.
    Evaluation and testing similarly shifts back towards more subjective, qualitative inspection, as the adequacy of generated outputs can be often evaluated only through human qualitative analysis~\cite{DBLP:journals/pacmse/LiangLRM25}.
    \item \textit{Runtime Verification and Validation}: Architecturally, \aiware{} retrieves and reshapes persistent software engineering challenges.
    The dynamic and probabilistic nature of \ai{} revives the need for continuous runtime evaluation and adaptive verification and validation frameworks~\cite{11121725,DBLP:conf/csit/Joshi24,11127266}.
    \item \textit{{Interoperability Issues}}: The challenges of integrating opaque machine learning elements return emphasis on transparency~\cite{DBLP:journals/pacmse/LiangLRM25}.
    \item \textit{Document Stores}: When using RAG (retrieval augmented generation), long-standing document stores and search infrastructures regain prominence, now replaced by vector databases that enable semantic search and provide external, trusted, and verifiable knowledge to complement internal \ai model memories~\cite{DBLP:conf/cain/BarnettKTB024}.
    In this way, \aiware{} not only revives but also reconfigures older practices and concerns of information retrieval systems.
    \item \textit{Goal-Oriented Requirements}: Concerning requirements engineering, rigid and deterministic specifications no longer align with the stochastic and unpredictable behavior of LLMs. Goal-oriented requirements engineering approaches, which focus on soft goals and acceptable behavior ranges, could be adapted to address the inherent uncertainty and flexibility required in the development of the ML system~\cite{10771003}. 
\end{itemize}

\subsubsection{Obsolesces}
\label{sec:obsolesces3}

\begin{itemize}
    \item \textit{Manual Annotation}: In information retrieval, the rise of RAG reduces the need for labor-intensive manual annotation, further eroding the role of static methods.
    This last effect extends to many cases that require human-in-the-loop approaches~\cite{DBLP:conf/cain/BarnettKTB024}.
    \item \textit{Traditional Static Testing}: In terms of quality assurance, deterministic test-driven development is challenged by the probabilistic and uncertain outputs of \aiware{}, whose evaluation goes beyond a simple binary pass/fail metric~\cite{11127266}.
    Moreover, static, quantitative evaluation techniques also lose relevance, as prompt programming and RAG systems demand runtime, qualitative, and operational assessment as the underlying \ai{} model and data change~\cite{DBLP:conf/cain/BarnettKTB024}.
    Advances in vector databases reinforce this shift, making traditional storage, indexing, and retrieval mechanisms less capable of supporting rapid changes in models and data~\cite{DBLP:conf/icse/Shao0S00025,DBLP:conf/icsa/BucaioniWHL025}.
    \item \textit{Static Code Development}: At the process and organizational level, the experimental nature of prompt programming makes the prompt engineering stage detached from the slow and rigid code review cycles, which turn out less suitable for practical development~\cite{DBLP:journals/pacmse/LiangLRM25}.
    \item \textit{Separation between Machine Learning Cycle and \sdlc{}}: The long-standing separation of machine learning development from the broader \sdlc{} is becoming obsolete as integration of AI-based elements becomes relevant across all stages, from requirements to deployment~\cite{DBLP:conf/icsa/KholkarTPR24}.
    \item \textit{Opacity of AI elements} the demand for responsible AI, transparency, and explainability makes purely opaque integration unsustainable, replacing it with approaches that emphasize explainers, guardrails, and continuous monitoring to ensure trustworthiness and accountability~\cite{DBLP:conf/csit/Joshi24,DBLP:conf/cain/00010XLX024,DBLP:conf/icsa/LuZXXHW24,DBLP:conf/icsa/BucaioniWHL025}.
\end{itemize}

\section{Impact of \robot{}s in SE}
\label{sec:robots}
This section presents the results of Cycle~2 of our methodology (see Section~\ref{sec:cycle2}) concerning the \robot{} form of \ai{} augmentation. 
The corresponding literature review examines how (semi-)autonomous and goal-driven agents deliver part of the functionality of software systems or applications. 
As discussed in Section~\ref{sec:taxonomy}, \robot{}s differ from \teammate{}s in that they are not involved in the development process itself but become integral elements of the resulting software product, operating as self-contained AI components once deployed.

\subsection{Rapid Literature Review}
\label{sec:review4}

Given the wide range of application domains in which software systems can be realized to deliver parts of their functionality through \robot{}s, we opt for a broad search and review, as explained in the following.

\subsubsection{Review Approach}
\label{sec:review_approach4}
\begin{figure}[htb]
    \centering
    \includegraphics[width=1\linewidth]{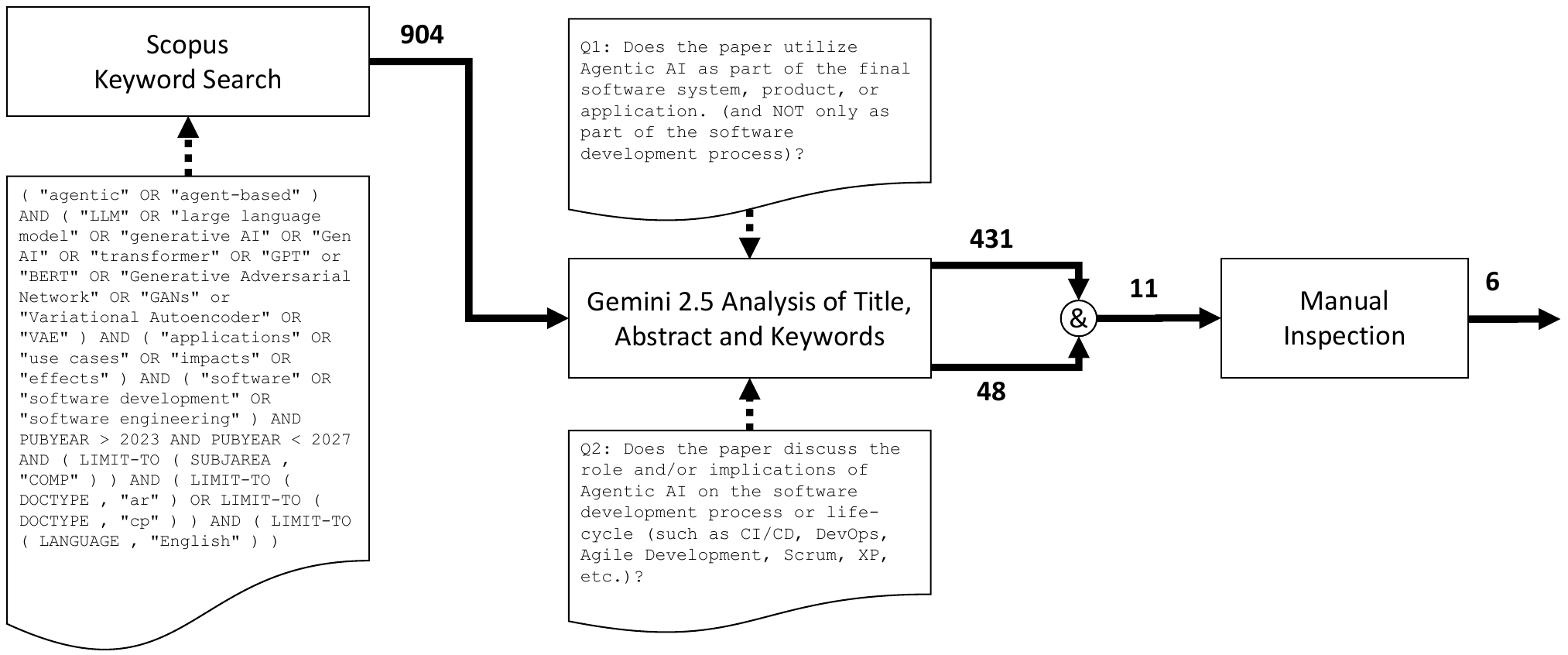}
    \caption{Identifying relevant publications concerning \robot{}s}
    \Description{Image shows the rapid literature review process to determine relevant publications.}
    \label{fig:bot-review}
\end{figure}

The rapid literature review approach we performed for \robot{}s is depicted in Fig.~\ref{fig:bot-review}.

We started with a keyword search using Elsevier's Scopus database (note that Scopus, since 2021, also includes arXiv preprints).
The keyword search gave us fine-grained control over the search terms (providing synonyms for agentic AI and requiring \ai{} as a technique), while also being able to explicitly state inclusion criteria (publications from 2024 and beyond) and exclusion criteria (only articles and conference publications in English).
This Scopus search resulted in 903 publications.

As a second step, we then employed Google's Gemini 2.5 Flash via its programmatic API to aid us in the analysis process.
We prompted Gemini to scrutinize the title, abstract and keywords to answer two specific questions: (Q1) whether the paper actually covers the use of \ai{} as a Robot (and not as a \teammate, for instance), and (Q2) whether the paper explicitly discusses impacts or implications for the SDLC.
For each of these questions, we not only asked for a yes/no answer but also required Gemini to provide its rationale for giving the respective answer.
The full prompt is provided in Appendix~\ref{sec:bots-prompts}.
This resulted in 11 publications for which both answers were yes.

Finally, we manually inspected these 11 publications and eliminated the ones that were out of scope, resulting in 6 relevant publications.

\subsubsection{Review Results}
\label{sec:review_results4}
The relevant publications on \robot{}s are listed in Table~\ref{tab:bot-sota}.

\begin{table}[htb]
    \centering
      \caption{Relevant publications concerning \robot{}s}
   \label{tab:bot-sota}
   \rowcolors{2}{white}{gray!20} 
    \footnotesize
    \begin{tabularx}{\textwidth}{Xlcll}

    \toprule
    Paper & Authors & Year & Application Domain & SDLC Aspects \\
    \midrule
    A Conversational Assistant Framework for Automation~\cite{DuesterwaldIJKM24} & Duesterwald et al. & 2024 & BPM & Overall: domain-specific lifecycle\\
    OMACS Based Adaptive Intelligent Tutoring System Development~\cite{Gowri25} & Gowri & 2025 & Education & Overall: domain-specific lifecycle\\
    When and How to Use AI in the Design Process? Implications for Human-AI Design Collaboration~\cite{LeeLH25} & Lee et al. & 2025 & (Industrial) Design & Overall: domain-specific lifecycle \\
    Retail Resilience Engine: An Agentic AI Framework for Building Reliable Retail Systems With Test-Driven Development Approach~\cite{MishraS25a} & Mishra et al. & 2025 & Retail & Single Task: testing \\
    Business Compliance Detection of Smart Contracts in Electricity and Carbon Trading Scenarios~\cite{WuWZLWFL24} & Wu et al. & 2024 & Electricity & Single Task: compliance verification\\

    \end{tabularx}

\end{table}

We observe that given the very low number of relevant publications, \robot{}s are clearly an emerging area.
Even though, \robot{}s were used in several different application domains.
Two of the papers focus on individual tasks, i.e., on isolated aspects of the lifecycle. 
For example, Mishra et al. focus on the task of testing retail systems.
Three of the papers address broader lifecycle aspects by introducing domain-specific development processes.
For example, Lee et al. integrate the use of \robot{}s into the Double Diamond lifecycle from (industrial) design\footnote{\url{https://www.designcouncil.org.uk/our-resources/the-double-diamond/}}.
However, these papers lack alignment and integration with an overall software development lifecycle.

\subsection{McLuhan's Tetrad}
\label{sec:roadmap4}
\input{figures/mcluhans_tetrad_robot}

As seen above, there is only little existing work on how \robot{}s may impact SE, indicating that this is a nascent research area. 
What this means is that the below McLuhan's tetrad and roadmap items are necessarily more speculative than those for the other, more established forms. 
To accommodate the fewer existing work, we also explore more general challenges identified for \robot{} in the literature, leading to the following items of the tetrad, also shown in Figure~\ref{fig:tetrad-robots}.

\subsubsection{Enhances}
\label{sec:enhances4}
\begin{itemize}
    \item \textit{Novel forms of human-computer interaction (HCI):} 
    Having agents -- in the form of \robot{}s -- deliver parts of the software systems' functionality opens up novel avenues for HCI.
    Instead of humans initiating the interaction with the "computer" (i.e., software), it can now be the \robot{}s that initiate this interaction.
    This leads to several different forms of human-agent interactions that may be chosen to realize a software application.
    As an example, Sauvola et al. describe four different scenarios with increasing AI participation~\cite{10.1007/s10515-024-00426-z}.
    As another example, Stanford's Social and Language Technologies Lab (SALT) envisions five levels of such interaction, ranging from fully human to fully AI task completion\footnote{\url{https://futureofwork.saltlab.stanford.edu/}}.
    Altogether, this raises the question of where in the SDLC such a decision is taken, or whether there is a need to adapt and evolve the initial decision.

    \item \textit{Increased human-AI team performance:}
    Combining \robot{}s and humans in a complementary manner offers an increase in overall team performance, i.e., a level of performance that neither of them can individually achieve~\cite{Hemmer2025}.
    Or, in other words, we see the emergence of hybrid intelligence, where  humans and \ai  complement each other rather than compete.
    Of course, this raises the question of when and how to determine what such a human-AI team should look like.

\end{itemize}

\subsubsection{Reverses}
\label{sec:reverses4}
\begin{itemize}

    \item \textit{Human-agent balance:}
    One key concern is how best to leverage AI capabilities and human expertise, thus increasing efficiency while maintaining human judgment and creativity~\cite{Vu25}.
    This requires defining clear roles and responsibilities between humans and \robot{}s, including identifying situations where human intervention is necessary; for example, in the case of failure of \robot{} s~\cite{Vu25}.
    In general, human-agent teaming raises questions to be answered during the SDLC, such as whether, for a given task, human actors will collaborate with agents as peers or whether humans take on supervisory roles with selective intervention.

    \item \textit{Traditional separation of concerns:}
    The SDLC has to provide activities to determine which parts of the functionality of the application software will be carried out by \robot{}s and which parts of manual human tasks can be automated by \robot{}s.
    This may go significantly beyond the traditional SE concept of separation of concerns, such as object orientation, and may require a much deeper analysis.
    On the one hand, it may require analyzing historical data of how previous applications were used~\cite{Fournier2025}.
    On the other hand, it may require clear metrics and performance indicators such that the impact of \robot{}s on process efficiency (e.g., in terms of cost and time, decision quality, and broader organizational outcomes) can be measured at run time~\cite{Vu25}.

    \item \textit{Static assurance of legal and ethical norms:}
    It must be ensured that \robot{}s comply with legal norms, while considering the context and environment in which they operate~\cite{Vu25}.
    Given the greater degree of autonomy of \robot{}s, using static rules to ensure compliance with these norms may no longer be feasible, making \robot{} s find unexpected and potentially harmful ways to achieve their goals~\cite{Gabriel25}.
    We thus require "guardrails", which codify norms in such a way as to ensure compliance also in unspecified, unstructured situations~\cite{grisold2025}.
    As abstractions for managing such guardrails, the notion of (normative) frames may be explored.
    In contrast to more operational notions of declarative or procedural abstractions, frames rely on deontic logic, which specifies obligations, permissions, and prohibitions\footnote{\url{http://www.dagstuhl.de/25192}}.
    The definition of guardrails at design time should be complemented by safety measures that monitor \robot performance at run time, thus helping to observe and prevent unwanted adaptations or evolutions of \robot{}s~\cite{Vu25}.

\end{itemize}

\subsubsection{Retrieves}
\label{sec:retrieves4}
\begin{itemize}
    \item \textit{Agent-oriented Software Engineering (AOSE):}
    Now that powerful \ai{} makes the vision of agent-based systems realistic, it is worth revisiting SE research on building agent-based systems, which was conceived more than two decades ago~\cite{Jennings00,WooldridgeC00}.
    The principles, methods and processes of agent-oriented software engineering (AOSE) provide a strong foundation for novel approaches to developing, operating, and maintaining \robot{}s.
    An example is the Gaia methodology~\cite{wooldridge1999}, intended to allow a software engineer to go systematically from a statement of requirements to a design that is sufficiently detailed for it to be implemented.
    Gaia provides an agent-specific set of concepts through which a software engineer can understand and model a complex system, such as responsibilities, activities, interaction protocols, and permissions.
\end{itemize}

\subsubsection{Obsolesces}
\label{sec:obsolesces4}
\begin{itemize}
   \item \textit{Natural language AI-to-AI communication:}
   If \robot{}s communicate with other \robot{}s or traditional software, it is no longer necessary for them to converse in natural language.
   Other forms of communication may become much more effective.
   As an example, \robot{}s realized via LLM may directly exchange the tokens and not the original natural-language input.
   As a more extreme example, interacting \robot{}s may even come up with their own efficient language, as demonstrated by the Gibberlink project\footnote{\url{https://olafwitkowski.com/2025/03/11/ai-to-ai-communication-unpacking-gibberlink-secrecy-and-new-ai-communication-channels/}}.
    This raises the question when and how in the SDLC to balance efficiency with transparency, i.e., the capability for human understanding of the AI-to-AI language.
\end{itemize}

\section{Research Roadmap}
\label{sec:commonalities}

Resulting as an outcome of Cycle 3 of our methodology (see Section~\ref{sec:cycle1}), this section synthesizes an overall research roadmap on \ai augmentation in SE. 
To do this, we analyze the four McLuhan tetrads and identify transversal patterns, recurring themes, and complementary insights.
The roadmap includes the four forms of \ai augmentation in SE with the associated challenges and opportunities.
The mapping of these challenges and opportunities to their sources (i.e., the analysis via the McLuhan's tetrad) offers a valuable lens for understanding why certain phenomena emerge and how their side effects can be anticipated, for instance, how an innovation that enhances a practice may simultaneously reverse or obscure critical aspects, creating new needs for interoperability and compliance. 
This dual perspective, operational and theoretical, supports a multidisciplinary research agenda. 

The overall roadmap consists of five parts, Parts A--D (Sections~\ref{sec:A}--\ref{sec:D}) for each form of \ai augmentation, and Part X (Section~\ref{sec:E}) for cross-form aspects.

\subsection{Roadmap Part A: \copilot{}s}\label{sec:A}
Table~\ref{tab:copilot_challenges_opportunities} shows the challenges and opportunities in augmenting the process with \copilot{}s. The table systematizes key research challenges in the integration of \ai into software development workflows, along with corresponding opportunities for innovation. The challenges span both technical and social dimensions, ranging from the specification of reliable interaction interfaces (A1) and mechanisms for traceability and accountability (A2) to quality assurance of generated artifacts (A3) and the long-term maintenance of evolving AI elements (A4). Several entries highlight the need to embed \copilot{}s seamlessly into established development processes (A5), while maintaining developer agency and transparency (A6). Broader concerns include addressing bias and fairness in domain-specific contexts (A7) and enabling collaborative work across multiple stakeholders (A8). Finally, systemic risks are considered, such as dependency management, safety, and security (A9), as well as the fundamental question of how to establish trustworthiness in \ai-assisted development (A10). 
The outlined opportunities point to a research agenda encompassing formal abstractions, hybrid verification techniques, orchestration mechanisms, debiasing methods, lightweight inference strategies, and governance frameworks, all aimed at creating reliable, sustainable, and accountable \copilot{}s for software engineering.

\begin{table}[htbp]
  \centering
    \caption{Challenges and Opportunities for \copilot{}s}
  \label{tab:copilot_challenges_opportunities}
  \rowcolors{2}{gray!20}{white} 
 {\footnotesize 
 \renewcommand{\arraystretch}{1.2}
  \begin{tabular}{p{3.2cm}p{5.5cm}p{5.5cm}}
    \toprule
    \textbf{Form of \ai augmentation} & \textbf{Challenge} & \textbf{Opportunity} \\ \midrule

    A1. Prompt engineering and specification clarity & How to systematically design prompts or APIs for GenAI elements so outputs are reliable, traceable, and modifiable? & Create formal abstractions or domain-specific prompt languages that can be composed, validated, and versioned in software pipelines. \\ 
    
    A2. Traceability, provenance, rationale, and audit trails & When a GenAI tool produces, e.g., code snippets, designs, test cases, how do you trace which input tokens or model internals led to it? & Embedding provenance metadata or ``explanation traces'' in software artifacts to enable accountability, debugging, and compliance. \\ 
    
    A3. Quality control \& validation of AI-generated artifacts & Ensuring that AI-generated process artifacts (e.g., requirements drafts, design sketches, code suggestions) are correct, consistent, non‑contradictory, and align with system constraints. & Hybrid verification approaches (symbolic + statistical) to validate, constrain, or filter AI outputs before acceptance. \\ 
    
    A4. Model drift, update, version management, and maintenance of GenAI elements & Over time, the context, codebase, or domain evolves; the GenAI elements must be re‑aligned. How to manage this versioning, continuous learning, and validation? & Develop continuous integration pipelines for AI models (CI/AI), similar to CI/CD, to monitor performance, retrain, revoke, or rollback when degradation occurs. \\ 
    A5. Workflow integration and orchestration & Embedding GenAI process tools into conventional development workflows (CI, issue trackers, code review) without disrupting developer flow. & Research orchestration layers or microservices that interleave GenAI calls transparently in Dev pipelines, handling context, caching, and incremental updates. \\ 
    
    A6. Cognitive load, transparency, and user control & Even as GenAI offers suggestions, developers must remain in control. Too much opacity or ``black‑box'' behavior may erode trust or lead to overreliance. & Design user interfaces and interaction modalities (e.g., ``explain why this suggestion'') that balance automation and human oversight effectively. \\ 
    
    A7. Bias, fairness, and domain alignment & The training data or model biases may propagate into process artifacts (requirements, testcases, design decisions). & Techniques to detect, mitigate, or ``debias'' generated outputs in domain‑specific engineering settings (e.g., safety‑critical, regulated systems). \\ 
    
    A8. Collaborative work and multi‑user context merging & When multiple developers or stakeholders interact with GenAI suggestions in parallel, merging outputs can lead to conflicts. Individual contexts (e.g., changes, actions, perspectives) need to be properly merged or synchronized. & Techniques to merge multiple suggestion threads, version control for AI outputs, context-aware collaborative ``AI‑augmented'' tools capable of merging changes so that the final result makes sense to everyone, ensuring everyone sees a consistent view of the shared work. \\ 
    
    A9. Safety, security, \& dependency risk & If the GenAI element itself becomes a dependency (e.g., an external API), what happens if it fails, changes, or is compromised? & Formalizing fallback strategies, sandboxing, and verification for AI-assisted process tools. \\ 
    
    A10. Trustworthy \ai{} & How to build \ai{} elements so that outputs are trustworthy? & Identify even less powerful, generic, and constrained \ai{} that is trustworthy under specific assumptions and in specific contexts.\\ \bottomrule
    
  \end{tabular}
  }

\end{table}

\begin{table}
    \centering
        \caption{Mapping between forms of \ai augmentation and its sources for \copilot{}s}
    \label{tab:mappingGenAICopilot}
      \rowcolors{2}{gray!20}{white} 
     {\footnotesize 
     \renewcommand{\arraystretch}{1.2}
    \begin{tabular}{p{6.0cm} p{8.4cm}}
    \toprule
        {\bf Research Challenges and Opportunities} & {\bf Source (McLuhan's tetrad)} \\ 
        \midrule
        
        A1. Prompt engineering and specification clarity & Trustworthiness and Reliability of the Reverses quadrant. \\
        
        A2. Traceability, provenance, rationale, and audit trails &  Explainability of the Reverses quadrant.\\
        
        A3. Quality control \& validation of AI-generated artifacts & Compliance of the Reverses quadrant.\\
        
        A4. Model drift, update, and maintenance of GenAI elements & Compliance of the Reverses quadrant.\\
        
        A5. Workflow integration and orchestration & Strict boundaries between SDLC stages of the Reverses quadrant. 
        \\
        
        A6. Cognitive load, transparency, and user control & Accountability and Trustworthiness \& Reliability of the Reverses quadrant \\
        
        A7. Bias, fairness, and domain alignment & Fairness of the Reverses quadrant.\\
        
        A8. Collaborative editing and multi-user context merging & Project \& Process Management and Human \& Team Factors of the Enhances quadrant. \\
        
        A9. Safety, security, \& dependency risk & Security of the Reverses quadrant.\\
        A10. Trustworthy GenAI & Trustworthiness and Reliability of the Reverses quadrant. Runtime Verification and Validation of the Retrieves quadrant.\\
        
        \bottomrule
        
    \end{tabular}}

\end{table}

Table~\ref{tab:mappingGenAICopilot} maps the research challenges and opportunities to the source that triggered the form of \ai Copilot augmentation. The source can be one or more elements in the McLuhan's tetrad. 
A clear and pragmatic research framework naturally emerges from the mapping. The introduction of GenAI elements as copilots in software development processes brings both significant technical opportunities (formal prompt abstractions, provenance tracking, hybrid verification, CI/AI pipelines, orchestration tools) and systemic and social risks (opacity, bias, dependency, degradation over time, and collaborative conflicts). Entries A1--A10 entries are not isolated; they form an interconnected set of challenges that require both engineering and organizational solutions. Only through controlled experimentation, empirical evidence, and collaboration across software engineering, machine learning, HCI, ethics, and regulatory disciplines will it be possible to turn these opportunities into truly useful, reliable, and adoptable \copilot{}s for software development. This section therefore closes with a call to pursue the operational directions outlined above, balancing technical innovation with socio-technical governance to build copilots that assist, rather than replace, human judgment and responsibility.


\subsection{Roadmap Part B: \teammate{}s}\label{sec:B}

Table~\ref{tab:agent_management} outlines the central challenges and opportunities in augmenting the process via autonomous agents (\teammate{}s)
. 
The issues primarily concern the balance between autonomy and human oversight (B1), conflict resolution in multi-agent or human–agent teams (B2), and the ability to engage in long-horizon, multi-step planning across development phases (B3). Questions of accountability and responsibility attribution (B4) as well as mechanisms for recovery, rollback, and self-diagnosis (B5) highlight the need for formal governance and resilience strategies. Additional challenges include coordination across heterogeneous agents operating on different parts of the software lifecycle (B6) and the calibration of human trust in agent recommendations (B7). The opportunities point to the development of supervisory architectures, negotiation protocols, embedded planning capabilities, formal responsibility frameworks, self-monitoring mechanisms, and coordination protocols, all aimed at creating safe, transparent, and effective human-agent collaboration in complex engineering workflows.

\begin{table}[htb]
  \centering
    \caption{Challenges and Opportunities for \teammate{}s}
  \label{tab:agent_management}
  \rowcolors{2}{gray!20}{white} 
 {\footnotesize 
 \renewcommand{\arraystretch}{1.2}
  \begin{tabular}{p{3.2cm} p{5.5cm} p{5.5cm}}
    \toprule
    \textbf{Form of \ai augmentation} & \textbf{Challenge} & \textbf{Opportunity} \\ \midrule
    
    B1. Agent autonomy vs human oversight balance & Deciding when the agent should act autonomously versus requesting human confirmation; how to avoid undesirable emergent behavior. & Develop meta‑control policies or supervisory architectures that mediate trust, escalation, and safe boundaries for agent actions. \\ 
    
    B2. Conflict resolution and negotiation among agents \& humans & In a team setting, multiple agents (or agent + human) may propose conflicting plans or changes. How to mediate conflicts? & Research negotiation protocols, consensus algorithms, or hierarchical control structures for hybrid teams. \\ 
    
    B3. Long-horizon planning, multi-step decision-making & Agents must plan across multiple SDLC phases, not just local suggestions. This requires prediction, cost estimation, tradeoff reasoning. & Agents with embedded planning modules, reinforcement learning for planning over development workflows, and cost-benefit reasoning. \\ 
    
    B4. Accountability and responsibility attribution & When an agent autonomously makes a change that fails or introduces defects, who is accountable? & Formal frameworks for attributing responsibility in human-AI collaborative systems, perhaps with contract models or ``permission scoping''. \\ 
    
    B5. Recovery, rollback, and agent self-diagnosis & Agents must detect when their own suggestions degrade system quality or violate constraints, and autonomously rollback or correct. & Research self‑monitoring agents that propose and validate undo plans, safe failsafes, and ``what-if'' scenario testing. \\ 
    
    B6. Inter-agent coordination across modules & If multiple agents operate (e.g., one for requirements, one for testing, one for deployment), they must coordinate and exchange context. & Architectures for agent collaboration, shared knowledge bases, or protocols for consistency and dependency handling. \\

    B7. Trust calibration and human-agent teaming & Humans must calibrate when to rely on agent proposals. Too much trust leads to overreliance, too little undermines utility. & Techniques to signal confidence, uncertainty, or explain rationale to support human trust calibration. \\ \bottomrule
  \end{tabular}
  }

\end{table}

\begin{table}[htb]
    \centering
        \caption{Mapping between forms of \ai augmentation and its sources for \teammate{}s}
    \label{tab:mappingGenAITeam}
      \rowcolors{2}{gray!20}{white} 
     {\footnotesize 
     \renewcommand{\arraystretch}{1.2}
    \begin{tabular}{p{6.4cm} p{8cm}}
    \toprule
        {\bf Research Challenges and Opportunities} & {\bf Source (McLuhan's tetrad)} \\ 
        \midrule
        
        B1. Agent autonomy vs human oversight balance &  Ethics and autonomy of the Reverses quadrant.\\
         
        B2. Conflict resolution and negotiation among agents \& humans& Ethics and autonomy of the Reverses quadrant. \\
         
        B3. Long-horizon planning, multi-step decision-making & Ethics and autonomy of the Reverses quadrant.\\
         
        B4. Accountability and responsibility attribution & Accountability \& Code Ownership of the Reverses quadrant.\\
         
        B5. Recovery, rollback, and agent self-diagnosis & Code Quality and Technical Debt of the Retrieves quadrant. Static Analysis of the Retrieves quadrant.\\
         
        B6. Inter-agent coordination across modules & Ethics and autonomy of the Reverses quadrant. \\
         
        B7. Trust calibration and human-agent teaming & Ethics and autonomy of the Reverses quadrant. \\

        \bottomrule
        
    \end{tabular}}

\end{table}

Table~\ref{tab:mappingGenAITeam} maps the research challenges and opportunities to the source that triggered the form of \ai Teammate augmentation. As above, the source can be one or more elements in the corresponding tetrad. The mapping to McLuhan's tetrad highlights that ethical and autonomy concerns predominate and should shape design choices.
Research should therefore focus on supervisory architectures, meta-control policies, negotiation and consensus protocols, embedded planning modules, and contract-like responsibility models.
Complementary engineering work must deliver runtime verification, safe fallback strategies, and shared knowledge protocols for inter-agent consistency.
Finally, multidisciplinary evaluation, combining controlled experiments, field studies, and governance frameworks, is essential to produce human-agent teams that are effective, accountable, and socially acceptable.


\subsection{Roadmap Part C: \aiware{}}\label{sec:C}

Table~\ref{tab:genai_challenges_opportunities} describes the \aiware aspects by highlighting deployment-centric challenges and opportunities for integrating \ai into software development and operation. 

\begin{table}[htb]
  \centering
    \caption{Challenges and Opportunities for \aiware{}}
  \label{tab:genai_challenges_opportunities}
  \rowcolors{2}{gray!20}{white} 
 {\footnotesize 
 \renewcommand{\arraystretch}{1.2}
  \begin{tabular}{p{3.2cm} p{5.5cm} p{5.5cm}}
    \toprule
    \textbf{Form of \ai augmentation} & \textbf{Challenge} & \textbf{Opportunity} \\ \midrule
    C1. Dynamic prompt engineering in deployed systems & At runtime, code must generate appropriate prompts or adaptively modify them as user input context changes. & Self-tuning prompt modules or meta-prompting layers that adapt to usage distributions and context drift. \\ 
    
    C2. Testing, verification, and validation of GenAI-infused features & Ensuring that the GenAI-enabled portions of software meet functional, nonfunctional, and safety requirements. & New testing paradigms, formal specifications for GenAI behaviors. \\ 
    
    C3. Performance, latency, and resource constraints in deployment & Embedding GenAI into products often raises concerns of inference cost, latency, memory, and energy. & Edge-compatible GenAI modules, hybrid local/cloud partitioning, caching, and approximate inference. \\ 
    
    C4. User expectation, reliability, and backoff strategies & Users often expect predictable behavior; when GenAI fails or returns suboptimal output, the product must cope gracefully. & Fallback rules, ensemble strategies, confidence thresholds, graceful degradation paths, human-in-the-loop fallback. \\ 
    
    C5. Security, privacy, and data leakage in GenAI elements & The embedded GenAI might memorize or leak private or sensitive data. & Techniques like differential privacy, context-window isolation, on-device fine-tuning, or encryption to mitigate leakage. \\ 
    
    C6. ML Model lifecycle in-field updates and A/B testing & Updating or deploying new machine learning models inside a product, measuring the effect on users online, dynamically rolling back if negative side effects are detected. & Tools for safe online A/B testing of GenAI submodules, rollback strategies, incremental updates, and versioning at runtime to actively maintain and improve a model after deployment, using real-world data and real-time feedback to keep it performing well over time. \\ 
    
    C7. Explainability and transparency within features & Users and regulators may require explanations. & Embed explanation layers or rationale surfaces in the product, enabling ``why this response'' queries or tracebacks. \\ \bottomrule
  \end{tabular}
  }

\end{table}

The table emphasizes the need for adaptive prompt mechanisms that adjust dynamically to runtime contexts (C1), along with rigorous approaches to testing and validation to assure the correctness and safety of GenAI-infused features (C2). Performance considerations such as inference cost, latency, and resource efficiency (C3) must be balanced with user expectations for reliability and robust fallback behavior when outputs fail (C4). Security and privacy concerns are critical, as embedded models may inadvertently expose sensitive data (C5). Sustaining these systems further requires lifecycle management strategies, including safe in-field updates, controlled A/B testing, and dynamic rollback capabilities (C6). Finally, explainability and transparency (C7) are essential for both user trust and regulatory compliance. Collectively, the opportunities point toward adaptive prompting layers, new verification paradigms, efficient deployment strategies, privacy-preserving techniques, robust lifecycle tooling, and explanation features that enable reliable and trustworthy GenAIware in the software development lifecycle.

\begin{table}[htb]
    \centering
        \caption{Mapping between forms of \ai augmentation and its sources for \aiware{}}
    \label{tab:mappingGenAIware}
      \rowcolors{2}{gray!20}{white} 
     {\footnotesize 
     \renewcommand{\arraystretch}{1.2}
    \begin{tabular}{p{6.4cm}p{8cm}}
    \toprule
        {\bf Research Challenges and Opportunities} & {\bf Source (McLuhan's tetrad)} \\ 
        \midrule
        
        C1. Dynamic prompt engineering in deployed systems &  Prompt Programming of the Enhances quadrant. \\
        
        C2. Testing, verification, and validation of GenAI-infused features & Responsible and Secure AI of the Enhances quadrant. Runtime Verification and Validation of the Retrieves quadrant. \\
        
        C3. Performance, latency, and resource constraints in deployment & Efficiency and Sustainability of the Reverses quadrant. \\
        
        C4. User expectation, reliability, and backoff strategies & Reliability of the Reverses quadrant. \\
          
        C5. Security, privacy, and data leakage in GenAI elements & Responsible and Secure AI of the Enhances quadrant. \\
          
        C6. ML Model lifecycle in-field updates and A/B testing & Traditional Static Testing, and Separation between Machine Learning Cycle and SDLC of the Obsolesces quadrant. \\
          
        C7. Explainability and transparency within features & Opacity of AI elements of the Obsolesces quadrant. \\

        \bottomrule

    \end{tabular}}

\end{table}

As already done for the \ai Copilot and \ai Teammate augmentations, Table~\ref{tab:mappingGenAIware} maps the research challenges and opportunities to one or more tetrad elements in Section~\ref{sec:aiware} that triggered \ai GenAIware in SDLC. In short, Table~\ref{tab:genai_challenges_opportunities} and Table~\ref{tab:mappingGenAIware} frame GenAIware as an integration-, embedding- and deployment-centered challenge: adaptive prompting, rigorous validation, performance and fallback engineering, security/privacy, lifecycle updates, and explainability are all essential. Key opportunities include adaptive prompting layers, new verification paradigms, efficient deployment strategies, privacy-preserving techniques, robust lifecycle tooling, and rich explanation features. Also, for this type of augmentation, design should be informed by the tetrad mapping so that enhancements do not inadvertently reverse or obscure critical system properties.
Top research priorities are runtime-aware prompt controllers that adapt to context and user intent, and scalable testing/validation frameworks for generative features. Engineering work must address cost/latency trade-offs (model selection, quantization, caching) while guaranteeing reliable fallback behavior. Security and privacy require privacy-by-design, data minimization, and provenance controls to avoid sensitive exposure.
Operational tooling-observability, A/B pipelines, and automated rollback will be crucial for safe in-field evolution.
Finally, multidisciplinary evaluation (benchmarks, field studies, regulatory alignment) is needed to ensure GenAIware is performant, trustworthy, and compliant in real-world software products.


\subsection{Roadmap Part D: \robot{}s}\label{sec:D}

Table~\ref{tab:genaiRobot_challenges_opportunities} describes challenges and opportunities of \robot{}s in SE. The table focuses on the challenges and opportunities of engineering software systems where part of  the functionality is delivered by autonomous \ai agents. 
Key risks include emergent behavior and goal drift over time (D1), the need to preserve user trust and control through override or reversibility mechanisms (D2), and the stability of agents that adapt or learn in the field without violating constraints (D3). Additional concerns arise when multiple agents interact in shared environments, where competition or interference may disrupt outcomes (D4), and when scaling such systems to production settings introduces significant infrastructure and cost challenges (D5). The opportunities point toward research in alignment and auditing techniques, user-facing control interfaces, safe online learning methods, multi-agent coordination protocols, and efficient architectures that minimize computational overhead while maintaining robust performance.

\begin{table}[t]
  \centering
    \caption{Challenges and Opportunities for \robot{}s}
  \label{tab:genaiRobot_challenges_opportunities}
  \rowcolors{2}{gray!20}{white} 
 {\footnotesize 
 \renewcommand{\arraystretch}{1.2}
  \begin{tabular}{p{3.2cm} p{5.5cm} p{5.5cm}}
    \toprule
    \textbf{Form of \ai augmentation} & \textbf{Challenge} & \textbf{Opportunity} \\ \midrule
    D1. Emergent behavior, alignment, and goal drift &  Over time, autonomous product agents may stray from their intended goals or produce unexpected behavior under edge conditions. & Research alignment techniques, safe sandbox testing, reward shaping, and continual auditing of behavior drift. \\ 
    D2. User trust, control, and override mechanisms & As the product acts more autonomously, users must retain ultimate control; the agent must allow override, reversibility, and ``undo.'' & User interfaces for control granularity, adjustable autonomy levels, interactive consent, and transparency. \\ 
   D3. Evolution, self-improvement, and adaptation & Agents that learn in the field must adapt without compromising stability or violating constraints. & Techniques for safe online learning, constraint enforcement, minimizing ``catastrophic forgetting'' and ensuring consistency. \\ 
   D4. Inter-agent competition or collaboration in a shared environment  & If multiple intelligent agents act and interact (e.g., multiple AI users on same platform), coordination, conflicts, or negative interference may arise. & Multi-agent coordination algorithms, negotiation protocols, hierarchical control, or regulatory ``protocols'' for agent behavior. \\ 
   D5. Scalability, infrastructure, and operational cost  & Running autonomous agents (with reasoning, evaluation, planning) at scale in production environments is expensive. & Efficient architectures, caching, hierarchical reasoning, selective activation, and event-triggered execution to reduce compute load. \\ \bottomrule
  \end{tabular}
  }

\end{table}

\begin{table}[htb]
    \centering
        \caption{Mapping between forms of \ai augmentation and its sources for \robot{}s}
    \label{tab:mappingGenAIRobots}
      \rowcolors{2}{gray!20}{white} 
 {\footnotesize 
 \renewcommand{\arraystretch}{1.2}
    \begin{tabular}{p{5.8cm}p{8.6cm}}
    \toprule
        {\bf Research Challenges and Opportunities} & {\bf Source (McLuhan's tetrad)} \\ 
        \midrule
           
           D1. Emergent behavior, alignment, and goal drift & Novel forms of human-computer interaction (HCI) of the Enhances quadrant. Human-agent balance and Traditional separation of concerns of the Reverses quadrant. \\
           
           D2. User trust, control, and override mechanisms & Human-agent balance and Traditional separation of concerns of the Reverses quadrant. \\
           
           D3. Evolution, self-improvement, and adaptation & Novel forms of human-computer interaction (HCI) of the Enhances quadrant. Static assurance of legal and ethical norms of the  Reverses quadrant. \\
           
           D4. Inter-agent competition or collaboration in a shared environment & Novel forms of human-computer interaction (HCI) of the Enhances quadrant. Human-agent balance of the Reverses quadrant.\\
           
           D5. Scalability, infrastructure, and operational cost & Traditional separation of concerns of the Reverses quadrant. \\
           
        \bottomrule
    \end{tabular}}

\end{table}

Table~\ref{tab:mappingGenAIRobots} completes the mapping of the research challenges and opportunities referring to the tetrad of the fourth form of augmentation in Section~\ref{sec:robots}. In summary, the integration of GenAI Robots into the SDLC brings tangible challenges and related risks, such as emergent behaviors and goal drift, loss of control, instability in field learning, multi-agent interference, and scalability costs, but also clear opportunities for software engineering research. This section highlights that, also in this case, solutions are not isolated: they require alignment and auditing methods, human control interfaces and reversibility, safe online learning techniques, multi-agent coordination protocols, and compute-efficient architectures. Progress demands a roadmap that combines theoretical research, field experiments, and engineering practices (testing, metrics, and governance) to measure and mitigate risk. Only a systemic and interdisciplinary approach can turn these autonomous agents from risk factors into reliable enablers of the software development process.


\subsection{Roadmap Part X: Cross-form Aspects}\label{sec:E}

Tables~\ref{tab:crosscutting_genai_challenges_opportunities} and ~\ref{tab:mappingResearchChallenges} describe cross-form challenges and opportunities (X1--X8) that extend beyond individual workflows or deployment contexts. X1 highlights the complexity of coordinating process-level and product-level agents, while X2 stresses the absence of robust benchmarks and evaluation methods for measuring GenAI's impact in engineering practice. Human factors emerge prominently in X3, where changes to team roles, skills, and dynamics require careful study, and in X4, where unresolved questions of intellectual property, licensing, and liability remain. Broader systemic concerns include sustainability and environmental costs (X5), resilience against adversarial manipulation (X6), and the limited transferability of GenAI methods across domains (X7). Finally, X8 underscores the need for continuous adaptation as software ecosystems evolve. Together, these themes point to a research agenda that blends technical innovation with organizational, legal, and ethical considerations to enable trustworthy 
GenAI augmentation.\\

\begin{table}[t]
  \centering
    \caption{Challenges and Opportunities for Cross-form GenAI Augmentation}
  \label{tab:crosscutting_genai_challenges_opportunities}
  \rowcolors{2}{gray!20}{white} 
 {\footnotesize 
 \renewcommand{\arraystretch}{1.2}
  \begin{tabular}{p{3.2cm}p{5.5cm}p{5.5cm}}
    \toprule
    \textbf{Form of \ai augmentation} & \textbf{Challenge} & \textbf{Opportunity} \\ \midrule
    X1. Hybrid systems: integrating process agents + product agents & In future systems, process-level agents (that plan development) and product-level agents (in software) may co-exist and influence each other. Managing their interactions is complex. & Architectures and protocols for cross-layer consistency, feedback loops, negotiation, and coordinated evolution. \\ 
    X2. Metrics, evaluation benchmarks, and empirical validation & How do we measure ``goodness'' of GenAI augmentation (for productivity, reliability, maintainability, human satisfaction)? & Define standard benchmarks, controlled experiments, and evaluation suites for hybrid human-AI engineering settings. \\ 
    X3. Human factors, team dynamics, education and roles & As GenAI changes workflows, roles and skills shift. Resistance, cognitive burden, misuse, or misalignment may arise. & Study human-AI teaming, change management, training curricula, and social/organizational adaptation strategies. \\ 
    X4. Intellectual property, licensing, and legal liability & Who owns code, designs, or artifacts generated by GenAI? What licenses apply? What liability when AI-generated code fails? & Legal and policy frameworks for IP attribution, shared licensing models, AI-generated artifact contracts, and compliance regimes. \\ 
    X5. Model robustness, adversarial attacks, and misuse & GenAI elements and agents can be manipulated via adversarial prompts or injection attacks to produce harmful outputs. & Robustness research, prompt sanitization, query filtering, anomaly detection, and secure input/output boundaries. \\ 
    X6. Updating of knowledge, evolving software landscapes & The software ecosystem (languages, frameworks, libraries) evolves rapidly; GenAI augmentation must keep up. & Continual learning, modular plugin adaptation, community-based model updates, and open model ecosystems in the software engineering domain. \\ \bottomrule
  \end{tabular}
  }

\end{table}

\begin{table}
    \centering
        \caption{Mapping between Cross-form Research Challenges and Opportunities}
    \label{tab:mappingResearchChallenges}
      \rowcolors{2}{gray!20}{white} 
 {\footnotesize 
 \renewcommand{\arraystretch}{1.2}
    \begin{tabular}{p{4cm}p{5cm}p{5cm}}
    \toprule
        {\bf Cross-cutting Research Challenges and Opportunities} & {\bf Source (Process)} & {\bf Source (Product)} \\ 
        \midrule
           X1. Hybrid systems: integrating process agents + product agents& A5. Workflow integration and orchestration. \newline B6. Inter-agent coordination across modules. &  D4. Inter-agent competition or collaboration in a shared environment.\\
           
           X2. Metrics, evaluation benchmarks, and empirical validation & A10. Trustworthy GenAI & C4. User expectation, reliability, and backoff strategies. \newline C6. ML Model lifecycle in-field updates and A/B testing. \\
            
           X3. Human factors, team dynamics, education and roles & A6. Cognitive load, transparency, and user control. \newline B2. Conflict resolution and negotiation among agents \& humans. \newline B7. Trust calibration and human-agent teaming. & C7. Explainability and transparency within features. \newline D2. User trust, control, and override mechanisms.\\
           
           X4. Intellectual property, licensing, and legal liability & A6. Cognitive load, transparency, and user control. \newline B4. Accountability and responsibility attribution. & C7. Explainability and transparency within features. \newline D2. User trust, control, and override mechanisms.\\
           
           %
           X5. Model robustness, adversarial attacks, and misuse & A9. Safety, security, \& dependency risk. \newline B5. Recovery, rollback, and agent self-diagnosis. & C2. Testing, verification, and validation of GenAI-infused features. \newline C5. Security, privacy, and data leakage in GenAI elements. \\
           
           %
           X6. Updating of knowledge, evolving software landscapes & A4. Model drift, update, and maintenance of GenAI elements. & D3. Evolution, self-improvement, and adaptation\\
           \bottomrule
    \end{tabular}}

\end{table}

\section{Threats to Validity}
\label{sec:validity}
This section discusses potential threats to the validity of our study and the measures taken to mitigate them.

\paragraph{Construct Validity.}
Construct validity concerns whether the key concepts investigated were correctly defined and operationalized. 
A potential threat lies in the subjective interpretation of what constitutes \ai{} augmentation in software engineering processes and artifacts. 
To address this, the fundamental forms of \ai{} in SE were initially elicited through collective discussions during the FSE 2025 “2030 Software Engineering” workshop and grounded in existing taxonomies, such as the one proposed by SEI. 
These constructs were subsequently refined and validated through multiple iterations of internal discussions and external feedback sessions involving peers. 
This triangulation of sources and perspectives aimed to ensure that the constructs reflect a shared and well-grounded understanding rather than the isolated view of a subset of authors. 

Another potential threat concerns the coverage of the Rapid Literature Reviews (RLRs) used in Design Cycle 2. 
The search strings adopted in each review may not have captured all relevant literature, potentially limiting the breadth of the investigated evidence. 
Moreover, the use of LLM-based filtering (using Qwen3 and Gemini-2.5 respectively) might have inadvertently excluded relevant publications due to the potential risks of bias and hallucination. 
To mitigate this validity risk, we manually inspected samples of the excluded papers to double check that no relevant literature was systematically missed.
We also provide open access to the data of our literature reviews~\cite{Zenodo}, thereby providing additional transparency and replicability of the RLR process. 

\paragraph{Internal Validity.}
Internal validity refers to the soundness of the study design and the correctness of the reasoning linking evidence to conclusions. 
A potential risk arises from interpretive bias during the construction of the McLuhan Tetrads and from the compressed time frame of the rapid reviews, which could have reduced the depth of evidence interpretation. 
To mitigate these risks, the research was organized into three design science cycles, each including explicit awareness, solution, and validation stages. 
Within Cycle 2, the RLRs followed a shared baseline protocol and were independently executed by different author teams. 
Subsequent cross-review by authors not involved in their initial construction helped reduce confirmation bias and ensured consistency and traceability across the four forms of \ai{}.

\paragraph{External Validity.}
External validity concerns the generalizability and transferability of the study results. 
Since the goal of this work was to design a conceptual roadmap rather than to produce statistically generalizable evidence, its external validity is inherently limited. 
However, its findings draw upon diverse forms of evidence, including literature, workshop discussions, and multi-author evaluations, which provide analytical rather than statistical generalization. 
Furthermore, the final roadmap was refined through iterative co-author reviews and three dedicated meetings, strengthening confidence in the broader relevance and applicability of the identified challenges and opportunities.

\paragraph{Conclusion Validity.}
Conclusion validity relates to the credibility, logical coherence, and consistency of the study’s outcomes. 
Potential threats include overgeneralization and interpretive divergence among authors. 
To address these, all results were progressively consolidated and validated across the three design science cycles. 
In the final Cycle 3, two authors synthesized the validated tetrads into the roadmap, while the remaining authors collectively reviewed and refined it until consensus was reached. 
This iterative synthesis and collective validation ensured that the conclusions drawn in the roadmap are well supported by the evidence gathered and consistently interpreted within the scope of the study.

\section{Conclusion and Perspectives}
\label{sec:conclusion}

GenAI implies a paradigm shift in SE, profoundly reshaping both development processes and the nature of software products. 
To navigate this complex transformation, we introduced a systematic framework for understanding and analyzing the impact of \ai on SE. 
We proposed a structuring of \ai augmentation along two principal dimensions: what is being augmented (process versus product) and the level of autonomy of the augmentation (passive versus active). 
This categorization yields four distinct and tangible forms of \ai augmentation: the GenAI Copilot, GenAIware, the GenAI Teammate, and the GenAI Robot.

Using this framework as our analytical lens, we conducted an examination of the state-of-the-art for each of the four forms. 
Our methodology -- grounded in a multi-cycle design science approach -- employed rapid literature reviews to gather evidence, which was then synthesized using McLuhan's tetrads. 
This approach enabled a holistic assessment of each form of \ai augmentation, identifying what it enhances, what established practices it reverses when pushed to extremes, what past concepts it retrieves, and what  it may render obsolete. 
This multidimensional analysis revealed a complex interplay of benefits and challenges, from enhanced productivity and retrieved formalisms and paradigms to issues such as trustworthiness, accountability, and the potential erosion of human expertise.

The culmination of this analysis is a comprehensive research roadmap that distills our findings into specific, actionable challenges and opportunities. 
This roadmap addresses not only the unique issues pertinent to each of the four forms but also identifies cross-cutting themes, including human-AI team dynamics, legal and ethical considerations, and the need for robust evaluation benchmarks. 

By offering a structured and evidence-based perspective, this work refines and updates existing roadmaps, serving as part of an overall prescriptive guide for the SE community contained within this special issue. 
As \ai continues its rapid evolution, this roadmap serves as a foundation for future research, encouraging a balanced approach that maximizes the transformative potential of \ai while proactively mitigating its inherent risks, thereby helping to shape the future principles, practices, and economics of software engineering.

\begin{center}
---
\end{center}

\noindent We like to close by offering our \textbf{ten predictions for software engineering in the year 2030}:

\textbf{Death of manual coding for routine tasks:} By 2030, 70\% of boilerplate code, CRUD operations, and standard API integrations will be autonomously generated by GenAI Teammates with zero human keystrokes. Human developers will spend less than 20\% of their time writing code from scratch. Our roadmap indicates that manual coding will become obsolete. The shift from developer-as-coder to developer-as-architect/orchestrator will be complete for routine functionality.

\textbf{Prompt Engineering as a core SE discipline:} By 2030, "Prompt Architect" will be a recognized job title with professional certifications. Universities will offer dedicated courses on prompt engineering, and 40\% of SE job postings will list prompt engineering as a required skill alongside traditional programming languages. Our roadmap indicates that prompt programming will be a new "programming" paradigm. As GenAIware proliferates, prompt quality directly determines system reliability.

\textbf{Emergence of "Software Compliance as Code":}
By 2030, legal compliance, ethical constraints, and safety requirements will be encoded as formal guardrails in machine-readable formats (e.g., "compliance specification languages") that GenAI systems must satisfy in real-time. 40\% of regulated software projects will use automated compliance verification that checks both human and AI contributions against these formal specifications. The EU AI Act and similar regulations will mandate this for high-risk AI systems. Our roadmap mentions static assurance of legal/ethical norms becoming insufficient together with IP, licensing, and legal liability, which will lead to widespread concerns about compliance, fairness, and responsible AI across all forms. The need for guardrails that work at runtime, not just design time, is critical.

\textbf{Rise of AI accountability standards:}
By 2030, at least three major organizations (ACM, IEEE, or ISO) will have published formal standards for AI accountability in software development, contributing to the aforementioned "software compliance as code". 50\% of Fortune 500 companies will require explicit "AI Attribution Metadata" in their codebases, documenting which code was human-written, AI-suggested, or AI-autonomous. This results from our roadmap indicating widespread concerns about accountability and code ownership.

\textbf{Emergence of "Responsible GenAI Engineering":}
By 2030, there will be dedicated "GenAI Responsibility Engineers" on software teams, who are in charge of implementing guardrails, detecting hallucinations, and ensuring ethical and sustainable AI behavior. Responsible engineering for GenAI will become distinct from traditional QA, with specialized tools and methodologies. At least 20\% of SE teams in regulated industries will have this role. This is backed by our roadmap that indicates the relevance of testing/verification of GenAI features, security and privacy, emergent behavior and goal drift detection, and adversarial attacks, which raises concerns about reliability and trustworthiness.

\textbf{Obsolescence of Classical IDEs:}
By 2030, traditional IDEs (such as VS Code, IntelliJ) will face existential competition from "conversational development environments" where 30\% of developers primarily interact with their codebase through natural language chat interfaces rather than file editors. The concept of "manually opening and editing files" becomes optional for many tasks. This becomes evident from our roadmap as GenAI Teammates will mean that traditional IDE editors fade, emphasizing natural language interaction as the primary interface between humans and AI.

\textbf{Integrated CI/CD/AI pipelines:}
By 2030, CI/CD will evolve into "CI/CD/AI" where 80\% of software projects include automated pipelines for monitoring, retraining, and rolling back GenAI components. Model drift detection and automated A/B testing of GenAI features will be as common as unit testing is today. Our roadmap clearly indicates the consideration of model drift and maintenance, AI model lifecycle updates, and the need for runtime verification and validation.

\textbf{Multi-Agent orchestration platforms:}
By 2030, at least 5 major platforms for orchestrating multiple GenAI agents (both Teammates and Robots) will have emerged, with at least one achieving "Kubernetes-level" adoption (used by 40\%+ of organizations). These platforms will standardize agent-to-agent communication protocols, making multi-agent systems as manageable as microservices are today. This is backed by our roadmap indicating emerging inter-agent coordination, multi-agent competition/collaboration, as well as hybrid systems integrating process and product agents.

\textbf{"10x Developer" becomes "10x Orchestrator":}
By 2030, the concept of the "10x developer" will shift from someone who writes 10x more code to someone who effectively orchestrates 10+ GenAI agents simultaneously. Performance metrics will measure "agents successfully coordinated" rather than "lines of code written." Top performers will manage complex ecosystems of specialized AI agents rather than large codebases. Our roadmap reflects this transformation across all forms, particularly with respect to agent autonomy balance, trust calibration, and the general shift from coding to orchestration.

\textbf{GenAI-induced technical debt crisis and recovery:}
By 2030, 60\% of software maintenance effort will focus on understanding and improving AI-generated code rather than human-written code. In particular, we will see a new generation of "AI Code Archaeology" tools that can analyze, explain, and refactor legacy AI-generated code. This is to address the development of previous years where industry will experience a "GenAI technical debt crisis", during which accumulated low-quality AI-generated code creates major maintenance problems. Our roadmap indicates this by mentioning quality control of AI artifacts and maintenance of GenAI elements, together with code comprehension, code quality and static analysis.

\begin{acks}
Some parts of our work were assisted by Generative AI.
We used Elsevier's ScopusAI, Google's Gemini-2.5, and Alibaba's Qwen3 during the rapid literature review  (also see Appendix~\ref{sec:prompts}).
We also used Anthopic's Claude Sonnet 4.5 to help us brainstorm the concluding predictions for software engineering in the year 2030.

We thank Abhik Roychoudhury and Mauro Pezze -- co-chairs of the FSE 2025 workshop on "Software Engineering 2030" -- for their brilliant organization of such an encouraging and stimulating workshop, as well as to all fellow workshop participants for the engaging discussions.

This work was partly funded by Core Informatics at KIT (KiKIT) and Topic Engineering Secure Systems of the Helmholtz Association (HGF) and partly supported by the German Research Foundation (DFG) - SFB 1608 - 501798263 and KASTEL Security Research Labs, as well as 
(a) the MUR (Italy) Department of Excellence 2023--2027, 
(b) the European HORIZON-KDT-JU research project MATISSE ``Model-based engineering of Digital Twins for early verification and validation of Industrial Systems" (101140216),
(c) the PRIN project P2022RSW5W - RoboChor: Robot Choreography, 
(d) the PRIN project 2022JKA4SL - HALO: etHical-aware AdjustabLe autOnomous systems, and
(e) the PRIN project 2022JAFATE - CAVIA: enabling the Cloud-to-Autonomous-Vehicles continuum for future Industrial Applications; (f) the European Union - NextGenerationEU under the Italian Ministry of University and Research (MUR) National Innovation Ecosystem grant ECS00000041 - VITALITY – CUP: D13C21000430001; (g) PNRR MUR (Italy) -- Centro Nazionale HPC, Big Data e Quantum Computing, Spoke9 - Digital Society \& Smart Cities (grant CN\_00000013).

\end{acks}

\bibliographystyle{ACM-Reference-Format}
\bibliography{tosem-refs}










\appendix
\section{Search strings used for literature review}
\label{sec:searchStrings}
\subsection{Search Strings for \copilot{}s}
\label{sec:copilot-searchstrings}
\begin{tcolorbox}[colback=lightgray!20, colframe=gray, boxrule=0.5pt, arc=2pt, title=Search String ACM, attach title to upper={\ }, fonttitle={\bfseries\color{black}}, after title={\\},]
"query": \{  Fulltext:(("software development life cycle" OR "SDLC" OR "life*cycle" OR "development process" OR "software engineering process") AND ("LLM?" OR "Large Language Model?" OR "GenAI" OR "Generative AI" OR "Generative Artificial Intelligence" OR "transformer architecture")) AND ContentGroupTitle:("ICSE:" OR "ASE:" OR "FSE:" OR "ACM Transactions on Software Engineering and Methodology" OR "IEEE Transactions on Software Engineering" OR "Empirical Software Engineering" OR "Journal of Systems and Software" OR "Information and Software Technology" OR "Automated Software Engineering") \}
"filter": \{ E-Publication Date: (01/01/2019 TO *) \}
\end{tcolorbox}

\begin{tcolorbox}[colback=lightgray!20, colframe=gray, boxrule=0.5pt, arc=2pt, title=Search String IEEE, attach title to upper={\ }, fonttitle={\bfseries\color{black}}, after title={\\},]
(("Full Text Only": "software development life cycle") OR ("Full Text Only": "SDLC")  OR ("Full Text Only":"life*cycle")  OR ("Full Text Only":"development process") OR ("Full Text Only":"software engineering process")) AND (("Full Text Only": "LLM?") OR ("Full Text Only":"Large Language Model?") OR ("Full Text Only":"GenAI") OR ("Full Text Only":"Generative AI") OR ("Full Text Only":"Generative Artificial IntelligenceI") OR ("Full Text Only":"transformer architecture"))
AND
(("Publication Title": "*International Conference on Software Engineering *ICSE") OR ("Publication Title": "*Automated Software Engineering *ASE") OR ("Publication Title": "*Foundations of Software Engineering") OR ("Publication Title": "IEEE Transactions on Software Engineering") OR ("Publication Title": "Empirical Software Engineering") OR ("Publication Title": "Journal of Systems and Software")) AND
(("Publication Year": "2019") OR ("Publication Year": "2020") OR ("Publication Year": "2021") OR ("Publication Year": "2022") OR ("Publication Year": "2023") OR ("Publication Year": "2024") OR ("Publication Year": "2025"))
\end{tcolorbox}

\subsection{Search strings for \teammate{}}
\label{sec:teammate-searchstrings}
\begin{tcolorbox}[colback=lightgray!20, colframe=gray, boxrule=0.5pt, arc=2pt, title=Search String Google Scholar/Scopus, attach title to upper={\ }, fonttitle={\bfseries\color{black}}, after title={\\},]
ABS ( Agent software development generative AI ) OR ABS ( Agentic software development generative AI ) OR ABS ( Agent software development lifecycle generative AI ) OR ABS ( Agentic software development lifecycle generative AI ) OR ABS ( Agent sdlc generative AI ) OR ABS ( Agentic sdlc generative AI ) OR ABS ( Agent software development genAI ) OR ABS ( Agentic software development genAI ) OR ABS ( Agent software development lifecycle genAI ) OR ABS ( Agentic software development lifecycle genAI ) OR ABS ( Agent sdlc genAI ) OR ABS ( Agentic sdlc genAI )
\end{tcolorbox}

\subsection{Search strings for \aiware{}}
\label{sec:genaiware-searchstrings}
\begin{tcolorbox}[colback=lightgray!20, colframe=gray, boxrule=0.5pt, arc=2pt, title=Search String ACM/IEEE/Scopus, attach title to upper={\ }, fonttitle={\bfseries\color{black}}, after title={\\},]
("software development life?cycle" OR "sdlc" OR "software development process" OR "software engineering process" OR "development process" OR "software system" OR "software application" OR "software architecture" OR "system design" OR "application development" OR "software component" OR "software service" OR "system development" OR "application design" OR "application" OR "software component" OR "software product") AND
("llm?" OR "large language model?" OR "fine?tuned llm?" OR "fine?tuned language model?" OR "fine?tuned lm?" OR "fine?tuned language model?" OR "gen?ai" OR "generative?ai" OR "generative artificial intelligence" OR "transformer model" OR "transformer language model" OR "transformer lm" OR "neural language model" OR "neural lm" OR "instruction tuning" OR "instruction fine?tuning" OR "instruction following" OR "retrieval?augmented generation" OR "rag" OR "assistant tuning"OR "assistant fine?tuning" OR "function calling" OR "tool use" OR "guardrails") AND
("use case" OR "functionalit*" OR "integrat*" OR "deploy*" OR "architectur*" OR "backend" OR "micro?service" OR "application" OR "orchestration" OR "agent orchestration" OR "safety filter" OR "filter" OR "pipeline" OR "workflow" OR "service" OR "framework") AND
("ai?ware" OR "gen?ai?ware" OR "semantic search" OR "gopher" OR "agent orchestration" OR "multi?agent orchestration" OR "inference engine" OR "dev?ops 2.0" OR "prompt?engineering" OR "prompt?ware")
\end{tcolorbox}

\subsection{Search string for \robot{}}
\label{sec:robot-searchstrings}
\begin{tcolorbox}[colback=lightgray!20, colframe=gray, boxrule=0.5pt, arc=2pt, title=Search String Scopus, attach title to upper={\ }, fonttitle={\bfseries\color{black}}, after title={\\},]
( "agentic" OR "agent-based" ) AND ( "LLM" OR "large language model" OR "generative AI" OR "Gen AI" OR "transformer" OR "GPT" or "BERT" OR "Generative Adversarial Network" OR "GANs" or "Variational Autoencoder" OR "VAE" ) AND ( "applications" OR "use cases" OR "impacts" OR "effects" ) AND ( "software" OR "software development" OR "software engineering" ) AND PUBYEAR > 2023 AND PUBYEAR < 2027 AND ( LIMIT-TO ( SUBJAREA , "COMP" ) ) AND ( LIMIT-TO ( DOCTYPE , "ar" ) OR LIMIT-TO ( DOCTYPE , "cp" ) ) AND ( LIMIT-TO ( LANGUAGE , "English" ) )
\end{tcolorbox}

\section{Prompts used for literature review}
\label{sec:prompts}
\lstset{
    basicstyle=\ttfamily\small, 
}

\subsection{\copilot{}}
\label{sec:copilot-promts}
Prompt invoked for Qwen3 version qwen3:8b\_q4\_K\_M (commit 500a1f067a9f)\footnote{model card: \url{https://ollama.com/library/qwen3:8b}} using Ollama.

\noindent\textbf{System prompt}:

\begin{lstlisting}
You are a helpful academic assistant. If you include reasoning, enclose it in <think> tags and give a clear 'Final Answer:' at the end.
\end{lstlisting}

\noindent\textbf{User message Q1}:

\begin{lstlisting}
Title: {title}
Abstract: {abstract}
Keywords: {keywords}
Question: Based on title, abstract and keywords provided is there a chance that the paper discusses the role and/or implications of GenAI or LLMs on the software development process or lifecycle? Answer only with yes, no, or maybe.
\end{lstlisting}

\noindent\textbf{User message Q2}:

\begin{lstlisting}
Title: {title}
Abstract: {abstract}
Keywords: {keywords}
Question: Based on title, abstract and keywords provided does the paper utilize GenAI or LLMs as a tool to support or automate tasks along the software development process or lifecycle (and not as part of the final software system, product, or application)? Answer only with yes or no.
\end{lstlisting}

\subsection{\teammate{}}
\label{sec:teammate-prompts}

In the context of \teammate{}, no AI model was involved in the analysis of the paper during the Rapid Literature Review.

\subsection{\aiware{}}
\label{sec:genaiware-prompts}
Prompt invoked for LLM Qwen3 version unsloth/Qwen3-4B-Instruct-2507-GGUF\footnote{model card: \url{https://huggingface.co/unsloth/Qwen3-4B-Instruct-2507-GGUF}} using the Llama CPP python library version 0.3.16.

\textbf{System prompt}:

\begin{lstlisting}
The following is a conversation between a human user and an AI assistant. The assistant is knowledgeable in several research areas including computer science, software engineering and artificial intelligence. In this dialogue, the assistant answers to the user's questions and requests straightforwardly and concisely, providing detailed explanations only when required.
\end{lstlisting}

\textbf{User message Q1}:

\begin{lstlisting}
Based on the following paper metadata (title, abstract, and keywords), assess whether this paper likely contains relevant information about using Generative AI (e.g., Large Language Models) as a component or a service of a software system (and not as a development tool), specifically from a software development lifecycle perspective.
Answer yes or no depending on whether the paper discusses ways to augment the software development life cycle when using Generative AI to realise software functionalities that would otherwise not be available, then explain your response.

Title: "{TITLE}"

Abstract:

{ABSTRACT}

Keywords: {KEYWORDS}
\end{lstlisting}

\textbf{User message Q2}:

\begin{lstlisting}
Based on the following paper metadata (title, abstract, and keywords), assess whether this paper likely contains relevant information about the design, implementation, or architectural considerations deriving from the use of Generative AI (e.g., Large Language Models) as a component or a service of a software system, specifically from a software development lifecycle perspective.
Answer yes or no depending on whether the paper discusses ways to augment the software development life cycle when using Generative AI to realise software functionalities that would otherwise not be available, then explain your response.

Title: "{TITLE}"

Abstract:

{ABSTRACT}

Keywords: {KEYWORDS}
\end{lstlisting}

\subsection{Prompts for \robot{}s}

\label{sec:bots-prompts}
Prompt invoked for LLM version  gemini-2.5-flash via Python code using the Google GenAI Python library version 1.24.0.

\begin{lstlisting}
<Role> You are a researcher of software engineering that analyses research papers found via a literature data base search.

<Context> The data provides the title, abstract and keywords of each research paper that is supposed to cover argentic AI as part of software systems and applications.

Agentic AI means that part of the system is realized by an AI system or model, which (a) operates with a greater degree of autonomy, (b) is capable of  undertaking (human) roles, (c) manages multi-step tasks, (d) proactively collaborates with human developers or other agents, (c) achieves higher-level goals.

<Task> Your task is to analyze the paper and provide an answer to the following research question.

<Output> Answer this question by providing your answers as bullet points. Use the character '*' as bullets. Only provide the answers, but nothing else, i.e., do not repeat the given question, your train of thought, or additional background information.

Before providing the bullet-style answer to the question, give a general assessment of whether the question can be answered with TRUE or FALSE.

<Question 1> Does the paper satisfy the following criteria?
(a) The paper utilizes Agentic AI as part of the final software system, product, or application.
(b) The paper does NOT only focus on using Agentic AI as part of the software development process.
If both criteria are met, answer with TRUE.
If at least one of the criteria is NOT met, answer with FALSE.

<Question 2> Does the paper discuss the role and/or implications of Agentic AI on the software development process or lifecycle (such as CI/CD, DevOps, Agile Development, Scrum, XP, etc.)?
\end{lstlisting}

\end{document}

%% file: figures/mcluhans_tetrad.tex
\begin{figure}[h]
\centering
\begin{tikzpicture}[
    quadrant/.style={draw, rounded corners, minimum width=5cm, text width = 0.45 \textwidth, minimum height=7em, align=left, inner sep=5pt},
    centerbox/.style={draw, circle, minimum size=2cm, align=center, fill=gray!15, font=\small\bfseries},
    keyword/.style={font=\small\bfseries}
]

\def\gap{0.5em}

\node[quadrant, anchor=south east] at (-\gap,\gap) (enhance) {
    \setlength{\columnsep}{2pt}

        \begin{itemize}[leftmargin=2em]
            \item "What does the technology \textbf{enhance} or intensify?" 
        \end{itemize}

};

\node[quadrant, anchor=south west] at (\gap,\gap) (reverse) {
\setlength{\columnsep}{2pt}

        \begin{itemize}[leftmargin=2em]
            \item "What does the technology \textbf{reverse} or flip into when pushed to its extreme?"
        \end{itemize}

};

\node[quadrant, anchor=north east] at (-\gap,-\gap) (retrieve) {
\vtop{
    \setlength{\columnsep}{2pt}

        \begin{itemize}[leftmargin=2em]
            \item "What does the technology \textbf{retrieve} or recover from the past?"
        \end{itemize}

  }
};

\node[quadrant, anchor=north west] at (\gap,-\gap) (obsolesce) {
\vtop{
\setlength{\columnsep}{2pt}

    \begin{itemize}[leftmargin=2em]
        \item "What does the technology make \textbf{obsolete} or displace?" 
    \end{itemize}

}
};

\node[keyword, below right=0.25em and 0.25em of enhance.north west] {Enhances};
\node[keyword, below left=0.25em and 0.25em of reverse.north east] {Reverses};
\node[keyword, above right=0.25em and 0.25em of retrieve.south west] {Retrieves};
\node[keyword, above left=0.25em and 0.25em of obsolesce.south east] {Obsolesces};

\node[centerbox, text width = 4em] at (0,0) {Technology};

\end{tikzpicture}
\caption{A blank tetrad diagram.}
\Description{Image shows the four quadrants of a blank tetrad diagram.}
\label{fig:tetrad}
\end{figure}

%% file: figures/mcluhans_tetrad_copilot.tex
\begin{figure}
\centering
\begin{tikzpicture}[
    quadrant/.style={draw, rounded corners, minimum width=5cm, text width = 0.45 \textwidth, minimum height=10em, align=left, font=\scriptsize, inner sep=5pt},
    centerbox/.style={draw, circle, minimum size=2cm, align=center, fill=gray!15, font=\small\bfseries},
    keyword/.style={font=\small\bfseries}
]

\def\gap{0.5em}

\node[quadrant, anchor=south east,minimum height=16.7em] at (-\gap,\gap) (enhance) {
    \setlength{\columnsep}{1pt}
    \begin{multicols}{2}
      Requirements
      \vspace{-0.5\baselineskip}
      \begin{itemize}[leftmargin=2em]
      \item Extraction, Analysis, Verification, Validation of requirements \cite{10.1016/j.infsof.2025.107697}
      \item Generation or refinement of requirements \cite{10.1145/3696630.3728718}
    \end{itemize}
    Design \& Documentation~\cite{10.1145/3696630.3728718,10.1145/3715003,10.1145/3678172}
    \vspace{-0.5\baselineskip}
    \begin{itemize}[leftmargin=2em]
      
      \item Conceptualization \cite{10.1016/j.infsof.2025.107751}
      \item Identifying relevant documents ~\cite{10.1016/j.infsof.2025.107751} 

    \end{itemize}
    Implementation 
    \vspace{-0.5\baselineskip}
    \begin{itemize}[leftmargin=2em]
      \item Coding (literate/NL) ~\cite{10.1007/s10515-024-00426-z,10.1145/3695988,10.1016/j.infsof.2025.107751}
      \item Code comprehension ~\cite{10.1007/s10515-024-00426-z,10.1016/j.infsof.2025.107751,10.1145/3708519}
      \item Code refinement/ optimization ~\cite{10.1145/3696630.3728718,10.1016/j.infsof.2025.107751}
      \item (real-time) Code analysis ~\cite{10.1145/3696630.3728718,10.1145/3708519}
      
    \end{itemize}
    Testing \& Quality Assurance~\cite{10.1007/s10515-024-00426-z,10.1145/3696630.3730538,10.1145/3696630.3728718,10.1145/3695988}
    \vspace{-0.5\baselineskip}
    \begin{itemize}[leftmargin=2em]
      \item Test design/generation ~\cite{10.1145/3696630.3728718,WangHCLWW24}
      \item Bug detection and fixing ~\cite{10.1145/3696630.3728718,WangHCLWW24,10.1016/j.infsof.2025.107751,10.1145/3709353}
      \item Vulnerability Detection \cite{10.1145/3695988}
      \item Debugging \cite{10.1145/3696630.3728718,WangHCLWW24}
      \item Program repair ~\cite{WangHCLWW24}
      \item Code review and quality \cite{10.1145/3695988,10.1145/3708519,10.1145/3678172,10.1145/3696630.3727251,10.1145/3696630.3730538,10.1145/3715003,AhmedACCHHPPX25}

    \end{itemize}
    Project \& Process Management
    \vspace{-0.5\baselineskip}
    \begin{itemize}[leftmargin=2em]
      \item Productivity \cite{10.1145/3696630.3727251,10.1145/3678172,10.1145/3695988}
      \item Development time \cite{10.1016/j.infsof.2025.107751}
      \item Effort estimation \cite{10.1145/3696630.3728718}
      \item Completeness \& Consistency  \cite{keim_automated_2025} 
        \item Knowledge Management ~\cite{10.1016/j.infsof.2025.107751,keim_automated_2025}
    \end{itemize}
    Human \& Team Factors
    \vspace{-0.5\baselineskip}
    \begin{itemize}[leftmargin=2em]

      \item Onboarding/training ~\cite{10.1145/3696630.3728718}
      \item Skills Learning ~\cite{10.1145/3709353}
      \item Team coordination ~\cite{10.1145/3709353}
      \item Communication \cite{10.1145/3696630.3727251}
      \item Mental Health ~\cite{10.1145/3709353}
      
      \end{itemize}
    \end{multicols}
};

\node[quadrant, anchor=south west, minimum height=17.5em] at (\gap,\gap) (reverse) {
\setlength{\columnsep}{2pt}
  \begin{multicols}{2}
    \begin{itemize}[leftmargin=2em]
        \item Trustworthiness~\cite{10.1145/3695988, 10.1145/3696630.3728718,10.1145/3696630.3730538,10.1145/3709360,keim_automated_2025}
        \item Reliability~\cite{WangHCLWW24, 10.1145/3696630.3728718,10.1145/3696630.3730538,10.1145/3708519}
        \item Explainability\cite{10.1145/3696630.3730538,keim_automated_2025}
        \item Fairness ~\cite{AhmedACCHHPPX25}
        \item Compliance ~\cite{10.1145/3709360}
        \item Sustainability ~\cite{10.1145/3709360}
        \item Accountability~\cite{10.1145/3709360}
        \item Individual Understanding Capabilities~\cite{10.1145/3678172}
        \item Strict boundaries between \sdlc stages~\cite{10.1007/s10515-024-00426-z} 
        \item Clear code ownership (IP) \cite{10.1145/3696630.3728718}
        \item Security \cite{10.1145/3695988,10.1145/3708519}
    \end{itemize}
  \end{multicols}
};

\node[quadrant, anchor=north east] at (-\gap,-\gap) (retrieve) {
\vtop{
    \setlength{\columnsep}{2pt}
      \begin{multicols}{2}
        \begin{itemize}[leftmargin=2em]
            \item Formal Requirements Specification~\cite{10.1016/j.infsof.2025.107697}
            \item Formal Verification ~\cite{10.1016/j.infsof.2025.107697} 
            \item Rapid Prototyping~\cite{10.1007/s10515-024-00426-z,10.1145/3715003}
            \item Natural Language explanation of code~\cite{10.1145/3678172}
        \end{itemize}
      \end{multicols}
  }
};

\node[quadrant, anchor=north west] at (\gap,-\gap) (obsolesce) {
\vtop{
\setlength{\columnsep}{2pt}
  \begin{multicols}{2}
    \begin{itemize}[leftmargin=2em]
        \item Manual debugging and bug reproduction \cite{10.1145/3696630.3728718,WangHCLWW24}
        \item Manual elicitation and formalization of requirements~\cite{10.1016/j.infsof.2025.107697,10.1145/3715003} 
        \item Need for Search of Code snippets~\cite{10.1145/3678172}
        \item Need for manually created documentation~\cite{10.1145/3678172}
        \item Expert-only use of formal methods~\cite{10.1016/j.infsof.2025.107697}
    \end{itemize}
  \end{multicols}
}
};

\node[keyword, below right=0.25em and 0.25em of enhance.north west] {Enhances};
\node[keyword, below left=0.25em and 0.25em of reverse.north east] {Reverses};
\node[keyword, above right=0.25em and 0.25em of retrieve.south west] {Retrieves};
\node[keyword, above left=0.25em and 0.25em of obsolesce.south east] {Obsolesces};

\node[centerbox, text width = 4em] at (0,0) {\copilot{}s in \sdlc};

\end{tikzpicture}
\caption{McLuhan’s tetrad derived from the rapid literature review of \copilot publications.}
\Description{Image shows the four quadrants of the tetrad diagram resulting from the rapid literature review and analysis.}
\label{fig:tetrad-copilots}
\end{figure}

%% file: figures/mcluhans_tetrad_teammate.tex
\begin{figure}[h]
\centering
\begin{tikzpicture}[
    quadrant/.style={draw, rounded corners, minimum width=5cm, text width = 0.45 \textwidth, minimum height=12.5em, align=left, font=\scriptsize, inner sep=5pt},
    centerbox/.style={draw, circle, minimum size=2cm, align=center, fill=gray!15, font=\small\bfseries},
    keyword/.style={font=\small\bfseries}
]

\def\gap{0.5em}

\node[quadrant, anchor=south east] at (-\gap,\gap) (enhance) {
    \setlength{\columnsep}{2pt}
    \begin{multicols}{2}
        \begin{itemize}[leftmargin=2em]
            \item Speed \cite{prophetAgent}
            \item Time To Market \cite{sami2024experimenting}
            \item Debugging Skills \cite{bouzenia2024repairagent, 10.1145/3657604.3664663}
            \item Software Automation \cite{sami2024experimenting}
            \item Fast Prototyping \cite{11030030}
        \end{itemize}
    \end{multicols}
};

\node[quadrant, anchor=south west] at (\gap,\gap) (reverse) {
\setlength{\columnsep}{2pt}
  \begin{multicols}{2}
    \begin{itemize}[leftmargin=2em]
            \item Git Versioning \cite{11030040, sami2024experimenting}
            \item Code Comprehension \cite{10.1145/3696630.3728493}
            \item Ethics and autonomy \cite{bouzenia2024repairagent,khan2025ai,10.1145/3696630.3728717}
            \item Environmental/Electrical sustainability \cite{qian2023chatdev}
            \item Accountability \& Code Ownership \cite{10.1145/3696630.3728717, 10.1145/3696630.3728493}

    \end{itemize}
  \end{multicols}
};

\node[quadrant, anchor=north east] at (-\gap,-\gap) (retrieve) {
\vtop{
    \setlength{\columnsep}{2pt}
      \begin{multicols}{2}
        \begin{itemize}[leftmargin=2em]
            \item Software and Architectural Design \cite{10.1145/3696630.3728493,sami2024experimenting, DBLP:conf/satrends/BecattiniVV25}
            \item Requirement Engineering Elicitation and Modeling \cite{lin2024soen, DBLP:conf/satrends/BecattiniVV25}
            \item Linguistic Skills, Clarity and Transparent Communication \cite{qian2023chatdev, sami2024experimenting}
            \item Code Quality and Technical Debt \cite{qian2023chatdev, lin2024soen}
            \item Static Analysis \cite{lin2024soen, rondon2025evaluating, sami2024experimenting}
            \item Human-Computer Interactions Principles \cite{11052708}
            \item Dependency management \cite{zhao2024commit0, manish2024autonomous}
        \end{itemize}
      \end{multicols}
  }
};

\node[quadrant, anchor=north west] at (\gap,-\gap) (obsolesce) {
\vtop{
\setlength{\columnsep}{2pt}
  \begin{multicols}{2}
    \begin{itemize}[leftmargin=2em]
            \item Project Tracking Software Tools \cite{rasheed2023autonomousagentssoftwaredevelopment, sami2024experimenting}
            \item Traditional IDE editors \cite{11052708}
            \item Manual Coding \cite{multi-agentcollab,codepori,sami2024experimenting} 
            \item Conventional Quality Assurance Testing \cite{sami2024experimenting}
            
    \end{itemize}
  \end{multicols}
}
};

\node[keyword, below right=0.25em and 0.25em of enhance.north west] {Enhances};
\node[keyword, below left=0.25em and 0.25em of reverse.north east] {Reverses};
\node[keyword, above right=0.25em and 0.25em of retrieve.south west] {Retrieves};
\node[keyword, above left=0.25em and 0.25em of obsolesce.south east] {Obsolesces};

\node[centerbox, text width = 4em] at (0,0) {\teammate{} in \sdlc};

\end{tikzpicture}
\caption{McLuhan’s tetrad derived from the rapid literature review of \teammate{} publications.}
\Description{Image shows the four quadrants of the tetrad diagram resulting from the rapid literature review and analysis.}
\label{fig:tetrad-teammates}
\end{figure}

%% file: figures/mcluhans_tetrad_genaiware.tex
\begin{figure}[htb]
\centering
\begin{tikzpicture}[
    quadrant/.style={draw, rounded corners, minimum width=5cm, text width = 0.45 \textwidth, minimum height=12.5em, align=left, font=\scriptsize, inner sep=5pt},
    centerbox/.style={draw, circle, minimum size=2cm, align=center, fill=gray!15, font=\small\bfseries},
    keyword/.style={font=\small\bfseries}
]

\def\gap{0.5em}

\node[quadrant, anchor=south east] at (-\gap,\gap) (enhance) {
    \setlength{\columnsep}{2pt}
    \begin{multicols}{2}
        \begin{itemize}[leftmargin=2em]
            \item Prompt Programming~\cite{DBLP:conf/chi/FengYY0ZL24,DBLP:conf/www/Bakharia25,11082094}
            \item Data Management~\cite{DBLP:conf/icsa/BucaioniWHL025,11121725}
            \item User Experience~\cite{DBLP:conf/icsa/BucaioniWHL025,DBLP:journals/access/AminiranjbarTWPV25,DBLP:conf/satrends/BecattiniVV25}
            \item Modular Software Design~\cite{DBLP:conf/icsa/HeckingSF25,11113067,11127266,DBLP:conf/csit/Joshi24}
            \item Responsible and Secure AI~\cite{DBLP:conf/icsa/LuZXXHW24,11121725}
        \end{itemize}
    \end{multicols}
};

\node[quadrant, anchor=south west] at (\gap,\gap) (reverse) {
\setlength{\columnsep}{2pt}
  \begin{multicols}{2}
    \begin{itemize}[leftmargin=2em]
        \item Reliability~\cite{DBLP:conf/cain/BarnettKTB024,10771003,11127266,DBLP:conf/icse/Shao0S00025,DBLP:conf/eit/LanKPPP24,DBLP:journals/pacmse/LiangLRM25}
        \item Standardization~\cite{DBLP:conf/icse/Shao0S00025,DBLP:conf/icsa/BucaioniWHL025,DBLP:conf/csit/Joshi24}
        \item Transparency and Trustworthiness~\cite{DBLP:conf/icsa/LuZXXHW24,DBLP:journals/pacmse/LiangLRM25,DBLP:conf/cain/00010XLX024,DBLP:conf/icsa/HeckingSF25,DBLP:conf/icsa/KholkarTPR24,DBLP:conf/eit/LanKPPP24}
        \item Efficiency and Sustainability~\cite{DBLP:conf/cain/BarnettKTB024,DBLP:journals/pacmse/LiangLRM25,11121725,DBLP:conf/icsa/LuZXXHW24,DBLP:conf/cain/00010XLX024}
    \end{itemize}
  \end{multicols}
};

\node[quadrant, anchor=north east] at (-\gap,-\gap) (retrieve) {
\vtop{
    \setlength{\columnsep}{2pt}
      \begin{multicols}{2}
        \begin{itemize}[leftmargin=2em]
            \item Manual Evaluation~\cite{DBLP:conf/icsa/LuZXXHW24,DBLP:conf/cain/00010XLX024,DBLP:journals/pacmse/LiangLRM25,11121725}
            \item Runtime Verification and Validation~\cite{11121725,DBLP:conf/csit/Joshi24,11127266}
            \item Interoperability Issues~\cite{DBLP:journals/pacmse/LiangLRM25}
            \item Document Stores~\cite{DBLP:conf/cain/BarnettKTB024}
            \item Goal-Oriented Requirements~\cite{10771003}
        \end{itemize}
      \end{multicols}
  }
};

\node[quadrant, anchor=north west] at (\gap,-\gap) (obsolesce) {
\vtop{
\setlength{\columnsep}{2pt}
  \begin{multicols}{2}
    \begin{itemize}[leftmargin=2em]
        \item Manual Annotation~\cite{DBLP:conf/cain/BarnettKTB024}
        \item Traditional Static Testing~\cite{11127266,DBLP:conf/cain/BarnettKTB024,DBLP:conf/icse/Shao0S00025,DBLP:conf/icsa/BucaioniWHL025}
        \item Static Code Development~\cite{DBLP:journals/pacmse/LiangLRM25}
        \item Separation between Machine Learning Cycle and \sdlc{}~\cite{DBLP:conf/icsa/KholkarTPR24}
        \item Opacity of AI Components~\cite{DBLP:conf/csit/Joshi24,DBLP:conf/cain/00010XLX024,DBLP:conf/icsa/LuZXXHW24,DBLP:conf/icsa/BucaioniWHL025}
    \end{itemize}
  \end{multicols}
}
};

\node[keyword, below right=0.25em and 0.25em of enhance.north west] {Enhances};
\node[keyword, below left=0.25em and 0.25em of reverse.north east] {Reverses};
\node[keyword, above right=0.25em and 0.25em of retrieve.south west] {Retrieves};
\node[keyword, above left=0.25em and 0.25em of obsolesce.south east] {Obsolesces};

\node[centerbox, text width = 4em] at (0,0) {\aiware{} in \sdlc};

\end{tikzpicture}
\caption{McLuhan’s tetrad derived from the rapid literature review of \aiware{} publications.}
\Description{Image shows the four quadrants of the tetrad diagram resulting from the rapid literature review and analysis.}
\label{fig:tetrad-genaiware}
\end{figure}

%% file: figures/mcluhans_tetrad_robot.tex
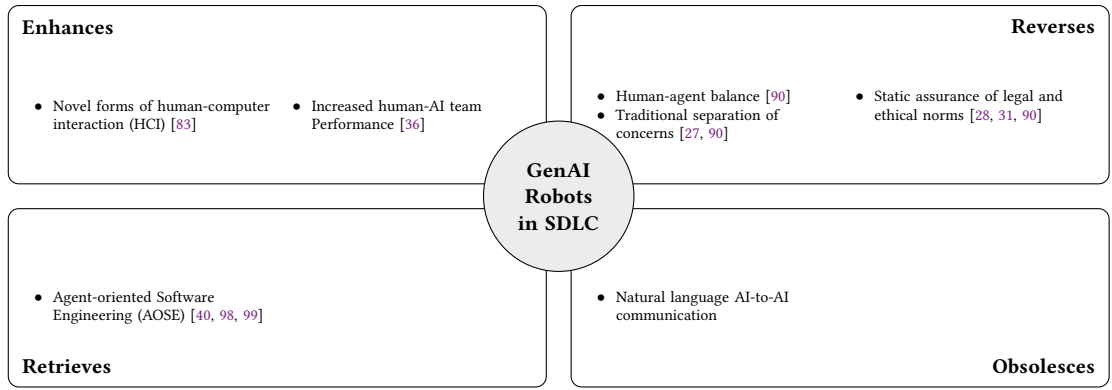
\begin{figure}[h]
\centering
\begin{tikzpicture}[
    quadrant/.style={draw, rounded corners, minimum width=5cm, text width = 0.45 \textwidth, minimum height=7.5em, align=left, font=\scriptsize, inner sep=5pt},
    centerbox/.style={draw, circle, minimum size=2cm, align=center, fill=gray!15, font=\small\bfseries},
    keyword/.style={font=\small\bfseries}
]

\def\gap{0.5em}

\node[quadrant, anchor=south east] at (-\gap,\gap) (enhance) {
    \setlength{\columnsep}{2pt}
    \begin{multicols}{2}
        \begin{itemize}[leftmargin=2em]
            \item Novel forms of human-computer interaction (HCI)~\cite{10.1007/s10515-024-00426-z}
            \item Increased human-AI team Performance~\cite{Hemmer2025}
        \end{itemize}
    \end{multicols}
};

\node[quadrant, anchor=south west] at (\gap,\gap) (reverse) {
\setlength{\columnsep}{2pt}
  \begin{multicols}{2}
    \begin{itemize}[leftmargin=2em]
        \item Human-agent balance~\cite{Vu25}
        \item Traditional separation of concerns~\cite{Fournier2025,Vu25}
        \item Static assurance of legal and ethical norms~\cite{Gabriel25,grisold2025,Vu25}
    \end{itemize}
  \end{multicols}
};

\node[quadrant, anchor=north east] at (-\gap,-\gap) (retrieve) {
\vtop{
    \setlength{\columnsep}{2pt}
      \begin{multicols}{2}
        \begin{itemize}[leftmargin=2em]
            \item Agent-oriented Software Engineering (AOSE)~\cite{Jennings00,WooldridgeC00,wooldridge1999}
        \end{itemize}
      \end{multicols}
  }
};

\node[quadrant, anchor=north west] at (\gap,-\gap) (obsolesce) {
\vtop{
\setlength{\columnsep}{2pt}
  \begin{multicols}{2}
    \begin{itemize}[leftmargin=2em]
        \item Natural language AI-to-AI communication
    \end{itemize}
  \end{multicols}
}
};

\node[keyword, below right=0.25em and 0.25em of enhance.north west] {Enhances};
\node[keyword, below left=0.25em and 0.25em of reverse.north east] {Reverses};
\node[keyword, above right=0.25em and 0.25em of retrieve.south west] {Retrieves};
\node[keyword, above left=0.25em and 0.25em of obsolesce.south east] {Obsolesces};

\node[centerbox, text width = 4em] at (0,0) {\robot{}s in \sdlc};

\end{tikzpicture}
\caption{McLuhan’s tetrad for \robot{}s.}
\label{fig:tetrad-robots}
\Description{Image shows the four quadrants of the tetrad diagram resulting from the rapid literature review and analysis.}
\end{figure}

%% file: tosem-refs.bib
@inproceedings{Fournier2025,
author = {Fournier, Fabiana and Limonad, Lior and David, Yuval},
year = {2025},
title = {Agentic AI Process Observability: Discovering Behavioral Variability},
booktitle    = {4th International Workshop on Process Management in the AI era (PMAI), collocated with ECAI, Bologna, Italy,October 25, 2025},
doi = {}
}

@article{TrinkenreichEtAl2025,
  author       = {Bianca Trinkenreich and
                  Fabio Calefato and
                  Geir Hanssen and
                  Kelly Blincoe and
                  Marcos Kalinowski and
                  Mauro Pezz{\`{e}} and
                  Paolo Tell and
                  Margaret{-}Anne D. Storey},
  title        = {Get on the Train or be Left on the Station: Using LLMs for Software
                  Engineering Research},
  journal      = {CoRR},
  volume       = {abs/2506.12691},
  year         = {2025},
  url          = {https://doi.org/10.48550/arXiv.2506.12691},
  doi          = {10.48550/ARXIV.2506.12691},
  eprinttype    = {arXiv},
  eprint       = {2506.12691},
  bibsource    = {dblp computer science bibliography, https://dblp.org}
}

@book{hevner:2021,
	title        = {{Externalities of Design Science Research: Preparation for Project Success}},
	author       = {Hevner, Alan R. and Storey, Veda C.},
	year         = 2021,
	booktitle    = {Lecture Notes in Computer Science (including subseries Lecture Notes in Artificial Intelligence and Lecture Notes in Bioinformatics)},
	publisher    = {Springer International Publishing},
	volume       = {12807 LNCS},
}

@article{baskerville:2018,
	title        = {{Design science research contributions: Finding a balance between artifact and theory}},
	author       = {Baskerville, Richard and Baiyere, Abayomi and others},
	year         = 2018,
	journal      = {Journal of the Association for Information Systems},
	volume       = 19,
	number       = 5,
}

@inproceedings{wieringa:2009,
	title        = {Design science as nested problem solving},
	author       = {Wieringa, Roel},
	year         = 2009,
	booktitle    = {Proceedings of the 4th international conference on design science research in information systems and technology},
}

@article{TOSEMPezze2025, author = {Abrah\~{a}o, Silvia and Grundy, John and Pezz\`{e}, Mauro and Storey, Margaret-Anne and Tamburri, Damian A.}, title = {Software Engineering by and for Humans in an AI Era}, year = {2025}, issue_date = {June 2025}, publisher = {Association for Computing Machinery}, address = {New York, NY, USA}, volume = {34}, number = {5}, issn = {1049-331X}, url = {https://doi.org/10.1145/3715111}, doi = {10.1145/3715111}, journal = {ACM Trans. Softw. Eng. Methodol.}, month = may, articleno = {129}, numpages = {46} }

@misc{ozkaya_2023,
author={Ozkaya, Ipek and Carleton, Anita and Robert, John and Schmidt, Douglas},
title={Application of Large Language Models (LLMs) in Software Engineering: Overblown Hype or Disruptive Change?},
month={{Oct}},
year={{2023}},
howpublished={Carnegie Mellon University, Software Engineering Institute's Insights (blog)},
url={https://doi.org/10.58012/6n1p-pw64},
}

@inproceedings{korn2025,
title={LLMREI: Automating Requirements Elicitation Interviews with LLMs},
author={Alexander Korn and Samuel Gorsch and Andreas Vogelsang},
booktitle    = {33rd {IEEE} International Requirements Engineering Conference, {RE}
                  2025, Valencia, Spain, Sep 1-5, 2025},
year={2025}
}

@article{ZhangFMSC24,
  author       = {Quanjun Zhang and
                  Chunrong Fang and
                  Yuxiang Ma and
                  Weisong Sun and
                  Zhenyu Chen},
  title        = {A Survey of Learning-based Automated Program Repair},
  journal      = {{ACM} Trans. Softw. Eng. Methodol.},
  volume       = {33},
  number       = {2},
  pages        = {55:1--55:69},
  year         = {2024},
  url          = {https://doi.org/10.1145/3631974},
  doi          = {10.1145/3631974},
  bibsource    = {dblp computer science bibliography, https://dblp.org}
}

@article{WangHCLWW24,
  author       = {Junjie Wang and
                  Yuchao Huang and
                  Chunyang Chen and
                  Zhe Liu and
                  Song Wang and
                  Qing Wang},
  title        = {Software Testing With Large Language Models: Survey, Landscape, and
                  Vision},
  journal      = {{IEEE} Trans. Software Eng.},
  volume       = {50},
  number       = {4},
  pages        = {911--936},
  year         = {2024},
  url          = {https://doi.org/10.1109/TSE.2024.3368208},
  doi          = {10.1109/TSE.2024.3368208},
  bibsource    = {dblp computer science bibliography, https://dblp.org}
}

@misc{jiang2024,
      title={A Survey on Large Language Models for Code Generation},
      author={Juyong Jiang and Fan Wang and Jiasi Shen and Sungju Kim and Sunghun Kim},
      year={2024},
      eprint={2406.00515},
      archivePrefix={arXiv},
      primaryClass={cs.CL},
      url={https://arxiv.org/abs/2406.00515},
}

@misc{weber2024,
      title={Large Language Models as Software Components: A Taxonomy for LLM-Integrated Applications},
      author={Irene Weber},
      year={2024},
      eprint={2406.10300},
      archivePrefix={arXiv},
      primaryClass={cs.SE},
      url={https://arxiv.org/abs/2406.10300},
}

@misc{li2025,
      title={The Rise of AI Teammates in Software Engineering (SE) 3.0: How Autonomous Coding Agents Are Reshaping Software Engineering},
      author={Hao Li and Haoxiang Zhang and Ahmed E. Hassan},
      year={2025},
      eprint={2507.15003},
      archivePrefix={arXiv},
      primaryClass={cs.SE},
      url={https://arxiv.org/abs/2507.15003},
}

@inproceedings{Fettke2025XABPs,
  title={{XABPs: Towards eXplainable Autonomous Business Processes}},
  author={Fettke, Peter and Fournier, Fabiana and Limonad, Lior and Metzger, Andreas and Rinderle-Ma, Stefanie and Weber, Barbara},
  booktitle={4th International Workshop on Process Management in the AI era (PMAI), collocated with the 28th European Conference on Artificial Intelligence (ECAI), Bologna, October 25-30, 2025},
  year={2025}
}

@article{10.1007/s10515-024-00426-z,
  title = {Future of Software Development with Generative {{AI}}},
  author = {Sauvola, Jaakko and Tarkoma, Sasu and Klemettinen, Mika and Riekki, Jukka and Doermann, David},
  year = {2024},
  month = mar,
  journal = {Automated Software Engg.},
  volume = {31},
  number = {1},
  publisher = {Kluwer Academic Publishers},
  address = {USA},
  issn = {0928-8910},
  doi = {10.1007/s10515-024-00426-z},
  issue_date = {May 2024}
}

@article{10.1016/j.infsof.2025.107697,
  title = {Formal Requirements Engineering and Large Language Models: {{A}} Two-Way Roadmap},
  author = {Ferrari, Alessio and Spoletini, Paola},
  year = {2025},
  month = apr,
  journal = {Inf. Softw. Technol.},
  volume = {181},
  number = {C},
  publisher = {Butterworth-Heinemann},
  address = {USA},
  issn = {0950-5849},
  doi = {10.1016/j.infsof.2025.107697},
  issue_date = {May 2025}
}

@article{10.1016/j.infsof.2025.107751,
  title = {Copiloting the Future: {{How}} Generative {{AI}} Transforms Software Engineering},
  author = {Banh, Leonardo and Holldack, Florian and Strobel, Gero},
  year = {2025},
  month = jun,
  journal = {Inf. Softw. Technol.},
  volume = {183},
  number = {C},
  publisher = {Butterworth-Heinemann},
  address = {USA},
  issn = {0950-5849},
  doi = {10.1016/j.infsof.2025.107751},
  issue_date = {Jul 2025}
}

@article{10.1145/3678172,
  title = {A Disruptive Research Playbook for Studying Disruptive Innovations},
  author = {Storey, Margaret-Anne and Russo, Daniel and Novielli, Nicole and Kobayashi, Takashi and Wang, Dong},
  year = {2024},
  month = nov,
  journal = {ACM Trans. Softw. Eng. Methodol.},
  volume = {33},
  number = {8},
  publisher = {Association for Computing Machinery},
  address = {New York, NY, USA},
  issn = {1049-331X},
  doi = {10.1145/3678172},
  articleno = {195},
  issue_date = {November 2024}
}

@article{Terragni24,
  author       = {Valerio Terragni and
                  Partha S. Roop and
                  Kelly Blincoe},
  title        = {The Future of Software Engineering in an AI-Driven World},
  journal      = {CoRR},
  volume       = {abs/2406.07737},
  year         = {2024},
  url          = {https://doi.org/10.48550/arXiv.2406.07737},
  doi          = {10.48550/ARXIV.2406.07737},
  eprinttype    = {arXiv},
  eprint       = {2406.07737},
  bibsource    = {dblp computer science bibliography, https://dblp.org}
}

@article{Duc2025,
  author       = {Anh Nguyen{-}Duc and
                  Beatriz Cabrero Daniel and
                  Adam Przybylek and
                  Chetan Arora and
                  Dron Khanna and
                  Tomas Herda and
                  Usman Rafiq and
                  Jorge Melegati and
                  Eduardo Guerra and
                  Kai{-}Kristian Kemell and
                  Mika Saari and
                  Zheying Zhang and
                  Huy Le and
                  Tho Quan and
                  Pekka Abrahamsson},
  title        = {Generative Artificial Intelligence for Software Engineering - {A}
                  Research Agenda},
  journal      = {Software Practice and Experience},
  year         = {2025},
  url          = { https://doi.org/10.1002/spe.70005},
}

@article{Esposito2025,
  author       = {Matteo Esposito and
                  Xiaozhou Li and
                  Sergio Moreschini and
                  Noman Ahmad and
                  Tom{\'{a}}s Cern{\'{y}} and
                  Karthik Vaidhyanathan and
                  Valentina Lenarduzzi and
                  Davide Taibi},
  title        = {Generative {AI} for Software Architecture. Applications, Trends, Challenges,
                  and Future Directions},
  journal      = {SSRN},
  year         = {2025},
  url          = { https://dx.doi.org/10.2139/ssrn.5196419},
}

@article{Jin2025,
  author       = {Haolin Jin and
                  Linghan Huang and
                  Haipeng Cai and
                  Jun Yan and
                  Bo Li and
                  Huaming Chen},
  title        = {From LLMs to LLM-based Agents for Software Engineering: {A} Survey
                  of Current, Challenges and Future},
  journal      = {CoRR},
  volume       = {abs/2408.02479},
  year         = {2024},
  url          = {https://doi.org/10.48550/arXiv.2408.02479},
  doi          = {10.48550/ARXIV.2408.02479},
  eprinttype    = {arXiv},
  eprint       = {2408.02479},
  bibsource    = {dblp computer science bibliography, https://dblp.org}
}

@article{LiZLWJT24,
  author       = {Jialong Li and
                  Mingyue Zhang and
                  Nianyu Li and
                  Danny Weyns and
                  Zhi Jin and
                  Kenji Tei},
  title        = {Generative {AI} for Self-Adaptive Systems: State of the Art and Research
                  Roadmap},
  journal      = {{ACM} Trans. Auton. Adapt. Syst.},
  volume       = {19},
  number       = {3},
  pages        = {13:1--13:60},
  year         = {2024},
  url          = {https://doi.org/10.1145/3686803},
  doi          = {10.1145/3686803},
  bibsource    = {dblp computer science bibliography, https://dblp.org}
}

@article{Alenezi2025,
  author       = {Mamdouh Alenezi and Mohammed Akour},
  title        = {{AI}-Driven Innovations in Software Engineering: A Review of Current Practices and Future Directions},
  journal      = {Applied Sciences},
volume = {15},
issue={3},
  year         = {2025},
  url          = {https://doi.org/10.3390/app15031344},
}

@article{Gu2025,
  author       = {Alex Gu and
                  Naman Jain and
                  Wen{-}Ding Li and
                  Manish Shetty and
                  Yijia Shao and
                  Ziyang Li and
                  Diyi Yang and
                  Kevin Ellis and
                  Koushik Sen and
                  Armando Solar{-}Lezama},
  title        = {Challenges and Paths Towards {AI} for Software Engineering},
  journal      = {CoRR},
  volume       = {abs/2503.22625},
  year         = {2025},
  url          = {https://doi.org/10.48550/arXiv.2503.22625},
  doi          = {10.48550/ARXIV.2503.22625},
  eprinttype    = {arXiv},
  eprint       = {2503.22625},
}

@article{SE2020V1,
editor = {Pezz\`{e}, Mauro and Roychoudhury, Abhik},
year = {2025},
issue_date = {June 2025},
publisher = {Association for Computing Machinery},
volume = {34},
number = {5},
issn = {1049-331X},
journal = {ACM Trans. Softw. Eng. Methodol.},
url          = {https://dl.acm.org/toc/tosem/2025/34/5},
}

@inproceedings{ChomatekSP25,
  author       = {Lukasz Chomatek and
                  Wojciech Slabosz and
                  Aneta Poniszewska{-}Maranda},
  editor       = {Leonardo Montecchi and
                  Jingyue Li and
                  Denys Poshyvanyk and
                  Dongmei Zhang},
  title        = {From Words to Wisdom: LLMs Summarizing Instructional Content},
  booktitle    = {Proceedings of the 33rd {ACM} International Conference on the Foundations
                  of Software Engineering, {FSE} Companion 2025, Clarion Hotel Trondheim,
                  Trondheim, Norway, June 23-28, 2025},
  pages        = {1623--1630},
  publisher    = {{ACM}},
  year         = {2025},
  url          = {https://doi.org/10.1145/3696630.3728697},
  doi          = {10.1145/3696630.3728697},
  bibsource    = {dblp computer science bibliography, https://dblp.org}
}

@article{AhmedACCHHPPX25,
  author       = {Iftekhar Ahmed and
                  Aldeida Aleti and
                  Haipeng Cai and
                  Alexander Chatzigeorgiou and
                  Pinjia He and
                  Xing Hu and
                  Mauro Pezz{\`{e}} and
                  Denys Poshyvanyk and
                  Xin Xia},
  title        = {Artificial Intelligence for Software Engineering: The Journey So Far
                  and the Road Ahead},
  journal      = {{ACM} Trans. Softw. Eng. Methodol.},
  volume       = {34},
  number       = {5},
  pages        = {119:1--119:27},
  year         = {2025},
  url          = {https://doi.org/10.1145/3719006},
  doi          = {10.1145/3719006},
  bibsource    = {dblp computer science bibliography, https://dblp.org}
}

@article{LiZH25,
  author       = {Hao Li and
                  Haoxiang Zhang and
                  Ahmed E. Hassan},
  title        = {The Rise of {AI} Teammates in Software Engineering {(SE)} 3.0: How
                  Autonomous Coding Agents Are Reshaping Software Engineering},
  journal      = {CoRR},
  volume       = {abs/2507.15003},
  year         = {2025},
  url          = {https://doi.org/10.48550/arXiv.2507.15003},
  doi          = {10.48550/ARXIV.2507.15003},
  eprinttype    = {arXiv},
  eprint       = {2507.15003},
  bibsource    = {dblp computer science bibliography, https://dblp.org}
}

@article{PezzeCGPQ24,
  author       = {Mauro Pezz{\`{e}} and
                  Matteo Ciniselli and
                  Luca Di Grazia and
                  Niccol{\`{o}} Puccinelli and
                  Ketai Qiu},
  title        = {The Trailer of the {ACM} 2030 Roadmap for Software Engineering},
  journal      = {{ACM} {SIGSOFT} Softw. Eng. Notes},
  volume       = {49},
  number       = {4},
  pages        = {31--40},
  year         = {2024},
  url          = {https://doi.org/10.1145/3696117.3696126},
  doi          = {10.1145/3696117.3696126},
  bibsource    = {dblp computer science bibliography, https://dblp.org}
}

@article{10.1145/3695988,
  title = {Large Language Models for Software Engineering: A Systematic Literature Review},
  author = {Hou, Xinyi and Zhao, Yanjie and Liu, Yue and Yang, Zhou and Wang, Kailong and Li, Li and Luo, Xiapu and Lo, David and Grundy, John and Wang, Haoyu},
  year = {2024},
  month = dec,
  journal = {ACM Trans. Softw. Eng. Methodol.},
  volume = {33},
  number = {8},
  publisher = {Association for Computing Machinery},
  address = {New York, NY, USA},
  issn = {1049-331X},
  doi = {10.1145/3695988},
  articleno = {220},
  issue_date = {November 2024}
}

@inproceedings{10.1145/3696630.3727251,
  title = {What Do Professional Software Developers Need to Know to Succeed in an Age of {{Artificial Intelligence}}?},
  booktitle = {Proceedings of the 33rd {{ACM}} International Conference on the Foundations of Software Engineering},
  author = {Kam, Matthew and Miller, Cody and Wang, Miaoxin and Tidwell, Abey and Lee, Irene A. and {Malyn-Smith}, Joyce and Perret, Beatriz and Tiwari, Vikram and Kenitzer, Joshua and Macvean, Andrew and Barrar, Erin},
  year = {2025},
  series = {{{FSE}} Companion '25},
  pages = {947--958},
  publisher = {Association for Computing Machinery},
  address = {Clarion Hotel Trondheim, Trondheim, Norway and New York, NY, USA},
  doi = {10.1145/3696630.3727251},
  isbn = {979-8-4007-1276-0}
}

@inproceedings{10.1145/3696630.3728718,
  title = {Challenges and Opportunities for Generative {{AI}} in Software Engineering: A Managerial View},
  booktitle = {Proceedings of the 33rd {{ACM}} International Conference on the Foundations of Software Engineering},
  author = {Rico, Sergio and {\"O}berg, Lena-Maria},
  year = {2025},
  series = {{{FSE}} Companion '25},
  pages = {1338--1344},
  publisher = {Association for Computing Machinery},
  address = {Clarion Hotel Trondheim, Trondheim, Norway and New York, NY, USA},
  doi = {10.1145/3696630.3728718},
  isbn = {979-8-4007-1276-0}
}

@inproceedings{10.1145/3696630.3730538,
  title = {{{AI}} in the Software Development Lifecycle: {{Insights}} and Open Research Questions},
  booktitle = {Proceedings of the 33rd {{ACM}} International Conference on the Foundations of Software Engineering},
  author = {Guimaraes, Everton and Nascimento, Nathalia},
  year = {2025},
  series = {{{FSE}} Companion '25},
  pages = {1353--1357},
  publisher = {Association for Computing Machinery},
  address = {Clarion Hotel Trondheim, Trondheim, Norway and New York, NY, USA},
  doi = {10.1145/3696630.3730538},
  isbn = {979-8-4007-1276-0}
}

@article{10.1145/3708519,
  title = {Automatic Programming: {{Large}} Language Models and Beyond},
  author = {Lyu, Michael R. and Ray, Baishakhi and Roychoudhury, Abhik and Tan, Shin Hwei and Thongtanunam, Patanamon},
  year = {2025},
  month = may,
  journal = {ACM Trans. Softw. Eng. Methodol.},
  volume = {34},
  number = {5},
  publisher = {Association for Computing Machinery},
  address = {New York, NY, USA},
  issn = {1049-331X},
  doi = {10.1145/3708519},
  articleno = {140},
  issue_date = {June 2025}
}

@article{10.1145/3709353,
  title = {From Today's Code to Tomorrow's Symphony: {{The AI}} Transformation of Developer's Routine by 2030},
  author = {Qiu, Ketai and Puccinelli, Niccol{\`o} and Ciniselli, Matteo and Di Grazia, Luca},
  year = {2025},
  month = may,
  journal = {ACM Trans. Softw. Eng. Methodol.},
  volume = {34},
  number = {5},
  publisher = {Association for Computing Machinery},
  address = {New York, NY, USA},
  issn = {1049-331X},
  doi = {10.1145/3709353},
  articleno = {121},
  issue_date = {June 2025}
}

@article{10.1145/3709360,
  title = {From Triumph to Uncertainty: {{The}} Journey of Software Engineering in the {{AI}} Era},
  author = {Mastropaolo, Antonio and {Escobar-Vel{\'a}squez}, Camilo and {Linares-V{\'a}squez}, Mario},
  year = {2025},
  month = may,
  journal = {ACM Trans. Softw. Eng. Methodol.},
  volume = {34},
  number = {5},
  publisher = {Association for Computing Machinery},
  address = {New York, NY, USA},
  issn = {1049-331X},
  doi = {10.1145/3709360},
  articleno = {131},
  issue_date = {June 2025}
}

@article{10.1145/3712005,
  title = {The Current Challenges of Software Engineering in the Era of Large Language Models},
  author = {Gao, Cuiyun and Hu, Xing and Gao, Shan and Xia, Xin and Jin, Zhi},
  year = {2025},
  month = may,
  journal = {ACM Trans. Softw. Eng. Methodol.},
  volume = {34},
  number = {5},
  publisher = {Association for Computing Machinery},
  address = {New York, NY, USA},
  issn = {1049-331X},
  doi = {10.1145/3712005},
  articleno = {127},
  issue_date = {June 2025}
}

@article{10.1145/3715003,
  title = {The Future of {{AI-driven}} Software Engineering},
  author = {Terragni, Valerio and Vella, Annie and Roop, Partha and Blincoe, Kelly},
  year = {2025},
  month = may,
  journal = {ACM Trans. Softw. Eng. Methodol.},
  volume = {34},
  number = {5},
  publisher = {Association for Computing Machinery},
  address = {New York, NY, USA},
  issn = {1049-331X},
  doi = {10.1145/3715003},
  articleno = {120},
  issue_date = {June 2025}
}

@inproceedings{10.1145/3657604.3664663,
author = {Wang, Tianjia and Ramanujan, Ramaraja and Lu, Yi and Mao, Chenyu and Chen, Yan and Brown, Chris},
title = {DevCoach: Supporting Students in Learning the Software Development Life Cycle at Scale with Generative Agents},
year = {2024},
isbn = {9798400706332},
publisher = {Association for Computing Machinery},
address = {New York, NY, USA},
url = {https://doi.org/10.1145/3657604.3664663},
doi = {10.1145/3657604.3664663},
booktitle = {Proceedings of the Eleventh ACM Conference on Learning @ Scale},
pages = {351–355},
numpages = {5},
location = {Atlanta, GA, USA},
series = {L@S '24}
}

@INPROCEEDINGS{11052708,
  author={Bellur, Abhiram and Batole, Fraol},
  booktitle={2025 IEEE/ACM Second IDE Workshop (IDE)}, 
  title={IDE Native, Foundation Model Based Agents for Software Refactoring}, 
  year={2025},
  volume={},
  number={},
  pages={42-45},
  doi={10.1109/IDE66625.2025.00013}}

@INPROCEEDINGS{11024270,
  author={Xu, Yisen},
  booktitle={2025 IEEE/ACM 47th International Conference on Software Engineering: Companion Proceedings (ICSE-Companion)}, 
  title={MUARF: Leveraging Multi-Agent Workflows for Automated Code Refactoring}, 
  year={2025},
  volume={},
  number={},
  pages={226-227},
  doi={10.1109/ICSE-Companion66252.2025.00071}}

@article{zhao2024commit0,
  title={Commit0: Library generation from scratch},
  author={Zhao, Wenting and Jiang, Nan and Lee, Celine and Chiu, Justin T and Cardie, Claire and Gall{\'e}, Matthias and Rush, Alexander M},
  journal={arXiv preprint arXiv:2412.01769},
  year={2024}
}

@misc{keim_automated_2025,
  title = {{Towards Automated Knowledge Management in the Software Life Cycle}},
  author = {Keim, Jan and Hey, Tobias and Scotti, Vincenzo and Mirandola, Raffaela and Koziolek, Anne},
  year = {2025},
  doi = {10.5445/IR/1000181618},
  urldate = {2025-08-27},
  isbn = {9781000181616},
}

@article{MishraS25a,
  author       = {Lalit Narayan Mishra and
                  Biswaranjan Senapati},
  title        = {Retail Resilience Engine: An Agentic {AI} Framework for Building Reliable
                  Retail Systems With Test-Driven Development Approach},
  journal      = {{IEEE} Access},
  volume       = {13},
  pages        = {50226--50243},
  year         = {2025},
  url          = {https://doi.org/10.1109/ACCESS.2025.3552592},
  doi          = {10.1109/ACCESS.2025.3552592},
  bibsource    = {dblp computer science bibliography, https://dblp.org}
}

@INPROCEEDINGS{Gowri25,
  author={Gowri, R.},
  booktitle={2025 International Conference on Data Science, Agents \& Artificial Intelligence (ICDSAAI)},
  title={OMACS Based Adaptive Intelligent Tutoring System Development},
  year={2025},
  volume={},
  number={},
  pages={1-6},
  doi={10.1109/ICDSAAI65575.2025.11011745}}

@article{LeeLH25,
  author       = {Seo{-}young Lee and
                  Matthew V. Law and
                  Guy Hoffman},
  title        = {When and How to Use {AI} in the Design Process? Implications for Human-AI
                  Design Collaboration},
  journal      = {Int. J. Hum. Comput. Interact.},
  volume       = {41},
  number       = {2},
  pages        = {1569--1584},
  year         = {2025},
  url          = {https://doi.org/10.1080/10447318.2024.2353451},
  doi          = {10.1080/10447318.2024.2353451},
  bibsource    = {dblp computer science bibliography, https://dblp.org}
}

@inproceedings{DuesterwaldIJKM24,
  author       = {Evelyn Duesterwald and
                  Vatche Isahagian and
                  K. R. Jayaram and
                  Ritesh Kumar and
                  Vinod Muthusamy and
                  Punleuk Oum and
                  Gegi Thomas and
                  Praveen Venkateswaran},
  title        = {A Conversational Assistant Framework for Automation},
  booktitle    = {Proceedings of the 25th International Middleware Conference Industrial
                  Track, Middleware Industrial Track 2024, Hong Kong, SAR, China, December
                  2-6, 2024},
  pages        = {1--7},
  publisher    = {{ACM}},
  year         = {2024},
  url          = {https://doi.org/10.1145/3700824.3701093},
  doi          = {10.1145/3700824.3701093},
  bibsource    = {dblp computer science bibliography, https://dblp.org}
}

@INPROCEEDINGS{11030030,
  author={Song, Hui and Goknil, Arda and Jiang, Xiaojun and Melum, Espen and Joe, Hyunwhan and Gazzotti, Caterina and Frascolla, Valerio and Videsjorden, Adela Nedisan and Nguyen, Phu},
  booktitle={2025 IEEE/ACM 4th International Conference on AI Engineering – Software Engineering for AI (CAIN)}, 
  title={Developing Multi-Agent LLM Applications Through Continuous Human-LLM Co-Programming}, 
  year={2025},
  volume={},
  number={},
  pages={42-47},
  doi={10.1109/CAIN66642.2025.00013}}

@inproceedings{WuWZLWFL24,
  author       = {Yin Wu and
                  Haijun Wang and
                  Yuanhui Zhang and
                  Xitao Li and
                  Hao Wu and
                  Ming Fan and
                  Ting Liu},
  title        = {Business Compliance Detection of Smart Contracts in Electricity and
                  Carbon Trading Scenarios},
  booktitle    = {35th {IEEE} International Symposium on Software Reliability Engineering,
                  {ISSRE} 2024 - Workshops, Tsukuba, Japan, October 28-31, 2024},
  pages        = {177--178},
  publisher    = {{IEEE}},
  year         = {2024},
  url          = {https://doi.org/10.1109/ISSREW63542.2024.00074},
  doi          = {10.1109/ISSREW63542.2024.00074},
  bibsource    = {dblp computer science bibliography, https://dblp.org}
}

@article{grisold2025,
  title={Guardrails for Human-AI Ecologies: Norm-Based Coordination and Design for Predictability},
  author={Grisold, Thomas and Berente, Nicholas and Seidel, Stefan},
  journal={Management Information Systems Quarterly},
  year={2025}
}

@article{Hemmer2025,
author = {Patrick Hemmer and Max Schemmer and Niklas Kühl and Michael Vössing and Gerhard Satzger},
title = {Complementarity in human-AI collaboration: concept, sources, and evidence},
journal = {European Journal of Information Systems},
volume = {0},
number = {0},
pages = {1--24},
year = {2025},
publisher = {Taylor \& Francis},
doi = {10.1080/0960085X.2025.2475962}
}

@book{Wooldridge2009,
  title={An introduction to multiagent systems},
  author={Wooldridge, Michael},
  year={2009},
  edition={Second},
  publisher={John wiley \& sons}
}

@inproceedings{WooldridgeC00,
  author       = {Michael J. Wooldridge and
                  Paolo Ciancarini},
  editor       = {Paolo Ciancarini and
                  Michael J. Wooldridge},
  title        = {Agent-Oriented Software Engineering: The State of the Art},
  booktitle    = {Agent-Oriented Software Engineering, First International Workshop,
                  {AOSE} 2000, Limerick, Ireland, June 10, 2000, Revised Papers},
  series       = {Lecture Notes in Computer Science},
  volume       = {1957},
  pages        = {1--28},
  publisher    = {Springer},
  year         = {2000},
  url          = {https://doi.org/10.1007/3-540-44564-1\_1},
  doi          = {10.1007/3-540-44564-1\_1},
  bibsource    = {dblp computer science bibliography, https://dblp.org}
}

@inproceedings{wooldridge1999,
  title={A methodology for agent-oriented analysis and design},
  author={Wooldridge, Michael and Jennings, Nicholas R and Kinny, David},
  booktitle={Proceedings of the third annual conference on Autonomous Agents},
  pages={69--76},
  year={1999}
}

@article{Jennings00,
  author       = {Nicholas R. Jennings},
  title        = {On agent-based software engineering},
  journal      = {Artif. Intell.},
  volume       = {117},
  number       = {2},
  pages        = {277--296},
  year         = {2000},
  url          = {https://doi.org/10.1016/S0004-3702(99)00107-1},
  doi          = {10.1016/S0004-3702(99)00107-1},
  bibsource    = {dblp computer science bibliography, https://dblp.org}
}

@inproceedings{Vu25,
      title={Agentic Business Process Management: Practitioner Perspectives on Agent Governance in Business Processes},
      author={Hoang Vu and Nataliia Klievtsova and Henrik Leopold and Stefanie Rinderle-Ma and Timotheus Kampik},
      year={2025},
booktitle    = {Responsible BPM Forum, 23rd International Conference Business Process Management ({BPM}), Sevilla, Spain, September 1-5, 2025, Proceedings},
  series       = {Lecture Notes in Computer Science},
publisher    = {Springer}
}

@article{Gabriel25,
    author = {Iason Gabriel and Geoff Keeling and Arianna Manzini and James Evans},
    title = {We need a new ethics for a world of AI agents},
    journal = {Nature},
    volume = {644},
    year = {2025}
}

@inproceedings{ErhaborUNA25,
  author       = {Daniel Erhabor and
                  Sreeharsha Udayashankar and
                  Meiyappan Nagappan and
                  Samer Al{-}Kiswany},
  title        = {Measuring the Runtime Performance of {C++} Code Written by Humans
                  Using Github Copilot},
  booktitle    = {47th {IEEE/ACM} International Conference on Software Engineering,
                  {ICSE} 2025, Ottawa, ON, Canada, April 26 - May 6, 2025},
  pages        = {2062--2074},
  publisher    = {{IEEE}},
  year         = {2025},
  url          = {https://doi.org/10.1109/ICSE55347.2025.00059},
  doi          = {10.1109/ICSE55347.2025.00059},
  bibsource    = {dblp computer science bibliography, https://dblp.org}
}

@inproceedings{HassanLRGC00TOL24,
  author       = {Ahmed E. Hassan and
                  Dayi Lin and
                  Gopi Krishnan Rajbahadur and
                  Keheliya Gallaba and
                  Filipe Roseiro C{\^{o}}go and
                  Boyuan Chen and
                  Haoxiang Zhang and
                  Kishanthan Thangarajah and
                  Gustavo Ansaldi Oliva and
                  Jiahuei (Justina) Lin and
                  Wali Mohammad Abdullah and
                  Zhen Ming (Jack) Jiang},
  editor       = {Marcelo d'Amorim},
  title        = {Rethinking Software Engineering in the Era of Foundation Models: {A}
                  Curated Catalogue of Challenges in the Development of Trustworthy
                  FMware},
  booktitle    = {Companion Proceedings of the 32nd {ACM} International Conference on
                  the Foundations of Software Engineering, {FSE} 2024, Porto de Galinhas,
                  Brazil, July 15-19, 2024},
  pages        = {294--305},
  publisher    = {{ACM}},
  year         = {2024},
  url          = {https://doi.org/10.1145/3663529.3663849},
  doi          = {10.1145/3663529.3663849},
  bibsource    = {dblp computer science bibliography, https://dblp.org}
}

@article{Sapkota25,
  author       = {Ranjan Sapkota and
                  Konstantinos I. Roumeliotis and
                  Manoj Karkee},
  title        = {{AI} Agents vs. Agentic {AI:} {A} Conceptual Taxonomy, Applications
                  and Challenges},
  journal      = {CoRR},
  volume       = {abs/2505.10468},
  year         = {2025},
  url          = {https://doi.org/10.48550/arXiv.2505.10468},
  doi          = {10.48550/ARXIV.2505.10468},
  eprinttype    = {arXiv},
  eprint       = {2505.10468},
  bibsource    = {dblp computer science bibliography, https://dblp.org}
}

@inproceedings{DBLP:conf/cain/BarnettKTB024,
  author       = {Scott Barnett and
                  Stefanus Kurniawan and
                  Srikanth Thudumu and
                  Zach Brannelly and
                  Mohamed Abdelrazek},
  title        = {Seven Failure Points When Engineering a Retrieval Augmented Generation
                  System},
  booktitle    = {{CAIN}},
  pages        = {194--199},
  publisher    = {{ACM}},
  year         = {2024}
}

@inproceedings{DBLP:conf/www/XuZRZYYXW25,
  author       = {Da Xu and
                  Danqing Zhang and
                  Chuanwei Ruan and
                  Lingling Zheng and
                  Bo Yang and
                  Guangyu Yang and
                  Shuyuan Xu and
                  Haixun Wang},
  title        = {Tutorial on Landing Generative {AI} in Industrial Social and E-commerce
                  Recsys},
  booktitle    = {{WWW} (Companion Volume)},
  pages        = {57--60},
  publisher    = {{ACM}},
  year         = {2025}
}

@inproceedings{DBLP:conf/cain/00010XLX024,
  author       = {Qinghua Lu and
                  Liming Zhu and
                  Xiwei Xu and
                  Yue Liu and
                  Zhenchang Xing and
                  Jon Whittle},
  title        = {A Taxonomy of Foundation Model based Systems through the Lens of Software
                  Architecture},
  booktitle    = {{CAIN}},
  pages        = {1--6},
  publisher    = {{ACM}},
  year         = {2024}
}

@inproceedings{DBLP:conf/www/Bakharia25,
  author       = {Aneesha Bakharia},
  title        = {Iterative Proof-Driven Development {LLM} Prompt},
  booktitle    = {{WWW} (Companion Volume)},
  pages        = {1596--1597},
  publisher    = {{ACM}},
  year         = {2025}
}

@article{DBLP:journals/pacmse/LiangLRM25,
  author       = {Jenny T. Liang and
                  Melissa Lin and
                  Nikitha Rao and
                  Brad A. Myers},
  title        = {Prompts Are Programs Too! Understanding How Developers Build Software
                  Containing Prompts},
  journal      = {Proc. {ACM} Softw. Eng.},
  volume       = {2},
  number       = {{FSE}},
  pages        = {1591--1614},
  year         = {2025}
}

@inproceedings{DBLP:conf/chi/FengYY0ZL24,
  author       = {Li Feng and
                  Ryan Yen and
                  Yuzhe You and
                  Mingming Fan and
                  Jian Zhao and
                  Zhicong Lu},
  title        = {CoPrompt: Supporting Prompt Sharing and Referring in Collaborative
                  Natural Language Programming},
  booktitle    = {{CHI}},
  pages        = {934:1--934:21},
  publisher    = {{ACM}},
  year         = {2024}
}

@inproceedings{DBLP:conf/iticse/Aveni0FH25,
  author       = {Timothy J. Aveni and
                  James Smith and
                  Armando Fox and
                  Bj{\"{o}}rn Hartmann},
  title        = {Supporting Students in Prototyping AI-backed Software with Hosted
                  Prompt Template APIs},
  booktitle    = {ITiCSE {(1)}},
  pages        = {65--71},
  publisher    = {{ACM}},
  year         = {2025}
}

@inproceedings{DBLP:conf/satrends/BecattiniVV25,
  author       = {Marco Becattini and
                  Roberto Verdecchia and
                  Enrico Vicario},
  title        = {{SALLMA:} {A} Software Architecture for LLM-Based Multi-Agent Systems},
  booktitle    = {SATrends@ICSE},
  pages        = {5--8},
  publisher    = {{IEEE}},
  year         = {2025}
}

@INPROCEEDINGS {11121725,
    author = { Nahar, Nadia and Kastner, Christian and Butler, Jenna and Parnin, Chris and Zimmermann., Thomas and Bird, Christian },
    booktitle = { 2025 IEEE/ACM 47th International Conference on Software Engineering: Software Engineering in Practice (ICSE-SEIP) },
    title = {{ Beyond the Comfort Zone: Emerging Solutions to Overcome Challenges in Integrating LLMs into Software Products }},
    year = {2025},
    volume = {},
    ISSN = {},
    pages = {516-527},
    doi = {10.1109/ICSE-SEIP66354.2025.00051},
    url = {https://doi.ieeecomputersociety.org/10.1109/ICSE-SEIP66354.2025.00051},
    publisher = {IEEE Computer Society},
    address = {Los Alamitos, CA, USA},
    month =May
}

@inproceedings{DBLP:conf/csit/Joshi24,
  author       = {Hrishikesh Joshi},
  title        = {Secure Software Architecture for Enterprise Generative Artificial
                  Intelligence},
  booktitle    = {{CSIT}},
  pages        = {1--5},
  publisher    = {{IEEE}},
  year         = {2024}
}

@INPROCEEDINGS {11127266,
    author = { Min, Ziran and Budnik, Christof J. },
    booktitle = { 2025 IEEE International Conference on Artificial Intelligence Testing (AITest) },
    title = {{ Verification and Validation of LLM-RAG for Industrial Automation }},
    year = {2025},
    volume = {},
    ISSN = {},
    pages = {50-53},
    doi = {10.1109/AITest66680.2025.00012},
    url = {https://doi.ieeecomputersociety.org/10.1109/AITest66680.2025.00012},
    publisher = {IEEE Computer Society},
    address = {Los Alamitos, CA, USA},
    month =Jul
}

@inproceedings{DBLP:conf/icsa/HeckingSF25,
  author       = {Tobias Hecking and
                  Thorsten Sommer and
                  Michael Felderer},
  title        = {An Architecture and Protocol for Decentralized Retrieval Augmented
                  Generation},
  booktitle    = {{ICSA} Companion},
  pages        = {31--35},
  publisher    = {{IEEE}},
  year         = {2025}
}

@inproceedings{DBLP:conf/eit/LanKPPP24,
  author       = {Qianlong Lan and
                  Anuj Kaul and
                  Nishant Kumar Das Pattanaik and
                  Piyush Pattanayak and
                  Vinothini Pandurangan},
  title        = {Securing Applications of Large Language Models: {A} Shift-Left Approach},
  booktitle    = {{EIT}},
  pages        = {1--2},
  publisher    = {{IEEE}},
  year         = {2024}
}

@inproceedings{DBLP:conf/icsa/BucaioniWHL025,
  author       = {Alessio Bucaioni and
                  Martin Weyssow and
                  Junda He and
                  Yunbo Lyu and
                  David Lo},
  title        = {A Functional Software Reference Architecture for LLM-Integrated Systems},
  booktitle    = {{ICSA} Companion},
  pages        = {1--5},
  publisher    = {{IEEE}},
  year         = {2025}
}

@inproceedings{DBLP:conf/icse/Shao0S00025,
  author       = {Yuchen Shao and
                  Yuheng Huang and
                  Jiawei Shen and
                  Lei Ma and
                  Ting Su and
                  Chengcheng Wan},
  title        = {Are LLMs Correctly Integrated into Software Systems?},
  booktitle    = {{ICSE}},
  pages        = {1178--1190},
  publisher    = {{IEEE}},
  year         = {2025}
}

@INPROCEEDINGS {11113067,
    author = { Fan, Wenqi and Wu, Pangjing and Ding, Yujuan and Ning, Liangbo and Wang, Shijie and Li, Qing },
    booktitle = { 2025 IEEE 41st International Conference on Data Engineering (ICDE) },
    title = {{ Towards Retrieval-Augmented Large Language Models: Data Management and System Design }},
    year = {2025},
    volume = {},
    ISSN = {},
    pages = {4509-4512},
    doi = {10.1109/ICDE65448.2025.00341},
    url = {https://doi.ieeecomputersociety.org/10.1109/ICDE65448.2025.00341},
    publisher = {IEEE Computer Society},
    address = {Los Alamitos, CA, USA},
    month =May
}

@inproceedings{DBLP:conf/icsa/LuZXXHW24,
  author       = {Qinghua Lu and
                  Liming Zhu and
                  Xiwei Xu and
                  Zhenchang Xing and
                  Stefan Harrer and
                  Jon Whittle},
  title        = {Towards Responsible Generative {AI:} {A} Reference Architecture for
                  Designing Foundation Model Based Agents},
  booktitle    = {{ICSA-C}},
  pages        = {119--126},
  publisher    = {{IEEE}},
  year         = {2024}
}

@INPROCEEDINGS{11082094,
  author={Chen, Jianhui and Peng, Yunchao},
  booktitle={2025 8th International Conference on Artificial Intelligence and Big Data (ICAIBD)},
  title={Development of AI Agent Based on Large Language Model Platforms},
  year={2025},
  volume={},
  number={},
  pages={864-868},
  doi={10.1109/ICAIBD64986.2025.11082094}
}

@inproceedings{DBLP:conf/icsa/KholkarTPR24,
  author       = {Deepali Kholkar and
                  Suraj Thapa and
                  Akhilesh Pal and
                  Suman Roychoudhury},
  title        = {Feature Model-based Integration of Machine Learning in Software Product
                  Lines},
  booktitle    = {{ICSA-C}},
  pages        = {295--302},
  publisher    = {{IEEE}},
  year         = {2024}
}

@INPROCEEDINGS{10771003,
  author={Elvira, Timothy and Procko, Tyler Thomas and Ochoa, Omar},
  booktitle={2024 Conference on AI, Science, Engineering, and Technology (AIxSET)},
  title={Requirements Elicitation for Machine Learning Applications: A Research Preview},
  year={2024},
  volume={},
  number={},
  pages={218-221},
  doi={10.1109/AIxSET62544.2024.00042}
}

@article{DBLP:journals/access/AminiranjbarTWPV25,
  author       = {Zahra Aminiranjbar and
                  Jianan Tang and
                  Qiudan Wang and
                  Shubha Pant and
                  Mahesh Viswanathan},
  title        = {{DAWN:} Designing Distributed Agents in a Worldwide Network},
  journal      = {{IEEE} Access},
  volume       = {13},
  pages        = {138795--138812},
  year         = {2025}
}

@article{DBLP:journals/csur/JiLFYSXIBMF23,
  author       = {Ziwei Ji and
                  Nayeon Lee and
                  Rita Frieske and
                  Tiezheng Yu and
                  Dan Su and
                  Yan Xu and
                  Etsuko Ishii and
                  Yejin Bang and
                  Andrea Madotto and
                  Pascale Fung},
  title        = {Survey of Hallucination in Natural Language Generation},
  journal      = {{ACM} Comput. Surv.},
  volume       = {55},
  number       = {12},
  pages        = {248:1--248:38},
  year         = {2023},
  url          = {https://doi.org/10.1145/3571730},
  doi          = {10.1145/3571730},
  bibsource    = {dblp computer science bibliography, https://dblp.org}
}

@inproceedings{cartaxo2018,
author = {Cartaxo, Bruno and Pinto, Gustavo and Soares, Sergio},
title = {The Role of Rapid Reviews in Supporting Decision-Making in Software Engineering Practice},
year = {2018},
isbn = {9781450364034},
publisher = {Association for Computing Machinery},
address = {New York, NY, USA},
url = {https://doi.org/10.1145/3210459.3210462},
doi = {10.1145/3210459.3210462},
booktitle = {Proceedings of the 22nd International Conference on Evaluation and Assessment in Software Engineering 2018},
pages = {24–34},
numpages = {11},
location = {Christchurch, New Zealand},
series = {EASE '18}
}

@article{Amalfitano_rapidreview,
title = {Alternatives for testing of context-aware software systems in non-academic settings: results from a Rapid Review},
journal = {Information and Software Technology},
volume = {149},
pages = {106937},
year = {2022},
issn = {0950-5849},
doi = {https://doi.org/10.1016/j.infsof.2022.106937},
url = {https://www.sciencedirect.com/science/article/pii/S0950584922000878},
author = {Santiago Matalonga and Domenico Amalfitano and Andrea Doreste and Anna Rita Fasolino and Guilherme Horta Travassos}
}

@article{sami2024experimenting,
  title={Experimenting with multi-agent software development: Towards a unified platform},
  author={Sami, Malik Abdul and Waseem, Muhammad and Rasheed, Zeeshan and Saari, Mika and Syst{\"a}, Kari and Abrahamsson, Pekka},
  journal={arXiv preprint arXiv:2406.05381},
  year={2024}
}

@inproceedings{10.1145/3696630.3728493,
author = {Zhang, Yiran and Li, Ruiyin and Liang, Peng and Sun, Weisong and Liu, Yang},
title = {Knowledge-Based Multi-Agent Framework for Automated Software Architecture Design},
year = {2025},
isbn = {9798400712760},
publisher = {Association for Computing Machinery},
address = {New York, NY, USA},
url = {https://doi.org/10.1145/3696630.3728493},
doi = {10.1145/3696630.3728493},
booktitle = {Proceedings of the 33rd ACM International Conference on the Foundations of Software Engineering},
pages = {530–534},
numpages = {5},
location = {Clarion Hotel Trondheim, Trondheim, Norway},
series = {FSE Companion '25}
}

@article{bouzenia2024repairagent,
  title={Repairagent: An autonomous, llm-based agent for program repair},
  author={Bouzenia, Islem and Devanbu, Premkumar and Pradel, Michael},
  journal={arXiv preprint arXiv:2403.17134},
  year={2024}
}

@article{qian2023chatdev,
  title={Chatdev: Communicative agents for software development},
  author={Qian, Chen and Liu, Wei and Liu, Hongzhang and Chen, Nuo and Dang, Yufan and Li, Jiahao and Yang, Cheng and Chen, Weize and Su, Yusheng and Cong, Xin and others},
  journal={arXiv preprint arXiv:2307.07924},
  year={2023}
}

@article{lin2024soen,
  title={Soen-101: Code generation by emulating software process models using large language model agents},
  author={Lin, Feng and Kim, Dong Jae and others},
  journal={arXiv preprint arXiv:2403.15852},
  year={2024}
}

@inbook{prophetAgent,
author = {Kong, Qichao and Lv, Zhengwei and Xiong, Yiheng and Wang, Dingchun and Sun, Jingling and Su, Ting and Li, Letao and Yang, Xu and Huo, Gang},
title = {ProphetAgent: Automatically Synthesizing GUI Tests from Test Cases in Natural Language for Mobile Apps},
year = {2025},
isbn = {9798400712760},
publisher = {Association for Computing Machinery},
address = {New York, NY, USA},
url = {https://doi.org/10.1145/3696630.3728543},
booktitle = {Proceedings of the 33rd ACM International Conference on the Foundations of Software Engineering},
pages = {174–179},
numpages = {6}
}

@article{khan2025ai,
  title={AI-Driven Automation in Agile Development: Multi-Agent LLMs for Software Engineering},
  author={Khan, Salman and Daviglus, Mendus},
  year={2025}
}

@inbook{10.1145/3696630.3728717,
author = {Ronanki, Krishna},
title = {Facilitating Trustworthy Human-Agent Collaboration in LLM-based Multi-Agent System oriented Software Engineering},
year = {2025},
isbn = {9798400712760},
publisher = {Association for Computing Machinery},
address = {New York, NY, USA},
url = {https://doi.org/10.1145/3696630.3728717},
booktitle = {Proceedings of the 33rd ACM International Conference on the Foundations of Software Engineering},
pages = {1333–1337},
numpages = {5}
}

@article{rondon2025evaluating,
  title={Evaluating agent-based program repair at google},
  author={Rondon, Pat and Wei, Renyao and Cambronero, Jos{\'e} and Cito, J{\"u}rgen and Sun, Aaron and Sanyam, Siddhant and Tufano, Michele and Chandra, Satish},
  journal={arXiv preprint arXiv:2501.07531},
  year={2025}
}

@inproceedings{10.1145/3718751.3718843,
author = {Jin, Wei and Chen, Hongzhi and Wang, Xiaoyan and Gong, Benru and Lin, Xiufeng},
title = {An AI-native application assemble platform for easy-integrating of AIGC based services},
year = {2025},
isbn = {9798400709753},
publisher = {Association for Computing Machinery},
address = {New York, NY, USA},
url = {https://doi.org/10.1145/3718751.3718843},
doi = {10.1145/3718751.3718843},
booktitle = {Proceedings of the 2024 4th International Conference on Big Data, Artificial Intelligence and Risk Management},
pages = {578–583},
numpages = {6},
location = {
},
series = {ICBAR '24}
}

@INPROCEEDINGS{11030040,
  author={Parthasarathy, Kannan and Vaidhyanathan, Karthik and Dhar, Rudra and Krishnamachari, Venkat and Kakran, Adyansh and Akshathala, Sreemaee and Arun, Shrikara and Karan, Amey and Muhammed, Basil and Dubey, Sumant and Veerubhotla, Mohan},
  booktitle={2025 IEEE/ACM 4th International Conference on AI Engineering – Software Engineering for AI (CAIN)}, 
  title={Engineering LLM Powered Multi-Agent Framework for Autonomous CloudOps}, 
  year={2025},
  volume={},
  number={},
  pages={201-211},
  doi={10.1109/CAIN66642.2025.00031}}

@article{manish2024autonomous,
  title={An autonomous multi-agent llm framework for agile software development},
  author={Manish, Sanwal},
  journal={International Journal of Trend in Scientific Research and Development},
  volume={8},
  number={5},
  pages={892--898},
  year={2024},
  publisher={IJTSRD}
}

@article{hu2024self,
  title={Self-evolving multi-agent collaboration networks for software development},
  author={Hu, Yue and Cai, Yuzhu and Du, Yaxin and Zhu, Xinyu and Liu, Xiangrui and Yu, Zijie and Hou, Yuchen and Tang, Shuo and Chen, Siheng},
  journal={arXiv preprint arXiv:2410.16946},
  year={2024}
}

@inproceedings{du2025multi,
  title={Multi-Agent Collaboration via Cross-Team Orchestration},
  author={Du, Zhuoyun and Qian, Chen and Liu, Wei and Xie, Zihao and Wang, Yifei and Qiu, Rennai and Dang, Yufan and Chen, Weize and Yang, Cheng and Tian, Ye and others},
  booktitle={Findings of the Association for Computational Linguistics: ACL 2025},
  pages={10386--10406},
  year={2025}
}

@inproceedings{multi-agentcollab,
author = {Wasif, Mubeen and Tunkel, David},
year = {2025},
month = {02},
pages = {},
title = {Multi-Agent Collaboration in AI: Enhancing Software Development with Autonomous LLMs},
doi = {10.13140/RG.2.2.31588.08328}
}

@misc{codepori,
      title={CodePori: Large-Scale System for Autonomous Software Development Using Multi-Agent Technology}, 
      author={Zeeshan Rasheed and Malik Abdul Sami and Kai-Kristian Kemell and Muhammad Waseem and Mika Saari and Kari Systä and Pekka Abrahamsson},
      year={2024},
      eprint={2402.01411},
      archivePrefix={arXiv},
      primaryClass={cs.SE},
      url={https://arxiv.org/abs/2402.01411}, 
}

@misc{rasheed2023autonomousagentssoftwaredevelopment,
      title={Autonomous Agents in Software Development: A Vision Paper}, 
      author={Zeeshan Rasheed and Muhammad Waseem and Kai-Kristian Kemell and Wang Xiaofeng and Anh Nguyen Duc and Kari Systä and Pekka Abrahamsson},
      year={2023},
      eprint={2311.18440},
      archivePrefix={arXiv},
      primaryClass={cs.SE},
      url={https://arxiv.org/abs/2311.18440}, 
}

@article{draxler2024ai,
   title={The AI ghostwriter effect: When users do not perceive ownership of AI-generated text but self-declare as authors},
   author={Draxler, Fiona and Werner, Anna and Lehmann, Florian and Hoppe, Matthias and Schmidt, Albrecht and Buschek, Daniel and Welsch, Robin},
   journal={ACM Transactions on Computer-Human Interaction},
   volume={31},
   number={2},
   pages={1--40},
   year={2024},
   publisher={ACM New York, NY}

}

@inproceedings{11052788,
  author       = {Minh Huynh Nguyen and
                  Thang Phan Chau and
                  Phong X. Nguyen and
                  Nghi D. Q. Bui},
  title        = {AgileCoder: Dynamic Collaborative Agents for Software Development
                  based on Agile Methodology},
  booktitle    = {{IEEE/ACM} Second International Conference on {AI} Foundation Models
                  and Software Engineering, Forge@ICSE 2025, Ottawa, ON, Canada, April
                  27-28, 2025},
  pages        = {156--167},
  publisher    = {{IEEE}},
  year         = {2025},
  url          = {https://doi.org/10.1109/Forge66646.2025.00026},
  doi          = {10.1109/FORGE66646.2025.00026},
  bibsource    = {dblp computer science bibliography, https://dblp.org}
}

@INPROCEEDINGS{10663037,
  author={Wei, Zhengyuan and Lee, Albert T.L. and Lee, Victor C.S. and Chan, Wing-Kwong},
  booktitle={2024 36th International Conference on Software Engineering Education and Training (CSEE\&T)}, 
  title={Toward AI-facilitated Learning Cycle in Integration Course Through Pair Programming with AI Agents}, 
  year={2024},
  volume={},
  number={},
  pages={1-5},
  doi={10.1109/CSEET62301.2024.10663037}}

@software{Zenodo,
  author       = {Andreas Metzger (contact person)},
  title        = {A Research Roadmap for Augmenting Software Engineering Processes and Software Products with Generative AI: Rapid Literature Review Results},
  month        = jan,
  year         = 2026,
  publisher    = {Zenodo},
  version      = {V1},
  doi          = {10.5281/zenodo.18345896},
}
